\documentclass{aa} 
\usepackage{graphicx}
\usepackage{lscape}
\usepackage{longtable}
\usepackage{arydshln}
\usepackage{multirow}
\usepackage{amsmath}
\usepackage{enumitem}
\usepackage{txfonts}
\usepackage{natbib}
\usepackage[htt]{hyphenat}

\bibpunct{(}{)}{;}{a}{}{,} 

\begin{document}

\title{AB Aur, a Rosetta stone for studies of planet formation (I): chemical study of a planet-forming disk} 
   \author{P. Rivi\`ere-Marichalar\inst{1}, A. Fuente\inst{1}, R. Le Gal\inst{2}, C. Baruteau\inst{3}, R. Neri\inst{4}, D. Navarro-Almaida\inst{1}, S. P. Trevi\~no-Morales\inst{5}, E. Mac\'ias\inst{6,7}, R. Bachiller\inst{1}, M. Osorio\inst{8}}
 \institute{Observatorio Astron\'omico Nacional (OAN,IGN), Calle Alfonso XII, 3. 28014 Madrid, Spain 
                   \email{p.riviere@oan.es}
 \and Center for Astrophysics, Harvard \& Smithsonian, 60 Garden St, Cambridge, MA 02138, USA
 \and CNRS / Institut de Recherche en Astrophysique et Plan\'etologie, 14 avenue Edouard Belin, F-31400 Toulouse, France 
 \and Institut de Radioastronomie Millim\'etrique, 300 rue de la Piscine, F-38406 Saint-Martin d'H\`eres, France 
 \and Chalmers University of Technology, Department of Space, Earth and Environment, SE-412 93 Gothenburg, Sweden 
 \and Joint ALMA Observatory, Alonso de C\'ordova 3107, Vitacura, Santiago 763-0355, Chile 
 \and European Southern Observatory,  Alonso de C\'ordova 3107, Vitacura, Santiago 763-0355, Chile 
 \and Instituto de Astrof\'isica de Andaluc\'ia, (CSIC) Glorieta de la Astronom\'ia s/n, E-18008 Granada, Spain 
   }
   \authorrunning{Rivi\`ere-Marichalar et al.}
   \date{}

 \abstract 
{\object{AB Aur} is a Herbig Ae star that hosts a prototypical transition disk. The disk shows a plethora of features connected with planet formation mechanisms, such as spiral arms, dust cavities, and dust traps. Understanding the physical and chemical characteristics of these features is crucial to advancing our knowledge of the planet formation processes.}
{We aim to characterize the gaseous disk around the Herbig Ae star AB Aur. A complete spectroscopic study was performed using NOEMA  to determine the physical and chemical conditions with high spatial resolution. }
{We present new NOrthern Extended Millimeter Array (NOEMA) interferometric observations of the continuum and  $^{12}$CO, $^{13}$CO, C$^{18}$O, H$_2$CO, and SO lines obtained at high resolution. We used the integrated intensity maps and stacked spectra to derive reliable estimates of the disk temperature. By combining our $^{13}$CO and C$^{18}$O observations, we computed the gas-to-dust ratio along the disk. We also derived column density maps for the different species and used them to compute abundance maps. The results of our observations were compared with a set of Nautilus astrochemical models to obtain insight into the disk properties.}
{We detected continuum emission in a ring that extends from 0.6$\arcsec$ to $\sim$2.0$\arcsec$, peaking at 0.97$\arcsec$ and with a strong azimuthal asymmetry. The molecules observed show different spatial distributions, and the peaks of the distributions are not correlated with the binding energy. Using H$_2$CO and SO lines, we derived a mean disk temperature of 39 K. We derived a gas-to-dust ratio that ranges from 10 to 40 along the disk. Abundance with respect to $^{13}$CO for SO ($\sim$2$\times 10^{-4}$) is almost one order of magnitude greater than the value derived for H$_2$CO (1.6$\times 10^{-5}$). The comparison with Nautilus models favors a disk with a low gas-to-dust ratio (40) and prominent sulfur depletion.}
{From a very complete spectroscopic study of the prototypical disk around AB Aur, we derived, for the first time, the gas temperature and the gas-to-dust ratio along the disk, providing information that is  essential to constraining hydrodynamical simulations. Moreover, we explored the  gas chemistry and, in particular, the sulfur depletion. The derived sulfur depletion is dependent on the assumed C/O ratio. Our data are better explained with C/O$\sim$0.7 and S/H=8$\rm \times 10^{-8}$. } 
{}

\titlerunning{AB Aur, a Rosetta stone for studies of planet formation (I)}
\maketitle

\section{Introduction} 

Planets are formed in circumstellar systems made of gas and dust, the so-called protoplanetary disks. The  precise mechanism that leads to planet formation remains an open question. We know that the gas in these systems is a key driver in their dynamical evolution. In particular, the dispersal of the disk gas determines the timescale for giant planet formation. However, key parameters such as the gas-to-dust ratio are to a large extent unknown. Therefore, characterizing the physical conditions and chemical composition of the gas is of paramount importance to understanding disk evolution. 

The chemical composition of the future planets will be inherited from that of the protoplanetary disk, and the chemical study of protoplanetary disks is a key strategy to understanding the diversity of planetary atmospheres.  Furthermore, gas is the primary material from which giant planets are made and its chemical composition sets the initial conditions for the planet composition. However, after decades of studying protoplanetary disks, their chemical composition is still poorly constrained.

While statistical studies are needed to tackle the question of planet formation, the detailed characterization of individual sources can provide deep insights into the physical properties and chemical reaction networks that are present in protoplanetary disks. The Herbig star AB Aur is a widely studied system hosting a transitional disk. Located at 162.9 $\pm$ 1.5 pc from the Sun \citep{GAIA2018}, it is well suited to studying the spatial distribution of gas and dust in the circumstellar environment in detail. The disk shows prominent emission from spiral arms at the near-infrared (NIR) and radio wavelength ranges, which could be explained by the presence of one or several forming giant planets \citep{Grady1999,Fukagawa2004,Hashimoto2011,Tang2012, Boccaletti2020}. The system also shows a cavity in continuum emission that extends from $\sim$70 to $\sim$100 au \citep{Pietu2005,Tang2012, Fuente2017}. Inside the cavity a compact source was detected by \cite{Tang2012}. The source could be an inner disk, needed to explain the high levels of accretion derived for the system \citep{GarciaLopez2006,Salyk2013}. The radio jet detected by \cite{Rodriguez2014} is also consistent with the observed levels of accretion. Finally, \cite{Tang2017} obtained images of the CO J = 2-1 line with ALMA and highlighted the presence of prominent spiral arms at a radius of 0$\arcsec .$3, which is within the dust cavity. However, no connection between the arms and the ring was observed in such images, most likely due to large structure filtering. 

This paper is part of a long-term study of which the main goal is to characterize the gas and dust of this prototypical disk. \cite{Fuente2017} presented high-spatial-resolution images of the continuum at 1.1 mm and 2.2 mm using the NOrthern Extended Millimeter Array (NOEMA). On the basis of the detailed comparison of such continuum images with two-fluid hydrodynamical simulations, it was suggested that the dust trap observed in AB Aur was associated with a decaying vortex. A similar model of a decaying vortex was proposed by \cite{Baruteau2019} to explain the asymmetric eccentric ring in the submillimeter continuum emission of the MWC 758 disk \citep{Dong2018}. Moreover, \cite{Fuente2017} provided an accurate measurement of the dust mass amounting to approximately 30 Earth masses. 

So far, a few molecular species have been detected in AB Aur, including CO, SO, HCO$^+$, HCN, and H$_2$CO \citep{Schreyer2008, Pietu2005, Tang2012, Tang2017, Fuente2010, Pacheco2015, Pacheco2016, Riviere2019}. It is remarkable that the first detection of SO in a protoplanetary disk was reported towards AB Aur \citep{Fuente2010, Pacheco2015, Pacheco2016}. Most species depicted ring-like emission co-spatial with the dust ring with a minimum towards the central dust gap, except $^{12}$CO and HCO$^+$. The higher angular resolution image of the HCO$^+$ 3$\rightarrow$2 line reported by \cite{Riviere2019} shows that within the dust cavity the HCO$^+$ emission is formed by a compact source towards the stellar position and at least one filament connecting the compact source with the outer ring. We concluded that the filamentary structure could be tracing an accretion flow from the outer ring into the inner disk.

In the present paper we report high-spatial-resolution NOEMA observations of  $^{12}$CO, $^{13}$CO, C$^{18}$O, H$_2$CO, and SO millimeter lines. The detection of several lines of H$_2$CO, and SO, allows us for the first time to constrain the physical conditions of the emitting gas. The focus of the present work is on the chemistry of the disk, and we leave a detailed study of the system kinematics for a future paper.

\section{Observations and data reduction}\label{Sec:obs_and_data_red}
The observations were performed using the NOEMA interferometer in its A and C configurations, with baselines in the range 16 to 375 m, aiming to achieve a maximum spatial resolution of $\rm \sim 0\arcsec.5$.

We used the PolyFiX correlator centered at 225 GHz with a channel width $\rm \Delta \nu$=2MHz for each sideband. Chunks with  $\rm \Delta \nu$=62.5 kHz were used to cover the requested transitions with higher spectral resolution, achieving spectral resolutions of $\sim$0.087 km s$\rm ^{-1}$. A total of 23 transitions from the different species were observed, including $\rm ^{12}$CO, $\rm ^{13}$CO, C$\rm ^{18}$O, $\rm ^{13}$C$\rm ^{17}$O, p-H$\rm _2$CO, SO, SiO, DCO$\rm ^+$, HC$\rm _3$N, OCS, CCS, $\rm ^{13}$CN,  $\rm ^{13}$CS, and CO$^+$ (see Table~\ref{Tab:transitions_observed}). We detected intense emission of the  J=2$\rightarrow$1 lines of $\rm ^{12}$CO, $\rm ^{13}$CO, C$\rm ^{18}$O, and the H$_2$CO 3$\rm _{03}$-2$\rm _{02}$ line and the SO  5$_6$-4$_5$ lines.  The H$_2$CO 3$\rm _{22}$-2$\rm _{21}$,  H$_2$CO 3$\rm _{21}$-2$\rm _{20}$ and SO  5$_5$-4$_4$ were detected at 3$\sigma$ level precluding synthesis imaging. Observations of the CO isotopologs $\rm ^{12}$CO, $\rm ^{13}$CO, and C$\rm ^{18}$O achieved sufficient S/N to allow for uniform weighting, reaching spatial resolutions of $\sim$0.$\arcsec$60$\times$0.$\arcsec$36 (98 au $\times$ 60 au at the distance to the source). The rest of the  observed species were mapped using natural weighting, resulting in resolutions $\sim$0.$\arcsec$73$\times$0.$\arcsec$47 (119 au $\times$ 77 au at the distance to the source).
Data reduction and map synthesis were done using the  \texttt{GILDAS}\footnote{See \texttt{http://www.iram.fr/IRAMFR/GILDAS} for  more information about the GILDAS  softwares.}\texttt{/CLASS} software.

\begin{table}
\caption{Overview of the different transitions observed}
\label{Tab:transitions_observed}
\centering
\begin{tabular}{lllll}
\hline \hline
Species & Trans. & Weight & Beam & RMS \\
-- & -- & -- & -- & mJy/beam \\
\hline
$\rm ^{12}$CO & 2-1 & Uniform & 0.$\arcsec$59$\times$0.$\arcsec$36 & 10.0 \\
$\rm ^{13}$CO & 2-1 & Uniform & 0.$\arcsec$63$\times$0.$\arcsec$39 & 9.5 \\
C$\rm ^{18}$O & 2-1 & Uniform & 0.$\arcsec$63$\times$0.$\arcsec$40 & 9.5 \\
H$\rm _2$CO & 3$\rm _{03}$-2$\rm _{02}$ & Natural & 0.$\arcsec$80$\times$0.$\arcsec$51 & 8.5 \\
-- & 3$\rm _{22}$-2$\rm _{21}$ & Natural & 0.$\arcsec$80$\times$0.$\arcsec$51 & 8.5 \\
-- & 3$\rm _{21}$-2$\rm _{20}$ & Natural & 0.$\arcsec$78$\times$0.$\arcsec$51 & 8.5 \\
SO & 5$_5$-4$_4$ & Natural & 0.$\arcsec$79$\times$0.$\arcsec$52& 9.0 \\
-- & 5$_6$-4$_5$ & Natural & 0.$\arcsec$79$\times$0.$\arcsec$51 & 8.5 \\
$\rm ^{13}$CN & 2-1 & Natural & 0.$\arcsec$80$\times$0.$\arcsec$52 & 8.5 \\
DCO$^+$ & 3-2 & Natural & 0.$\arcsec$79$\times$0.$\arcsec$52 & 9.1 \\
OCS & 18-17 & Natural & 0.$\arcsec$80$\times$0.$\arcsec$51 & 8.4 \\
SiO & 5-4 & Natural & 0.$\arcsec$80$\times$0.$\arcsec$51 & 8.6 \\
$\rm ^{13}$CS & 5-4 & Natural & 0.$\arcsec$72$\times$0.$\arcsec$43 & 9.0 \\
CO$^+$ & 2-1 & Natural & 0.$\arcsec$71$\times$0.$\arcsec$46 & 10.0 \\
OCS & 19-18 & Natural & 0.$\arcsec$72$\times$0.$\arcsec$47 & 9.0 \\
CCS & 8$_8$-7$_8$ & Natural & 0.$\arcsec$72$\times$0.$\arcsec$47 & 9.0 \\
HC$\rm _3$N & 24-23 & Natural & 0.$\arcsec$80$\times$0.$\arcsec$52 & 8.5 \\
\hline
\end{tabular}
\end{table} 

\section{Results}
Using the maps with the weightings as described in Sect. \ref{Sec:obs_and_data_red}, we study in the following the spatial distribution of the continuum and molecular emissions.

\subsection{Continuum emission}\label{Sec:cont_emission}
\begin{figure}[t!]
\begin{center}
 \includegraphics[width=0.4\textwidth,trim = 0mm 0mm 0mm 0mm,clip]{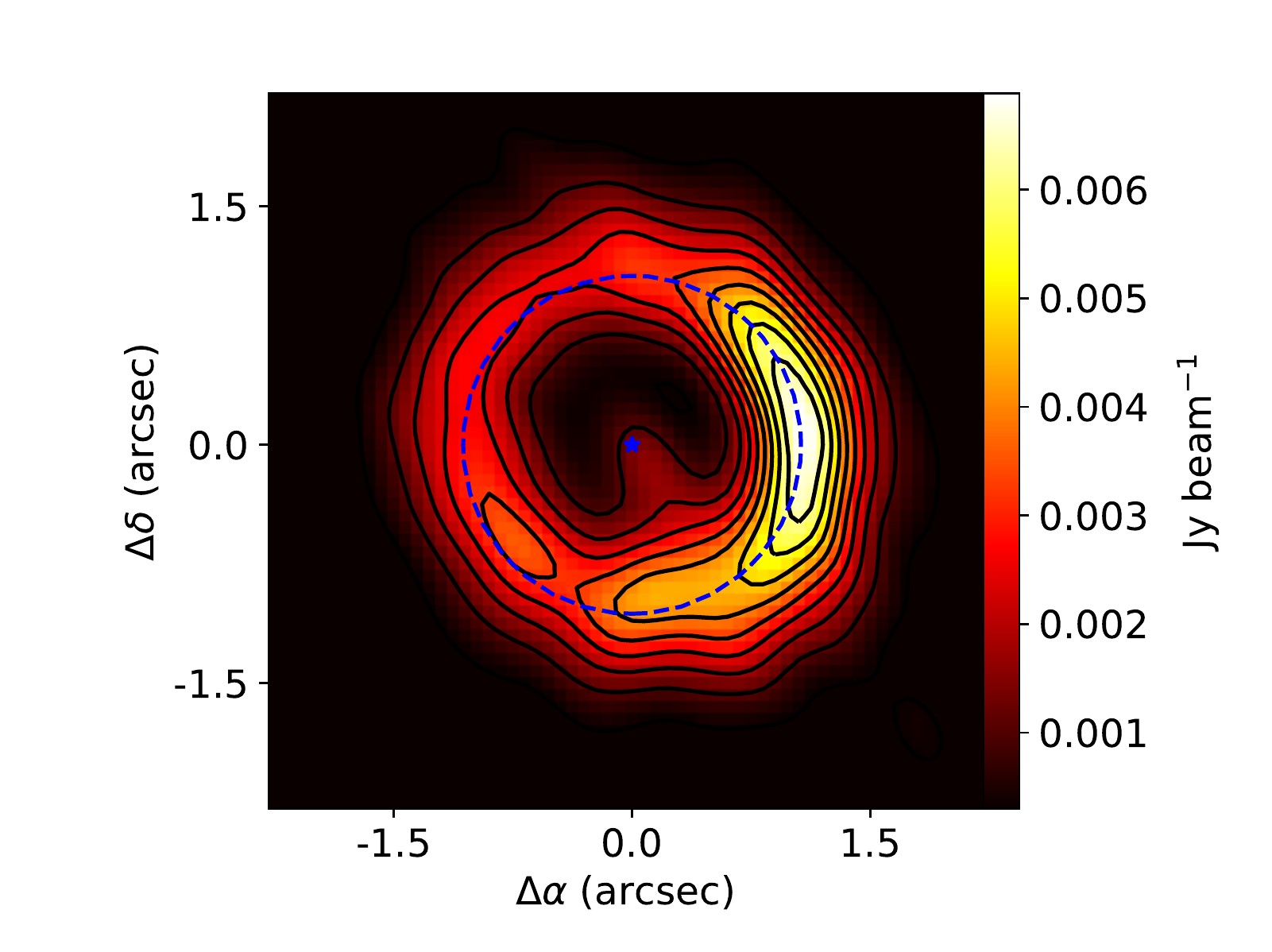}

 \includegraphics[width=0.4\textwidth,trim = 0mm 0mm 0mm 10mm,clip]{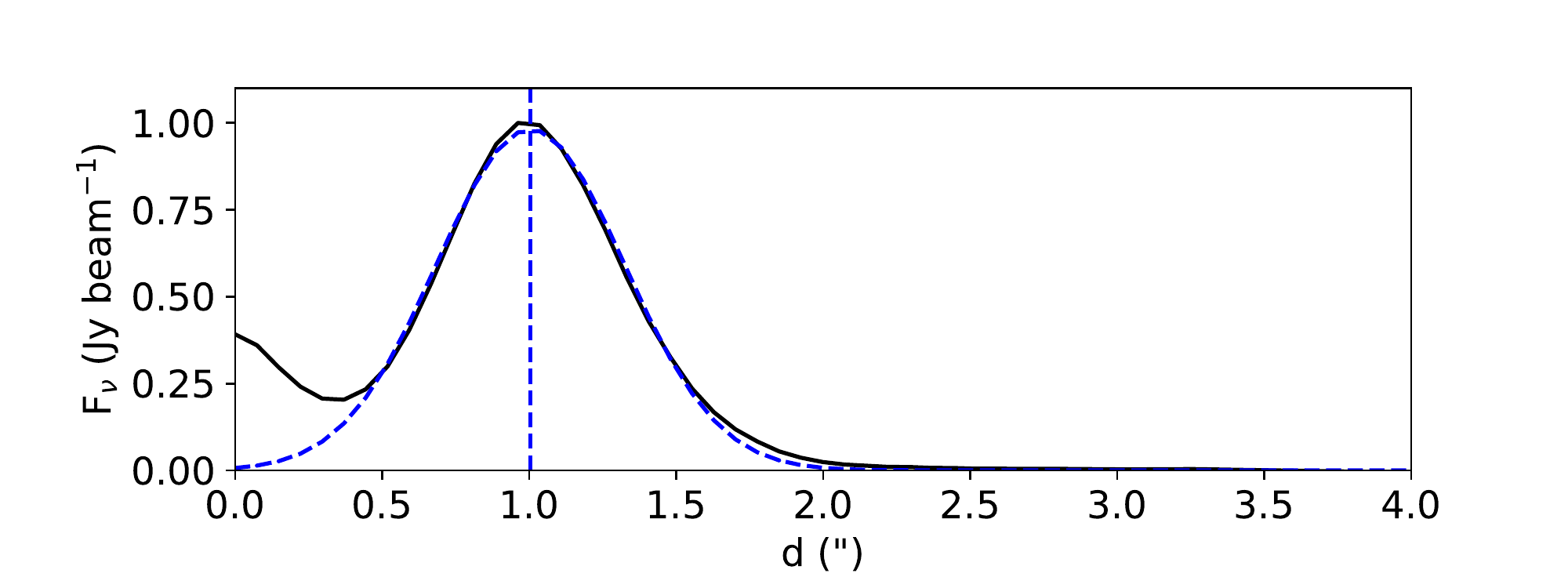}

 \includegraphics[width=0.4\textwidth,trim = 0mm 0mm 0mm 10mm,clip]{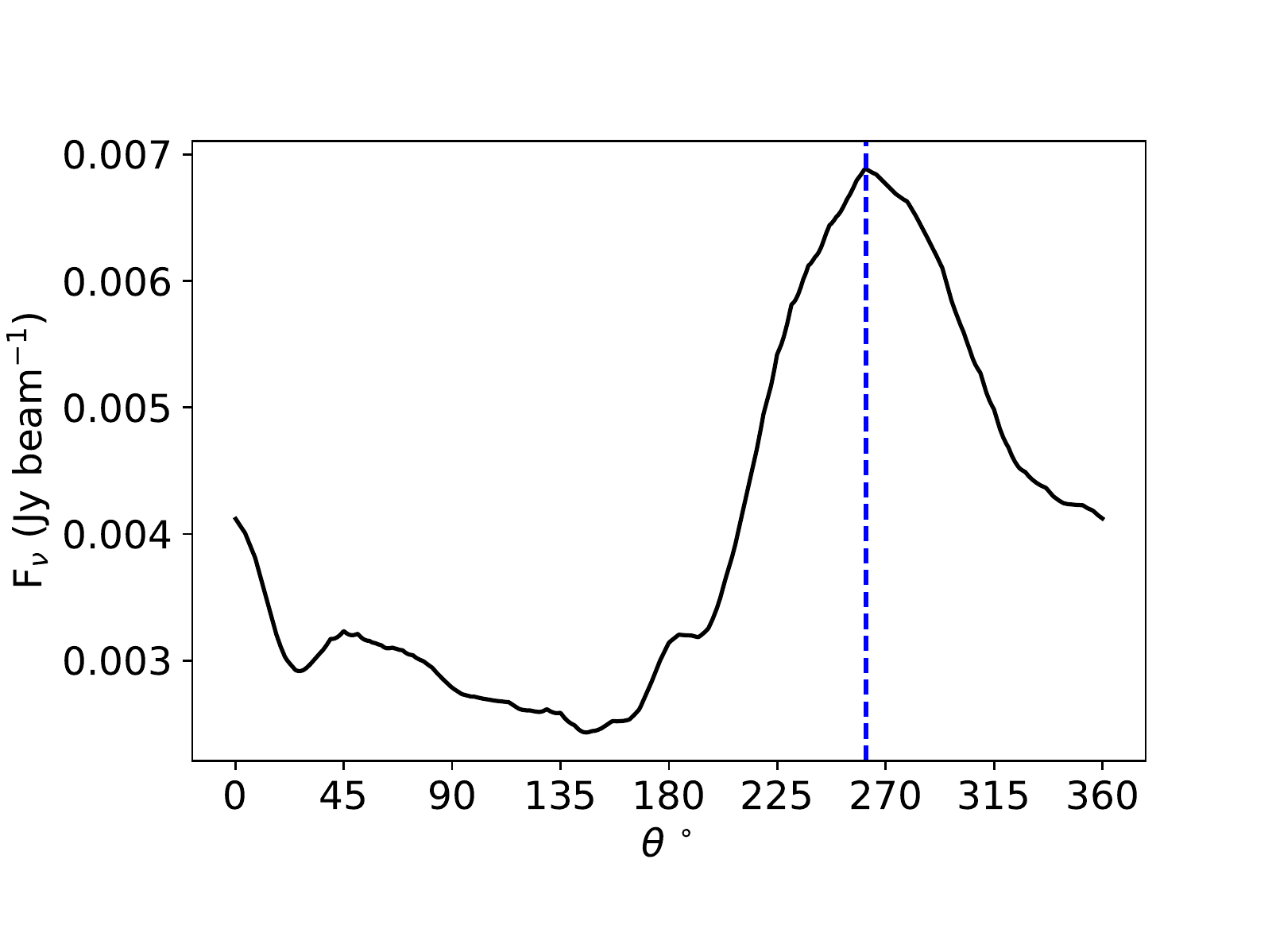}
  \caption{\textbf{Top}: Continuum intensity map. The blue dashed circle marks the position of the maximum in the radial direction, and the blue star marks the position of the centroid of the ellipse used to compute the inclination angle. \textbf{Middle:} Azimuthally averaged radial profile of continuum intensity map. The blue dashed curve shows a Gaussian fit to the profile. The blue dashed vertical line marks the position of the maximum in the radial direction. \textbf{Bottom:} Azimuthal cut along the radius of maximum intensity. The blue dashed vertical line marks the position of the maximum in the azimuthal direction.}
 \label{Fig:continuum_map}
\end{center}
\end{figure}

We show the continuum intensity map at 1mm in the top panel of Fig. \ref{Fig:continuum_map}. We measure a total flux of 97$\pm$19 mJy (assuming a calibration error of 20$\%$), close to the value reported by \cite{Tang2012} for their PdBI observations, and about six times brighter than the value reported by \cite{Tang2017}. As claimed by \cite{Tang2017}, their ALMA
data were not sensitive to emission extended on scales larger than 2$\arcsec$, which explains the difference in the recovered flux. The emission has a ring-like shape. By fitting an ellipse to the 10$\sigma$ emission level, we derive an inclination angle of 24.9$^\circ$. We detect 5$\sigma$ emission at distances as far as $\sim$2.3$\arcsec$ (373 au). The middle
panel  of Fig. \ref{Fig:continuum_map} shows the azimuthally averaged radial profile of the continuum emission at 1mm. The profile has an approximately Gaussian shape, with center at 0.97$\arcsec$ ($\sim$158 au) and FWHM $\sim$ 0.9$\arcsec$.  A cavity is observed starting at $r \sim 0\arcsec.6$ ($\sim$98 au). In the southern side of the disk ($\theta \sim$180$^{\circ}$) we observe emission extending from the central position to the ring at the 5$\sigma$ level. A strong azimuthal asymmetry in the form of a dust trap is clearly detected at PA $\theta \sim$ 270$^{\circ}$, as reported in previous studies \citep[see e.g.,][]{Tang2012}. The contrast ratio along the azimuthal direction in the position of the peak radius is $\rm \sim$ 2.6 (see Fig. \ref{Fig:continuum_map}, bottom panel).

\subsection{Molecular emission}  

\begin{figure}[t!]
\begin{center}
 \includegraphics[width=0.4\textwidth,trim = 0mm 10mm 0mm 0mm,clip]{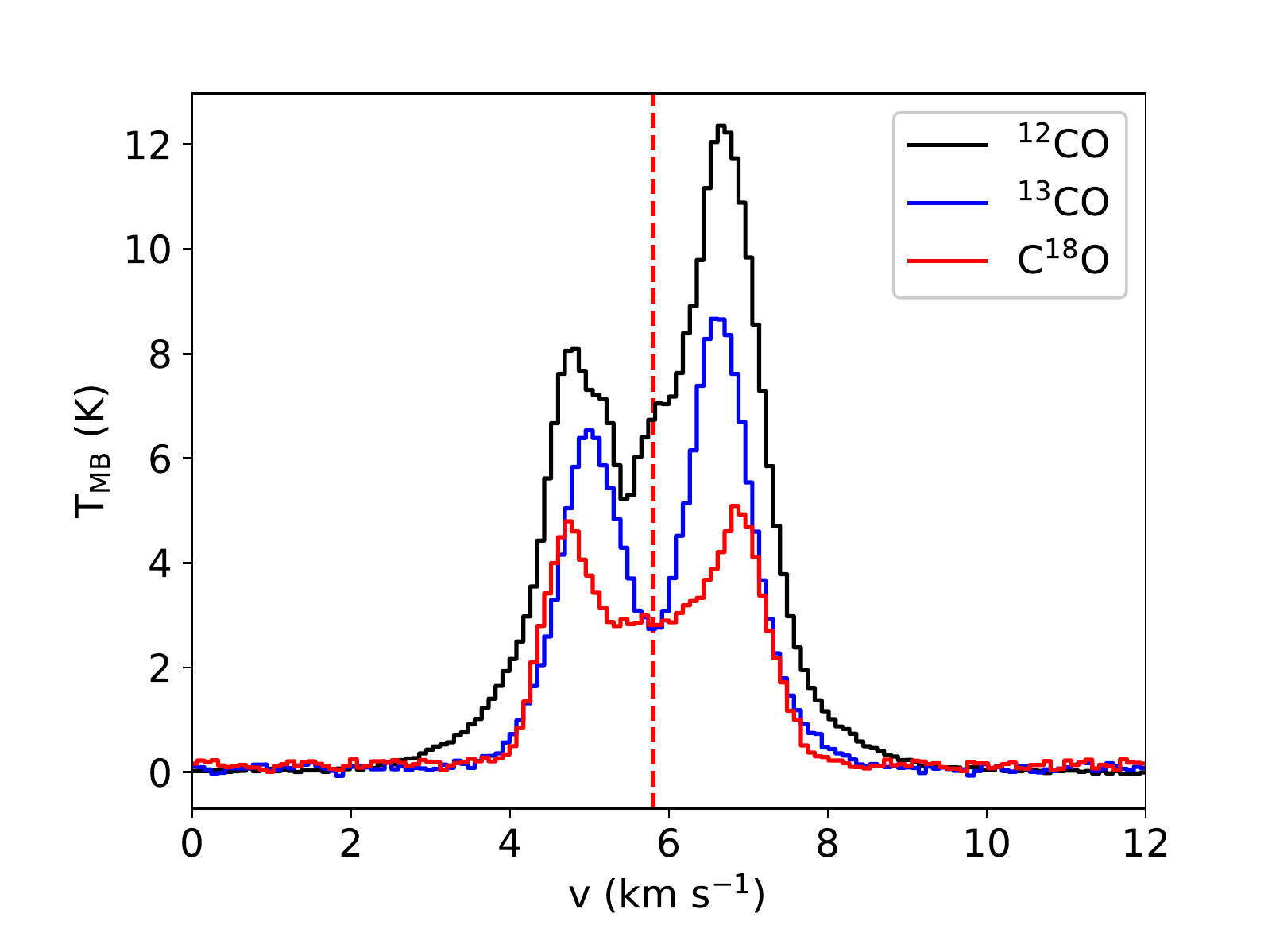}

 \includegraphics[width=0.4\textwidth,trim = 0mm 10mm 0mm 0mm,clip]{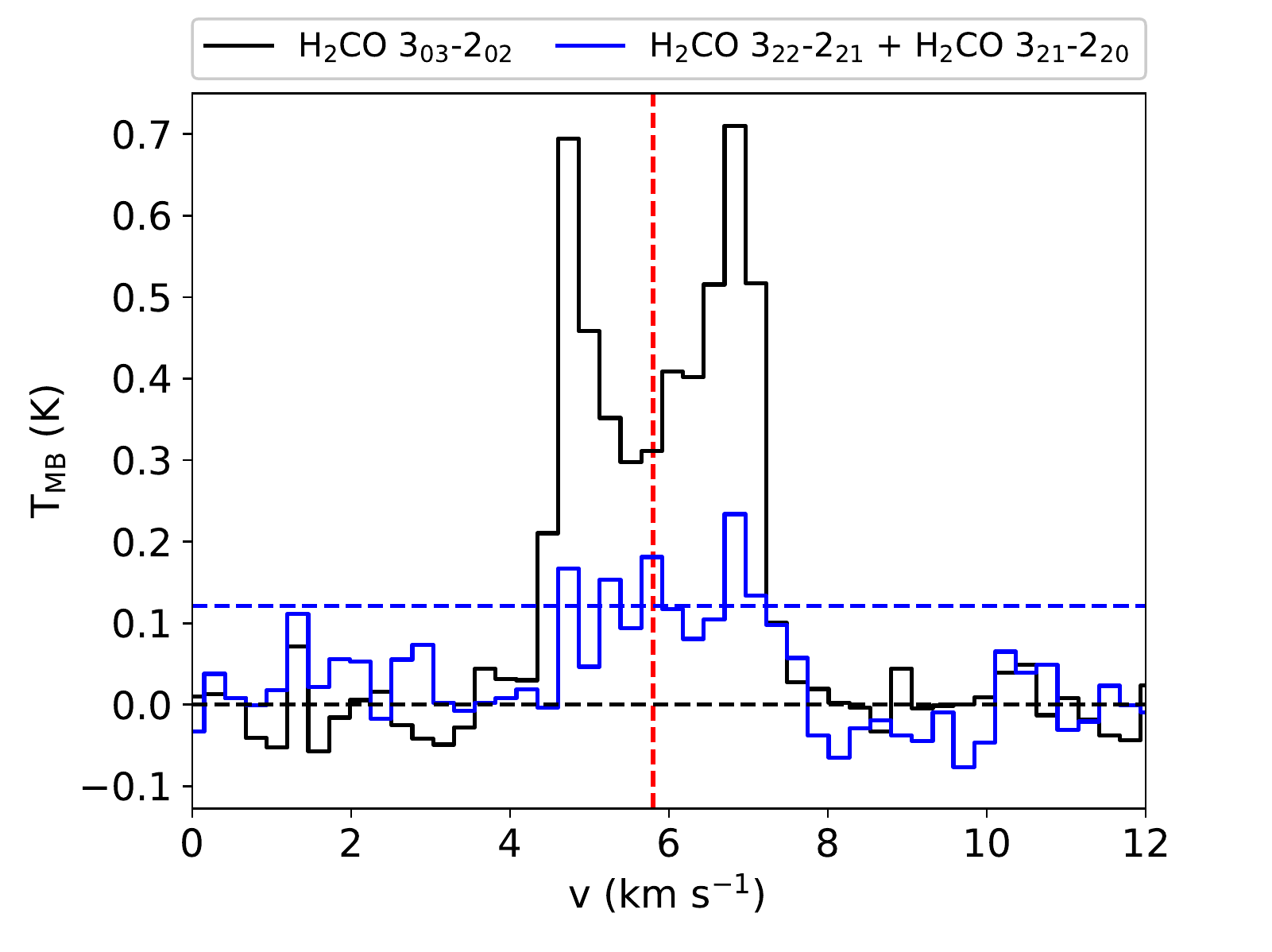}
 
 \includegraphics[width=0.4\textwidth,trim = 0mm 0mm 0mm 10mm,clip]{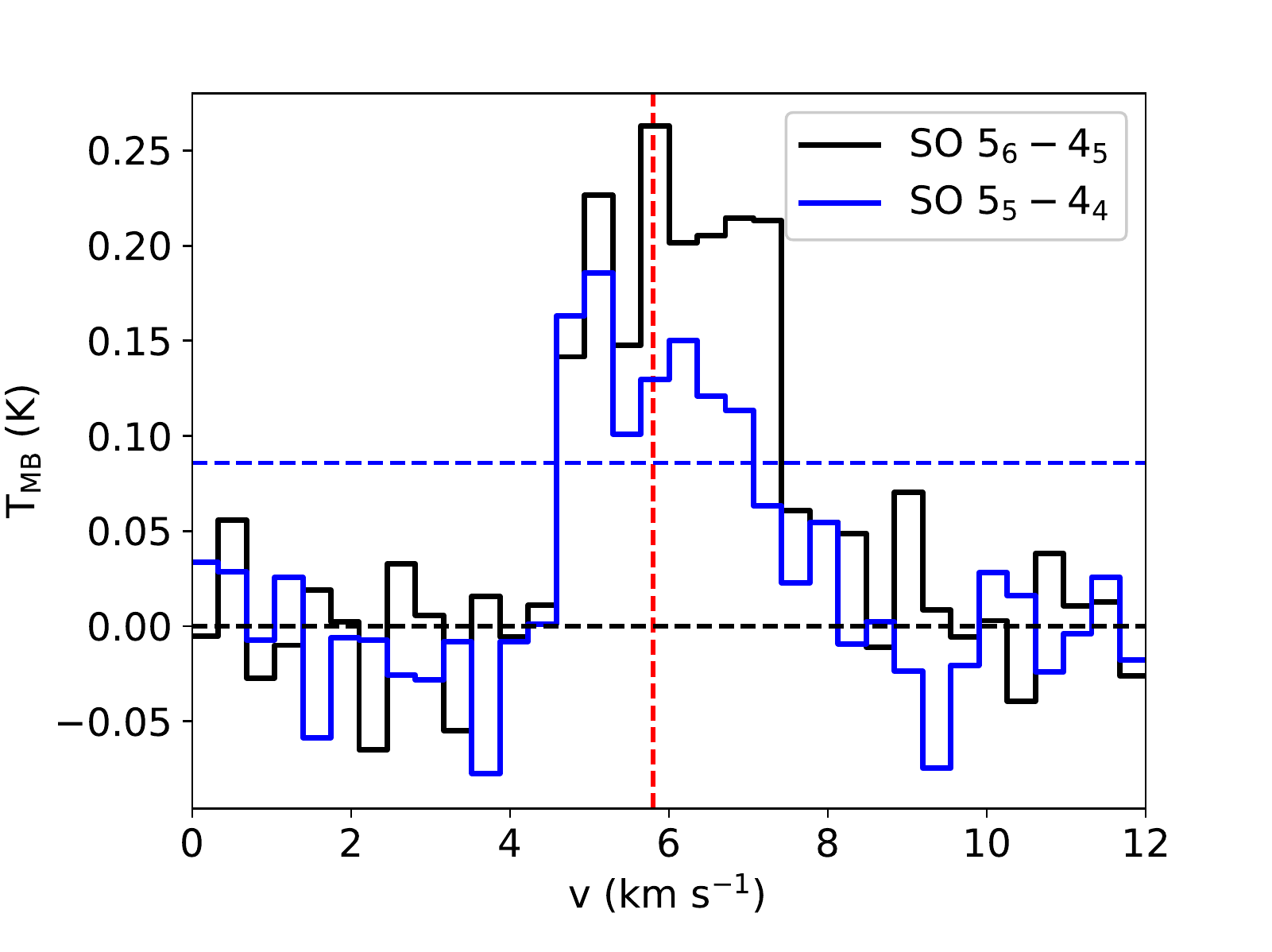} 
  \caption{{\bf Top:} $\rm ^{12}$CO, $\rm ^{13}$CO and C$\rm ^{18}$O spectra. {\bf Middle:} p-H$\rm _2$CO spectra. The black dashed horizontal line at T$\rm _{MB}$ = 0 and blue dashed line at 5$\sigma$ are also included. {\bf Bottom:} SO spectra after rebinning with n=4.}
 \label{Fig:stacked_spectra}
\end{center}
\end{figure}

\begin{figure*}[t!]
\begin{center}
 \includegraphics[width=0.25\textwidth, trim = 0mm 0mm 0mm 0mm,clip]{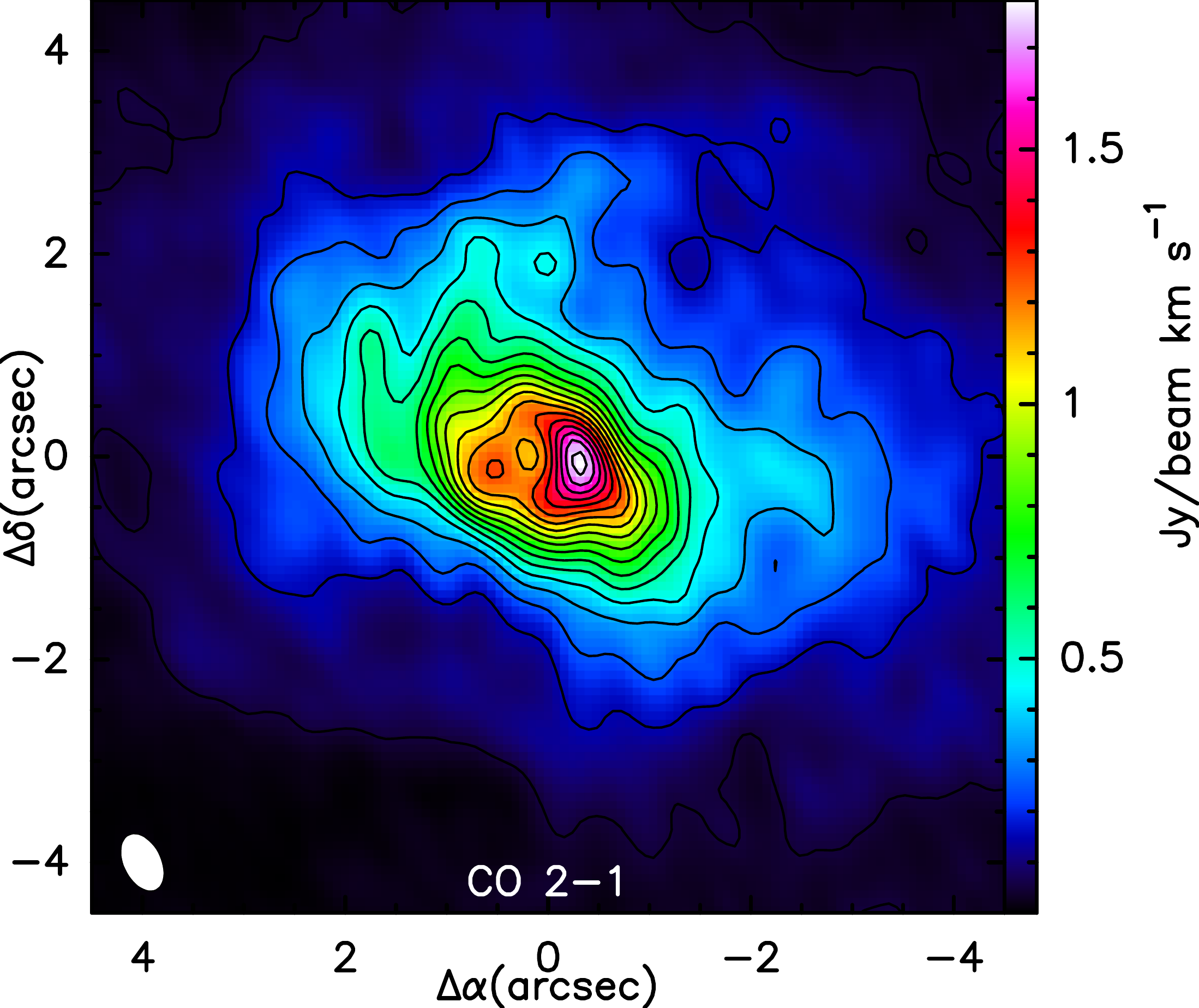}
 \includegraphics[width=0.25\textwidth, trim = 0mm 0mm 0mm 0mm,clip]{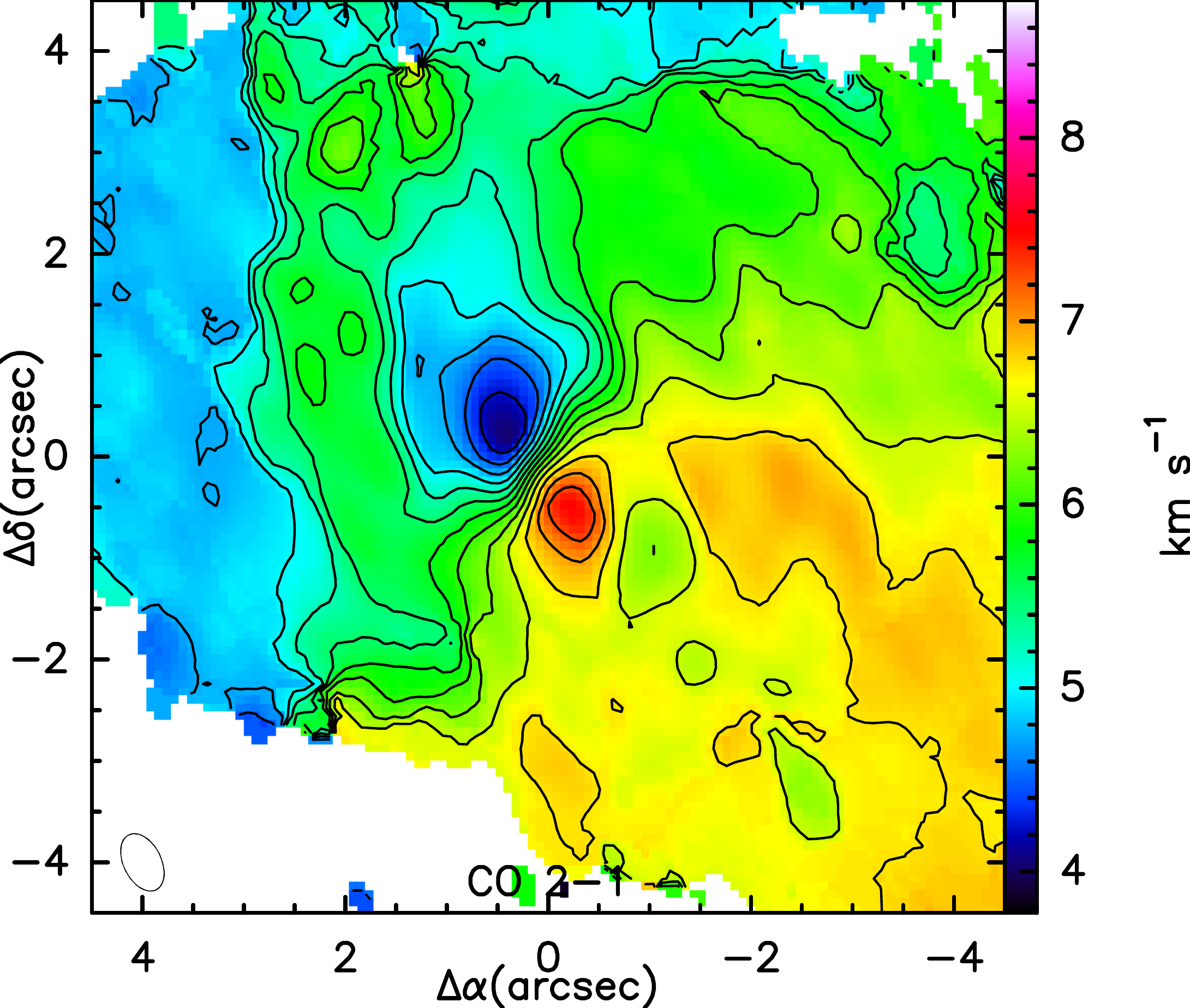}
 \includegraphics[width=0.25\textwidth, trim = 0mm 0mm 0mm 0mm,clip]{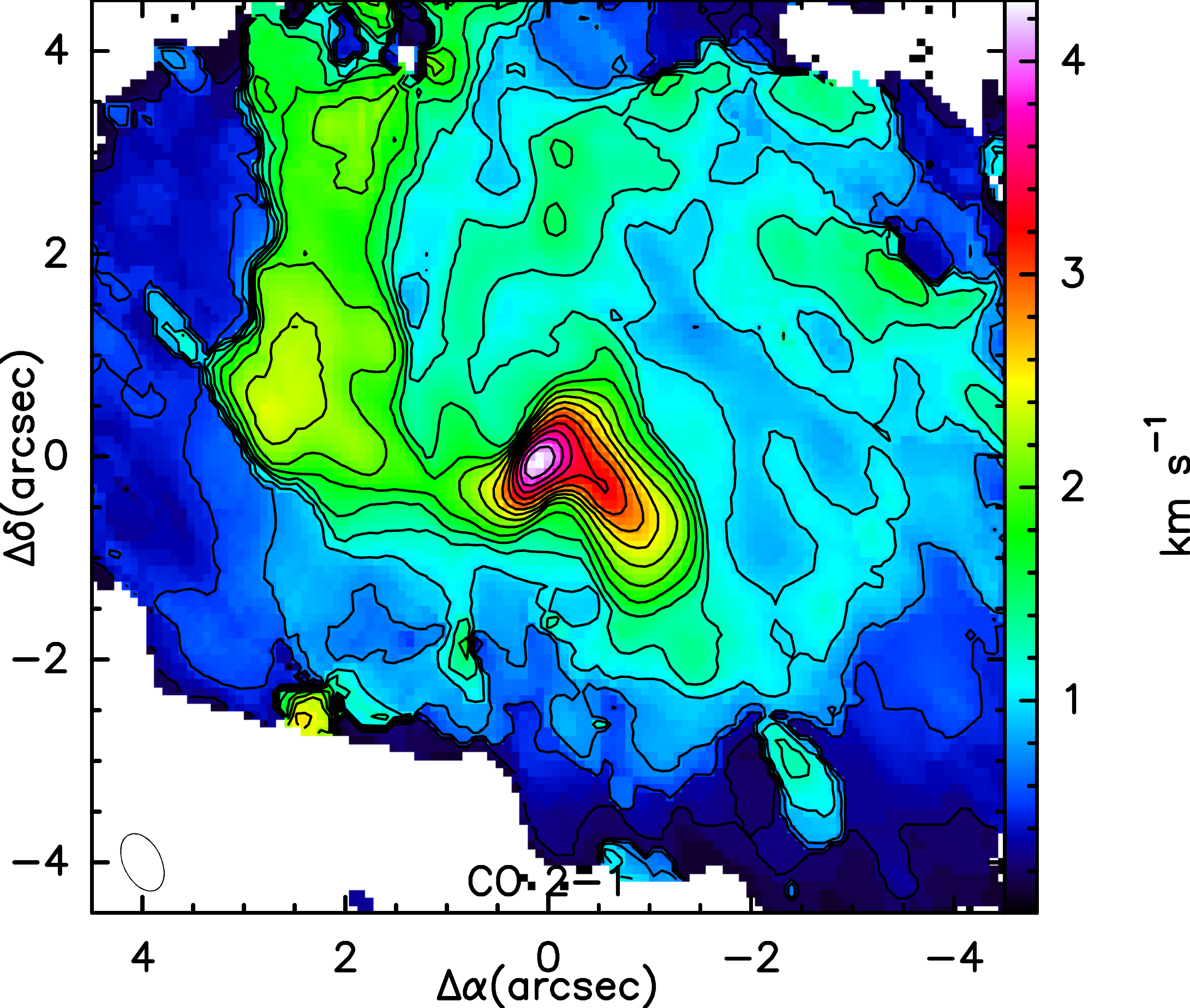}
  
 \includegraphics[width=0.25\textwidth, trim = 0mm 0mm 0mm 0mm,clip]{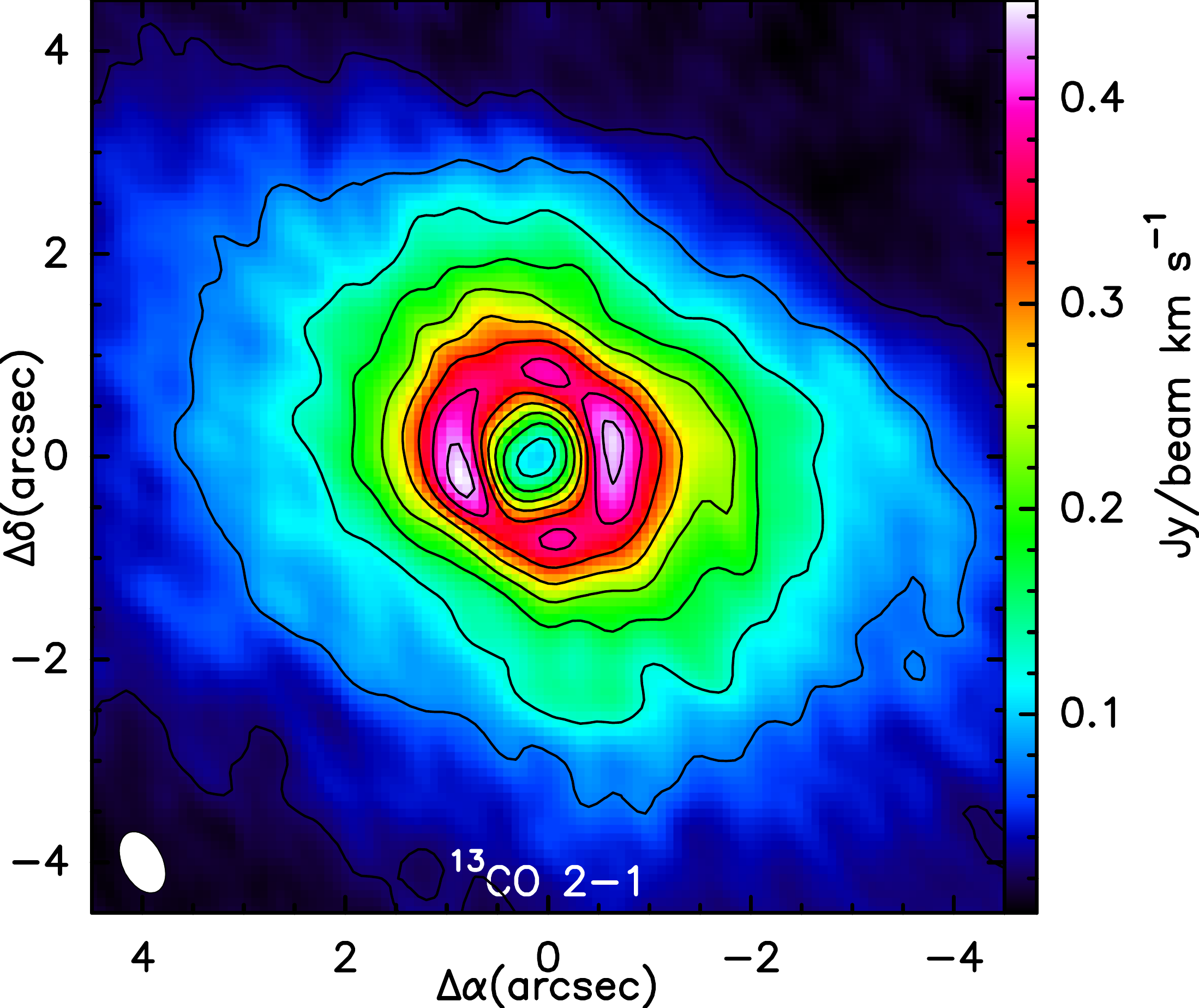}
 \includegraphics[width=0.25\textwidth, trim = 0mm 0mm 0mm 0mm,clip]{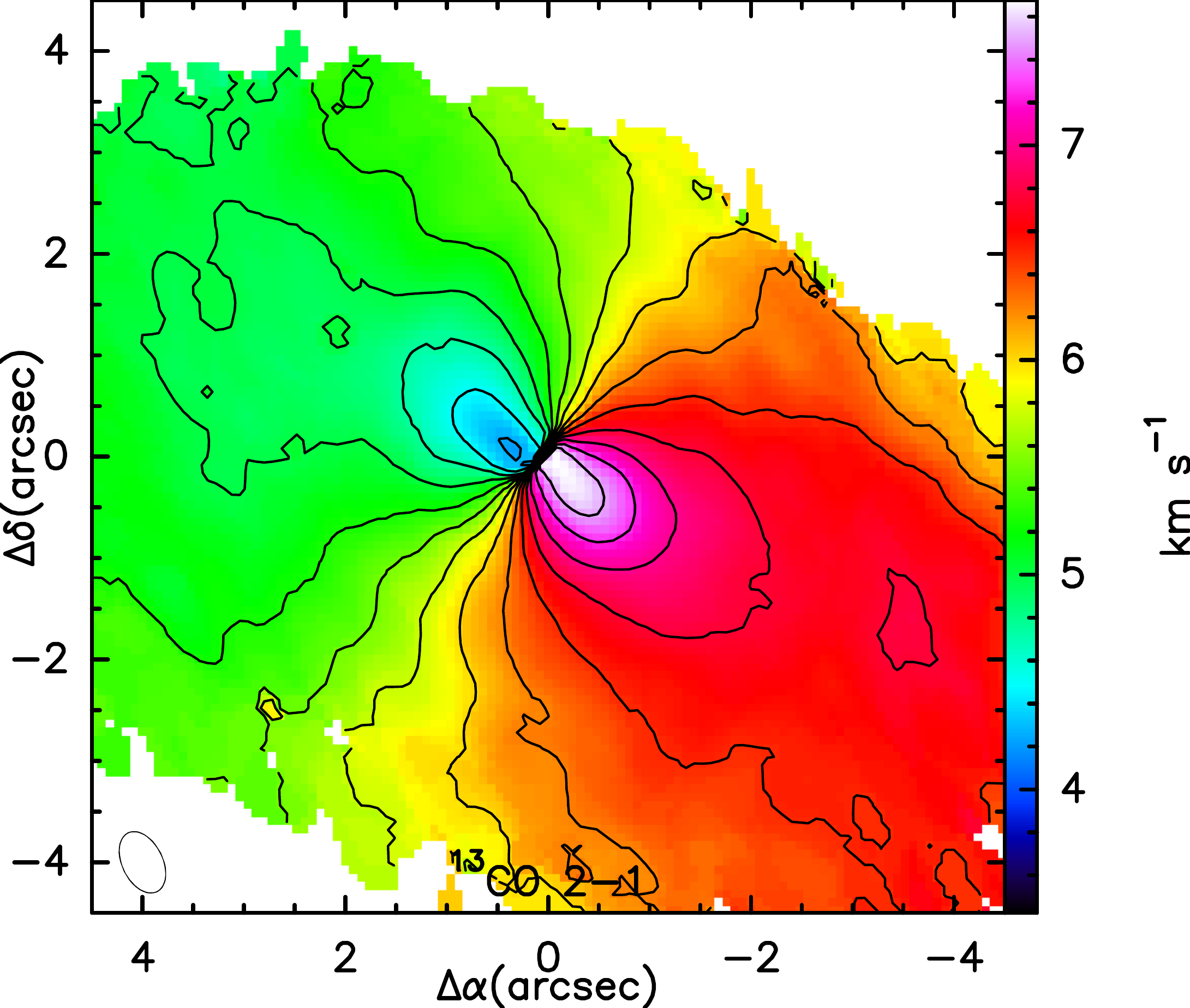} 
 \includegraphics[width=0.25\textwidth, trim = 0mm 0mm 0mm 0mm,clip]{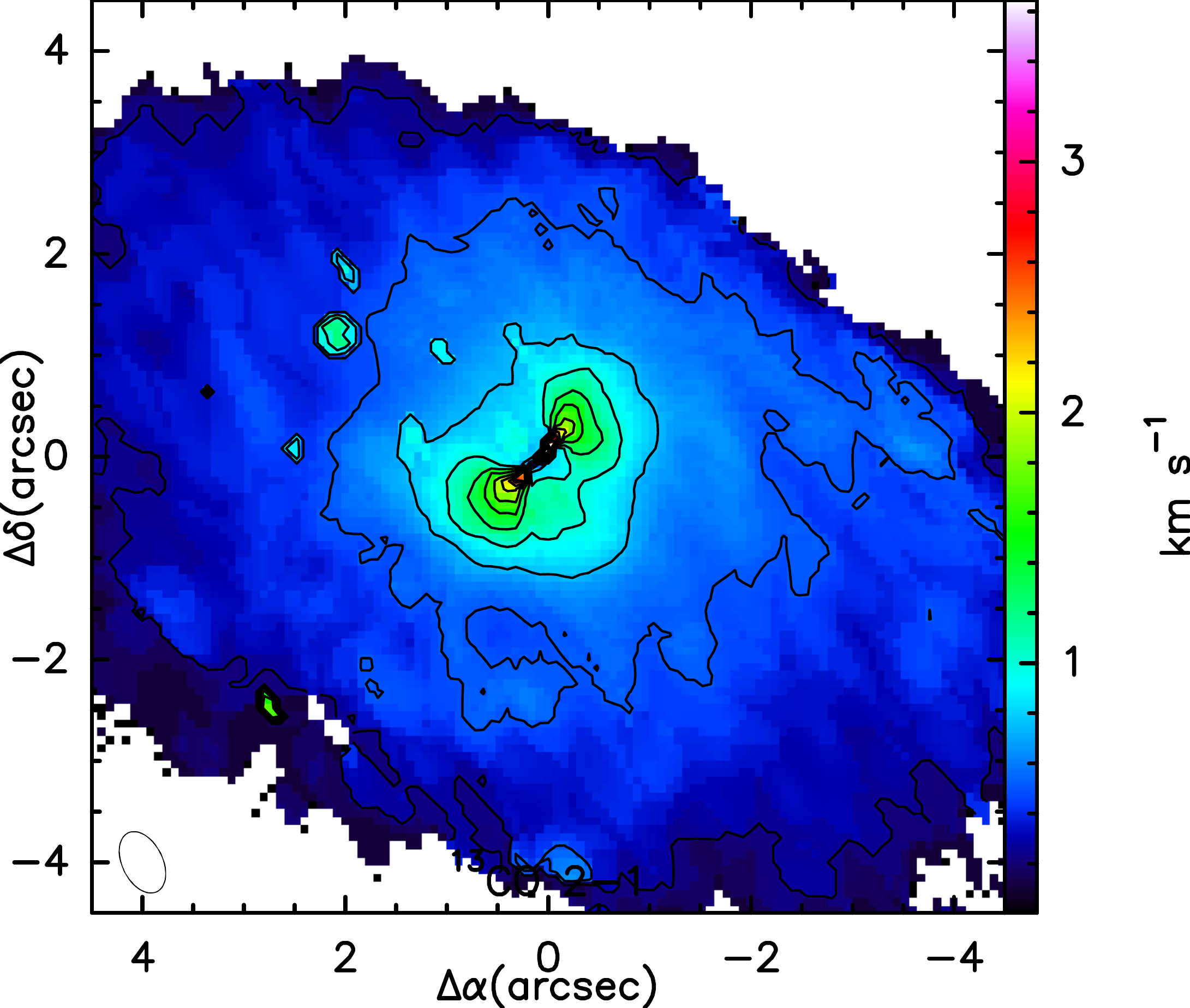}

 \includegraphics[width=0.25\textwidth, trim = 0mm 0mm 0mm 0mm,clip]{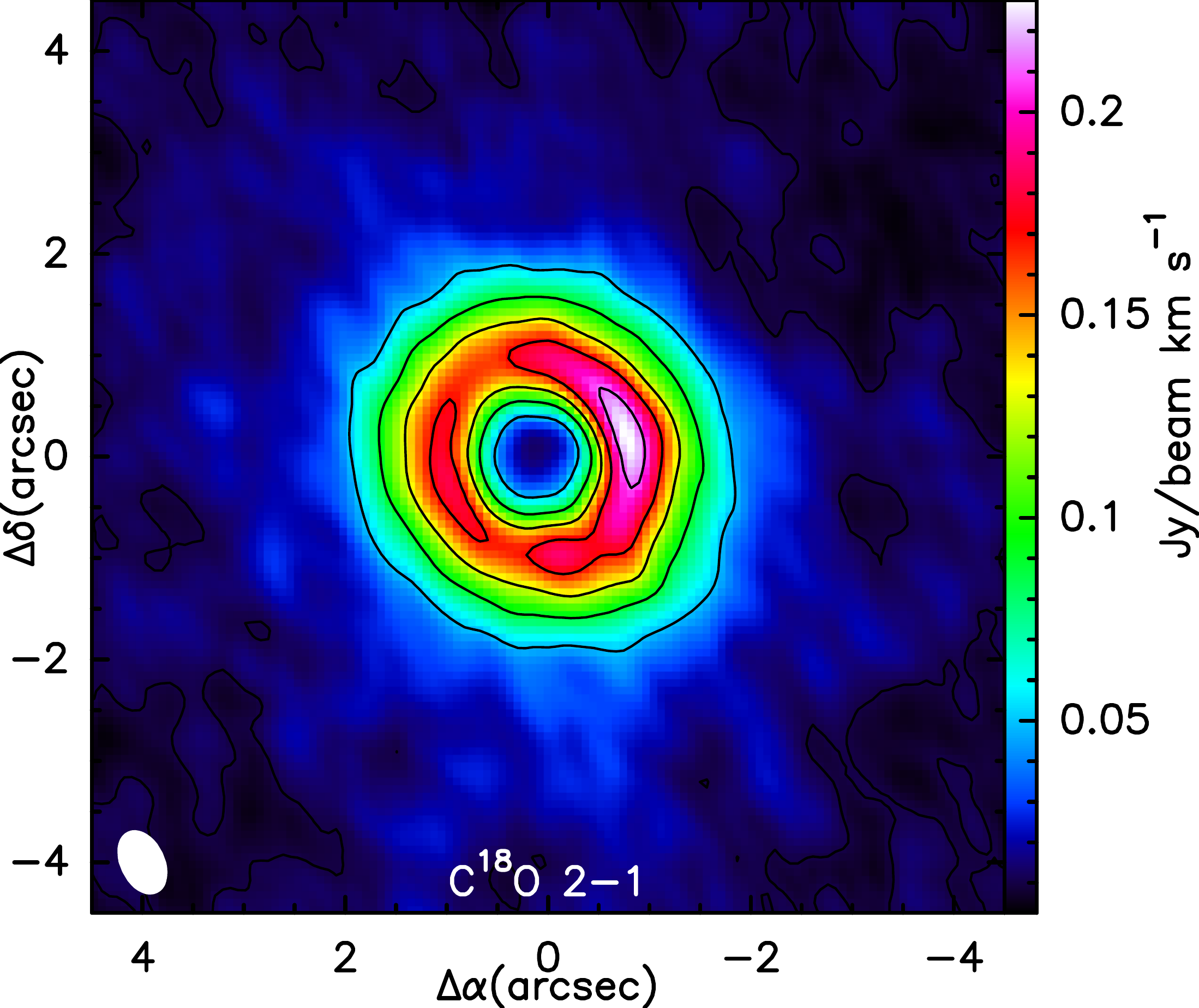}
 \includegraphics[width=0.25\textwidth, trim = 0mm 0mm 0mm 0mm,clip]{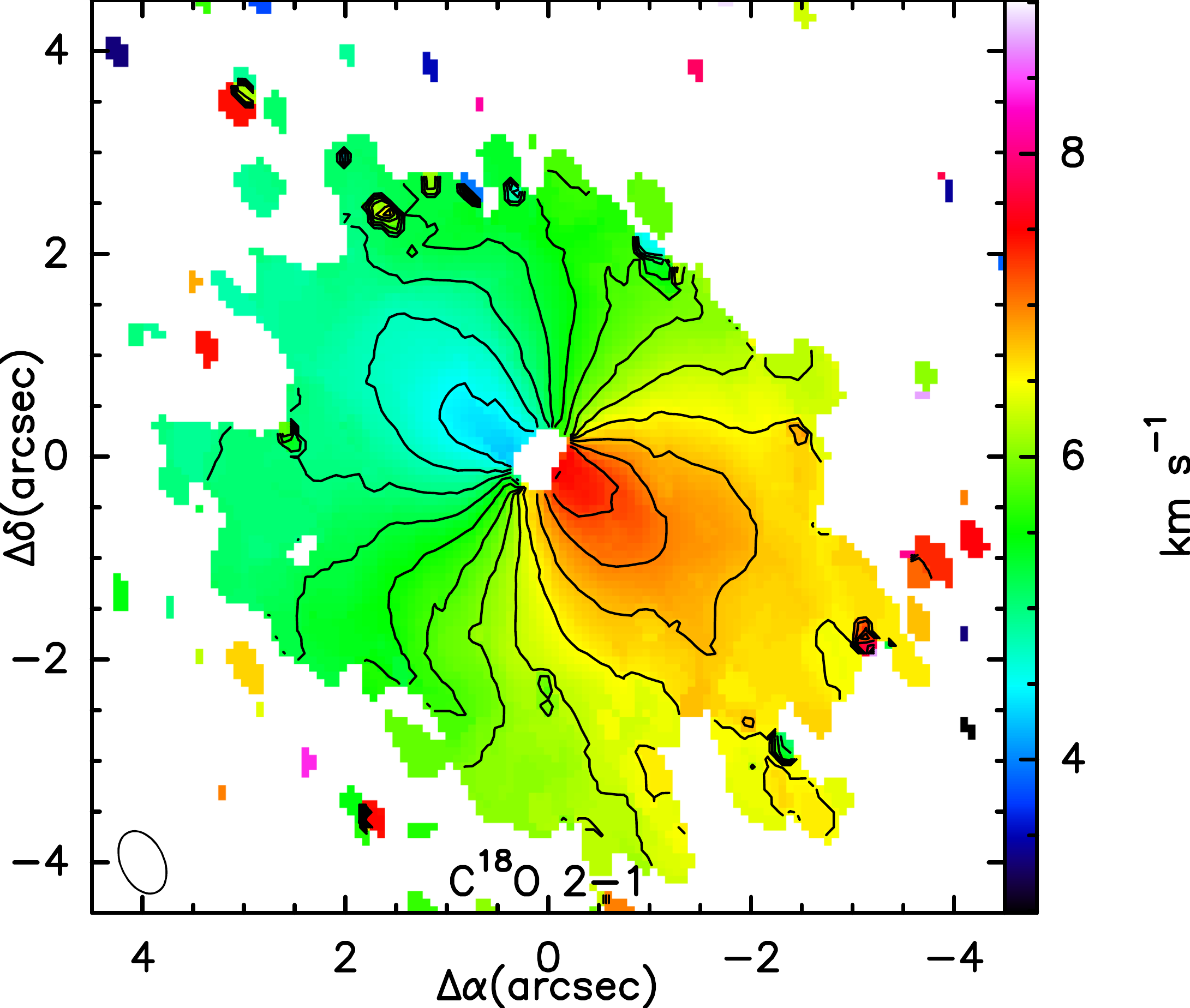}
 \includegraphics[width=0.25\textwidth, trim = 0mm 0mm 0mm 0mm,clip]{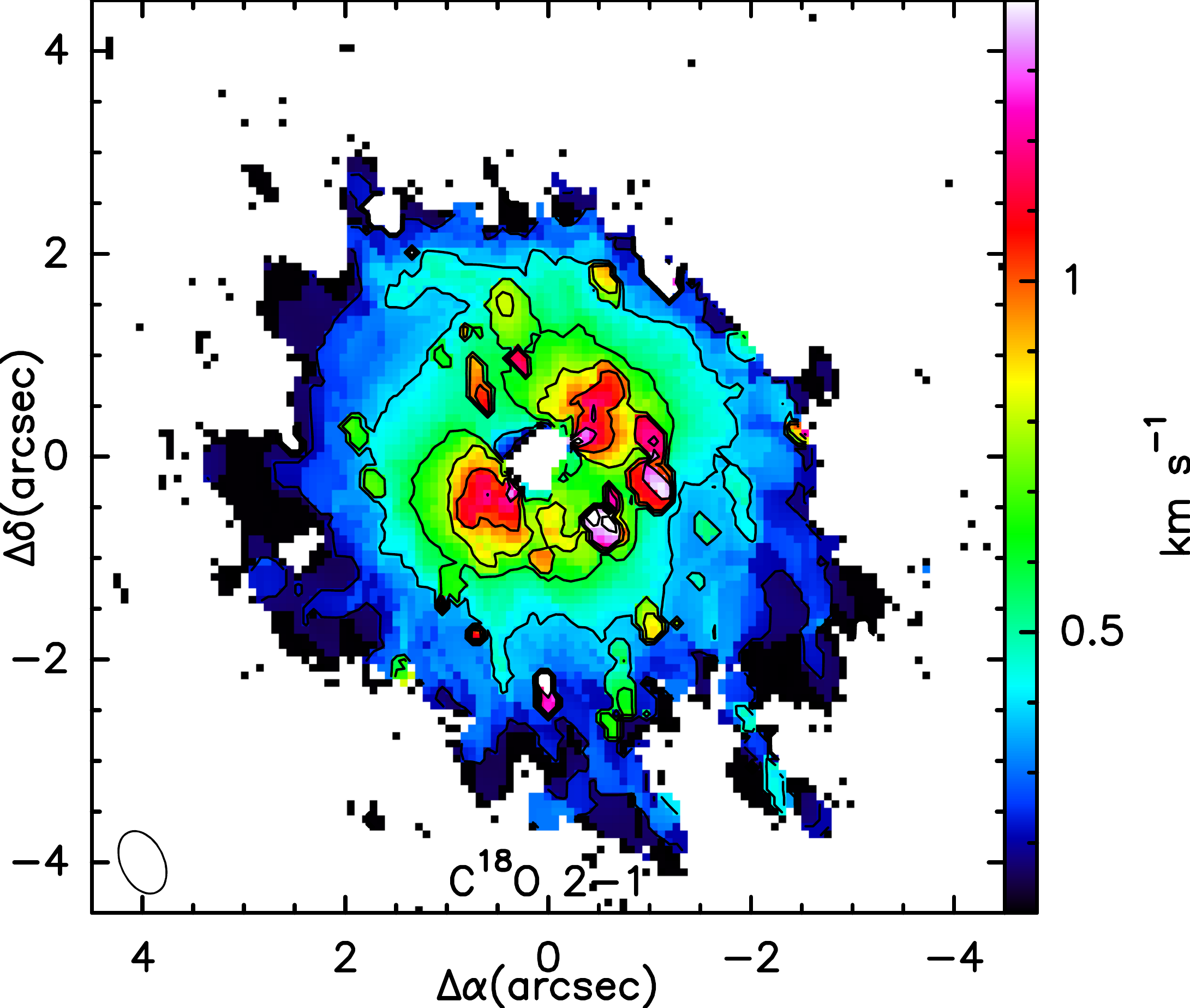} 

  \includegraphics[width=0.25\textwidth,trim = 0mm 0mm 0mm 0mm,clip]{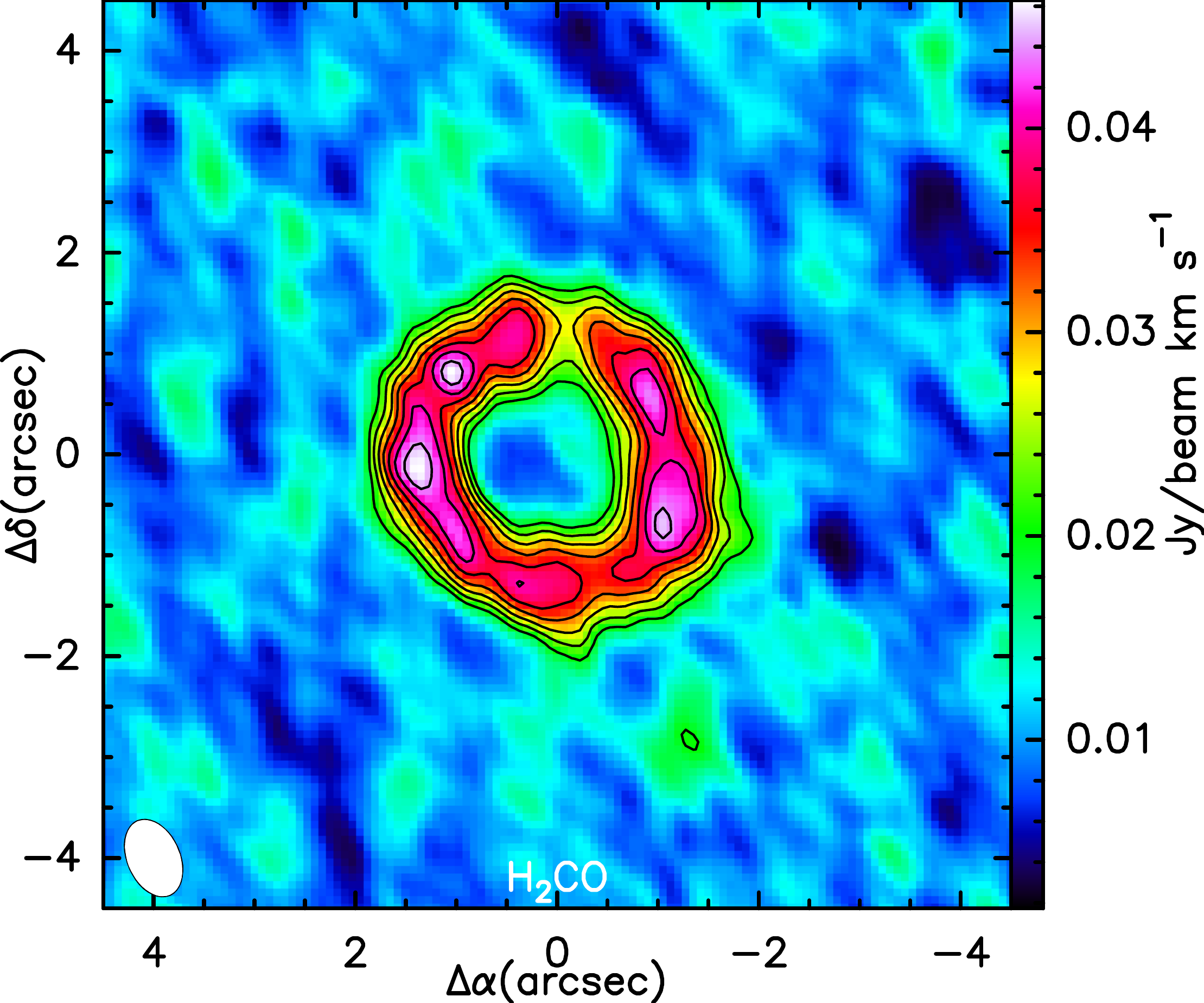} 
 \includegraphics[width=0.25\textwidth,trim = 0mm 0mm 0mm 0mm,clip]{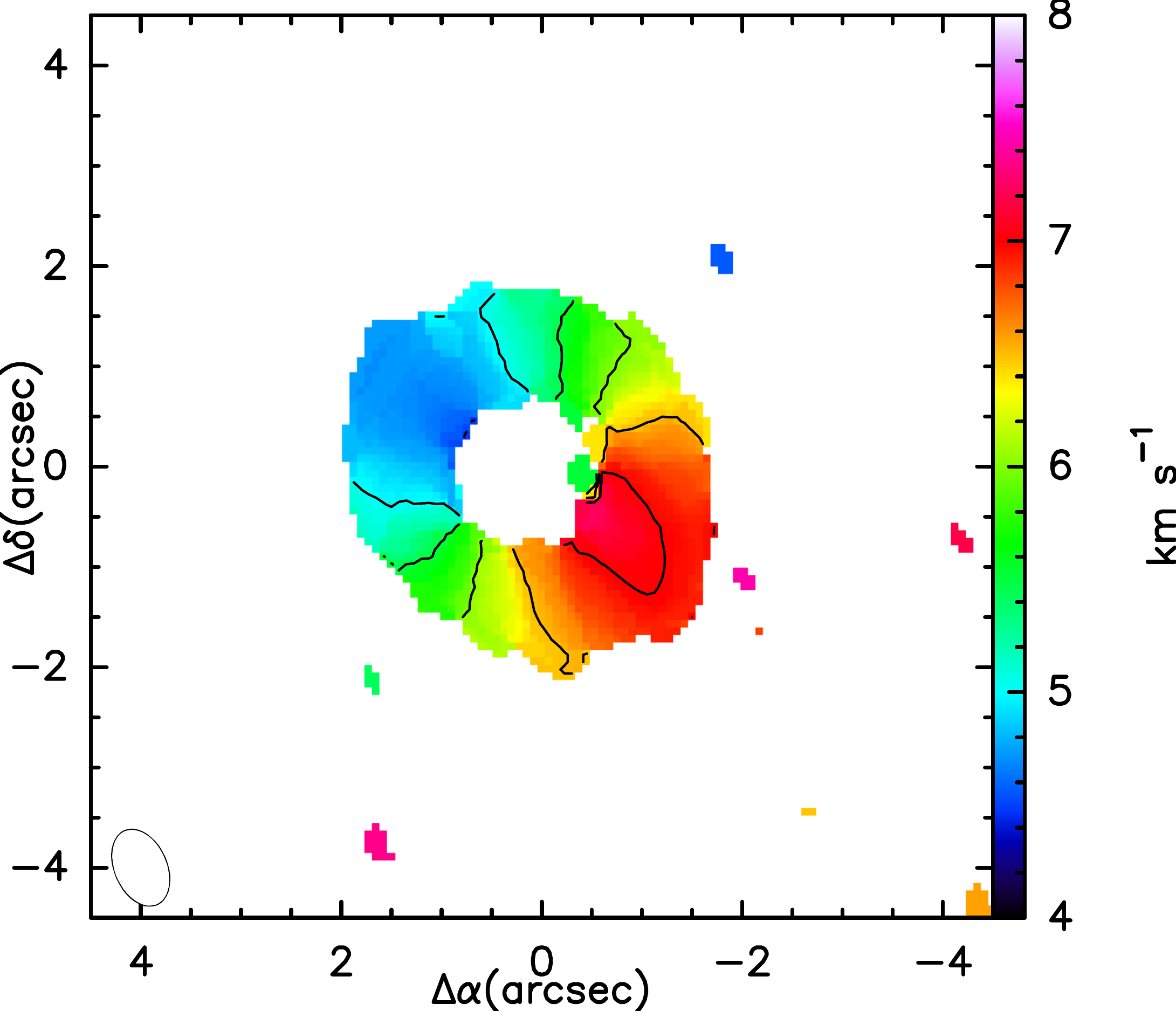}  
 \includegraphics[width=0.25\textwidth,trim = 0mm 0mm 0mm 0mm,clip]{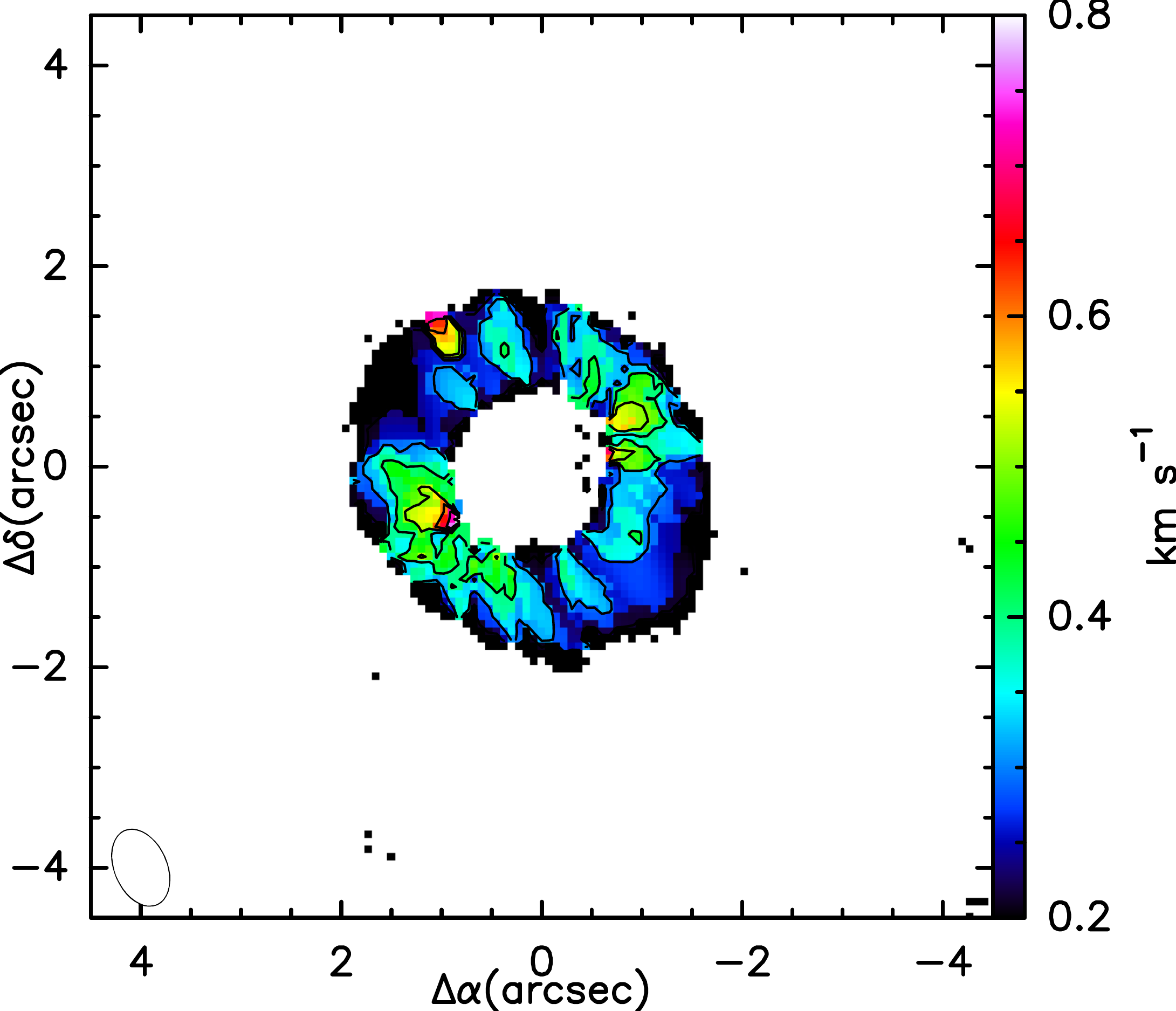} 
 
 \includegraphics[width=0.25\textwidth,trim = 0mm 0mm 0mm 0mm,clip]{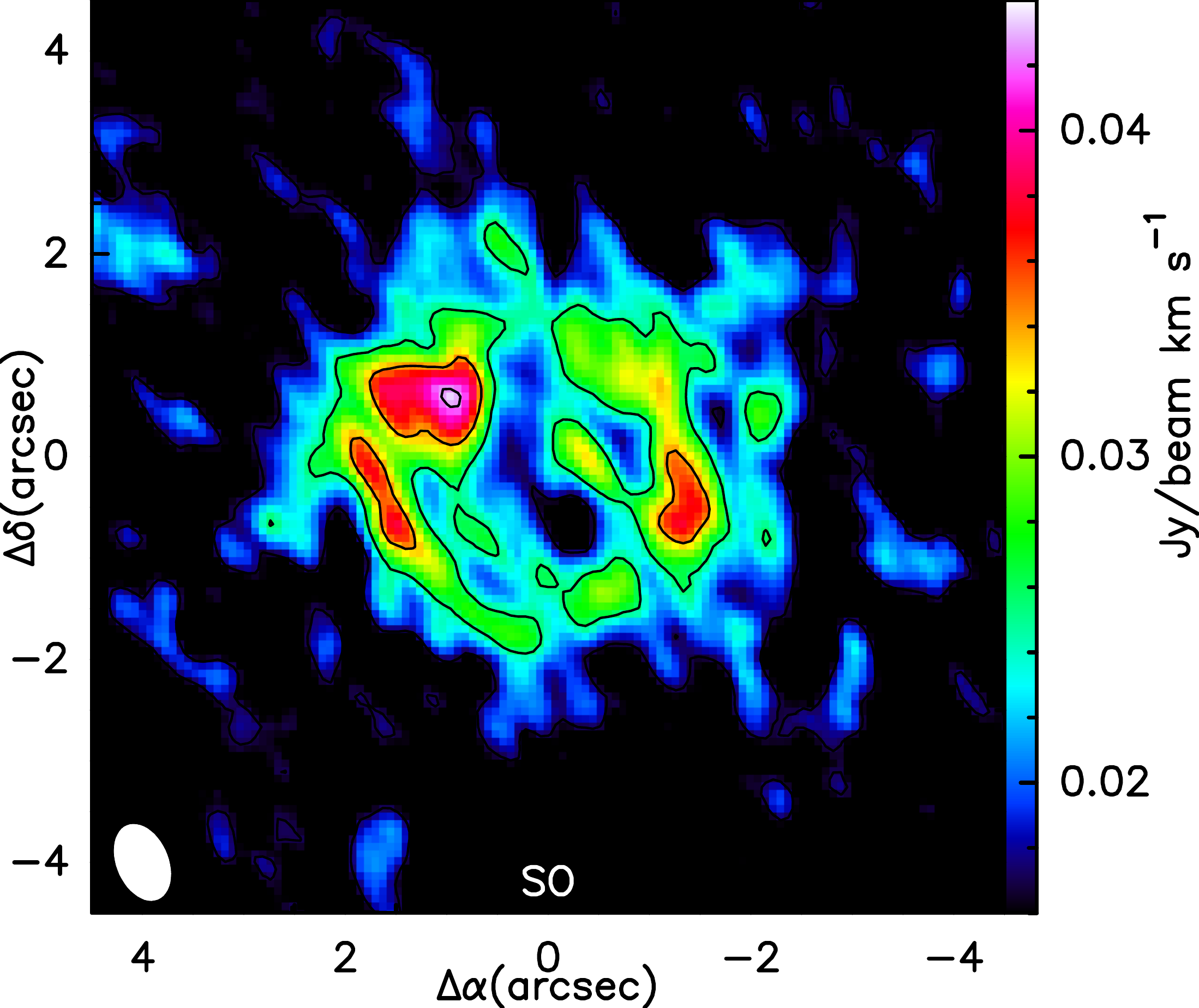} 
 \includegraphics[width=0.25\textwidth,trim = 0mm 0mm 0mm 0mm,clip]{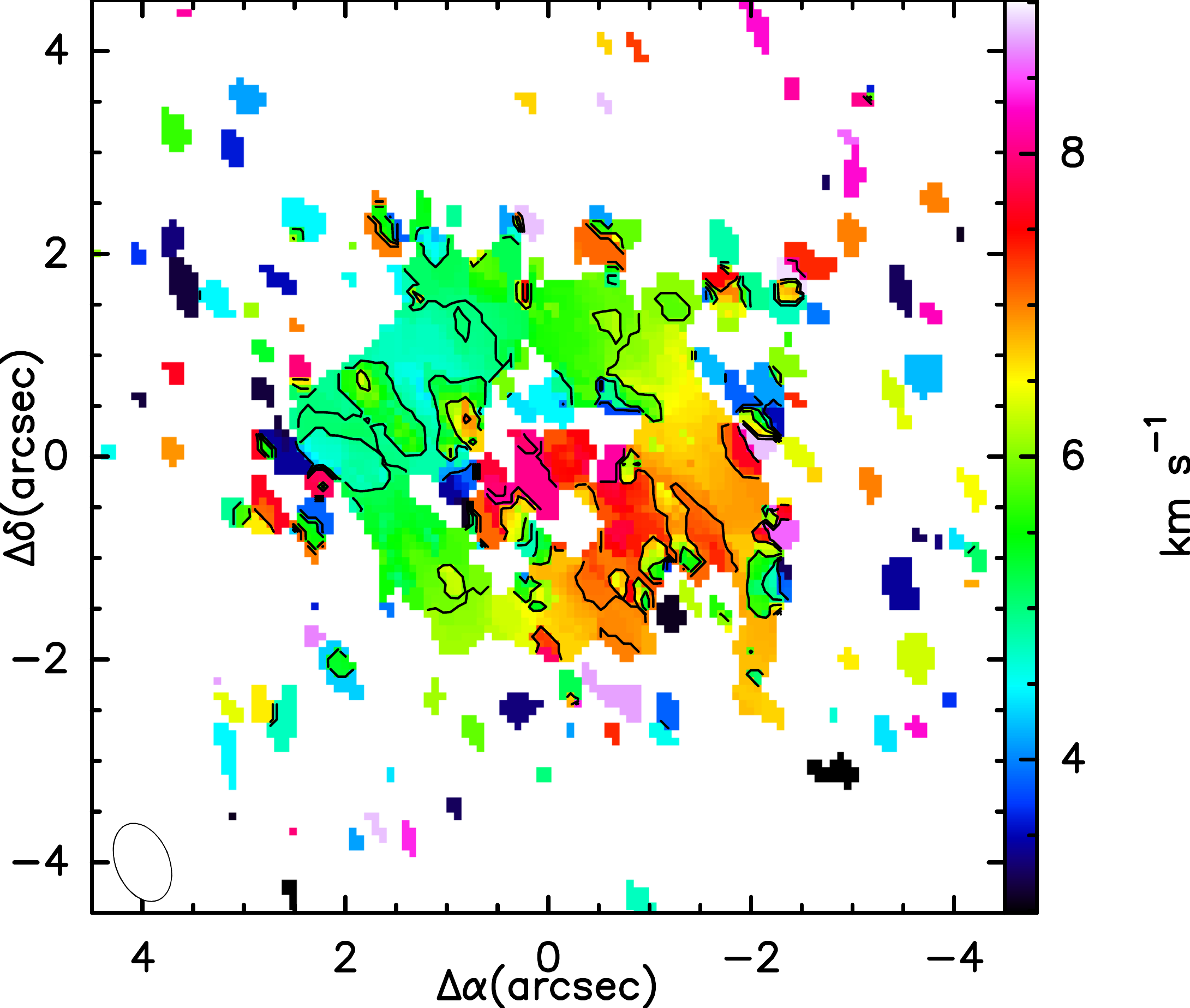}  
 \includegraphics[width=0.25\textwidth,trim = 0mm 0mm 0mm 0mm,clip]{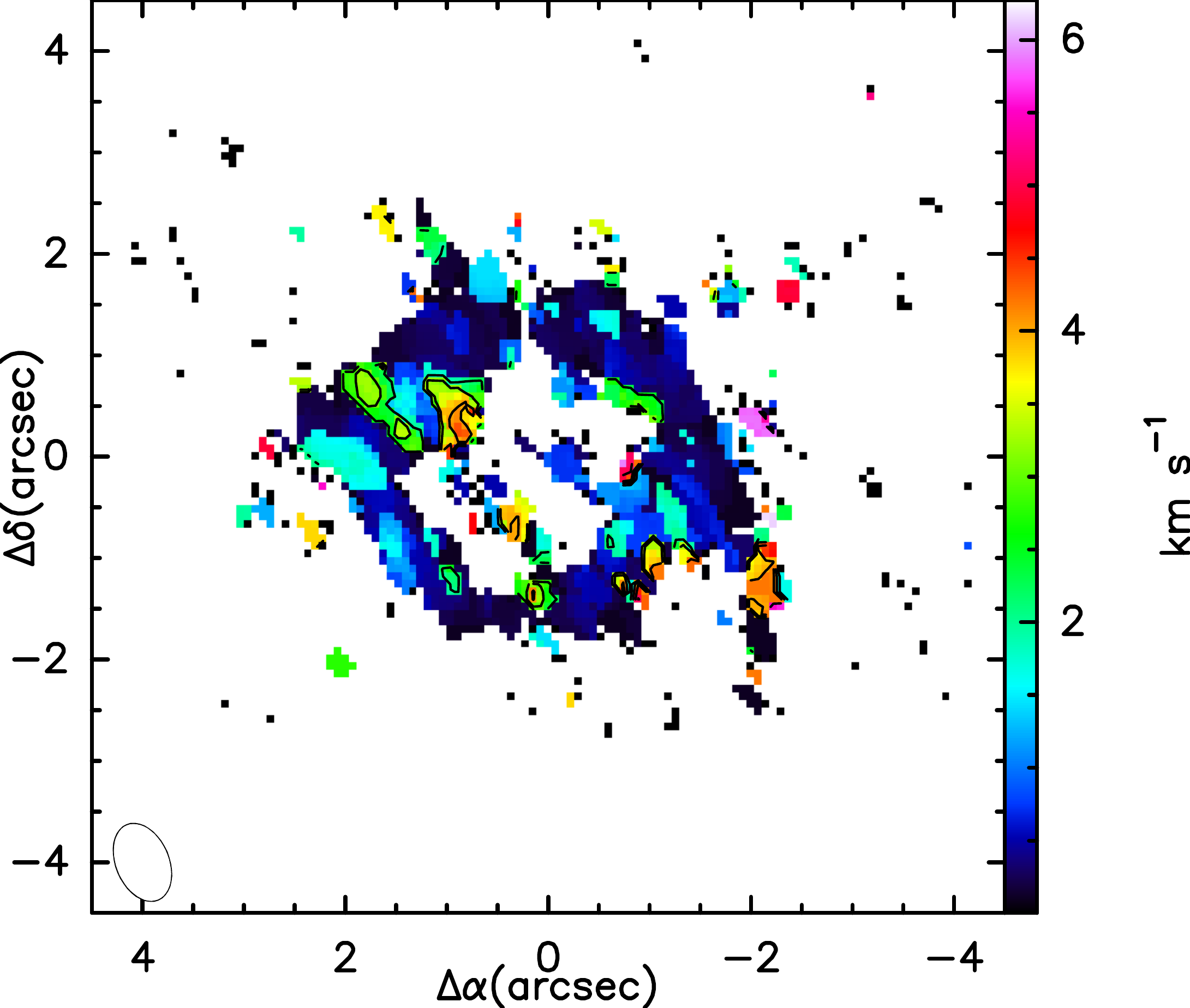} 
 
  \caption{Zeroth-, first-, and second-moment maps of the different species surveyed with NOEMA. {\bf Left:} Integrated intensity maps (zeroth-moment maps). {\bf Center:} Intensity-weighted velocity maps (first-moment maps). {\bf Right:} Velocity dispersion maps (second-moment maps). From top to bottom, species are $^{12}$CO, $^{13}$CO, C$^{18}$O, H$_{2}$CO, and SO. The maps were obtained after integrating channels in the range 3 to 9 km s$\rm ^{-1}$. For the first- and second-moment maps, only channels with a S/N $\rm >$ 5 were used. The white ellipses in the bottom left corner of each map depict the synthesized beam at each wavelength.}
 \label{Fig:moment_maps}
\end{center}
\end{figure*}

\begin{figure*}[h!]
\begin{center}
 \includegraphics[width=0.253\textwidth,trim = 0mm 0mm 0mm 0mm,clip]{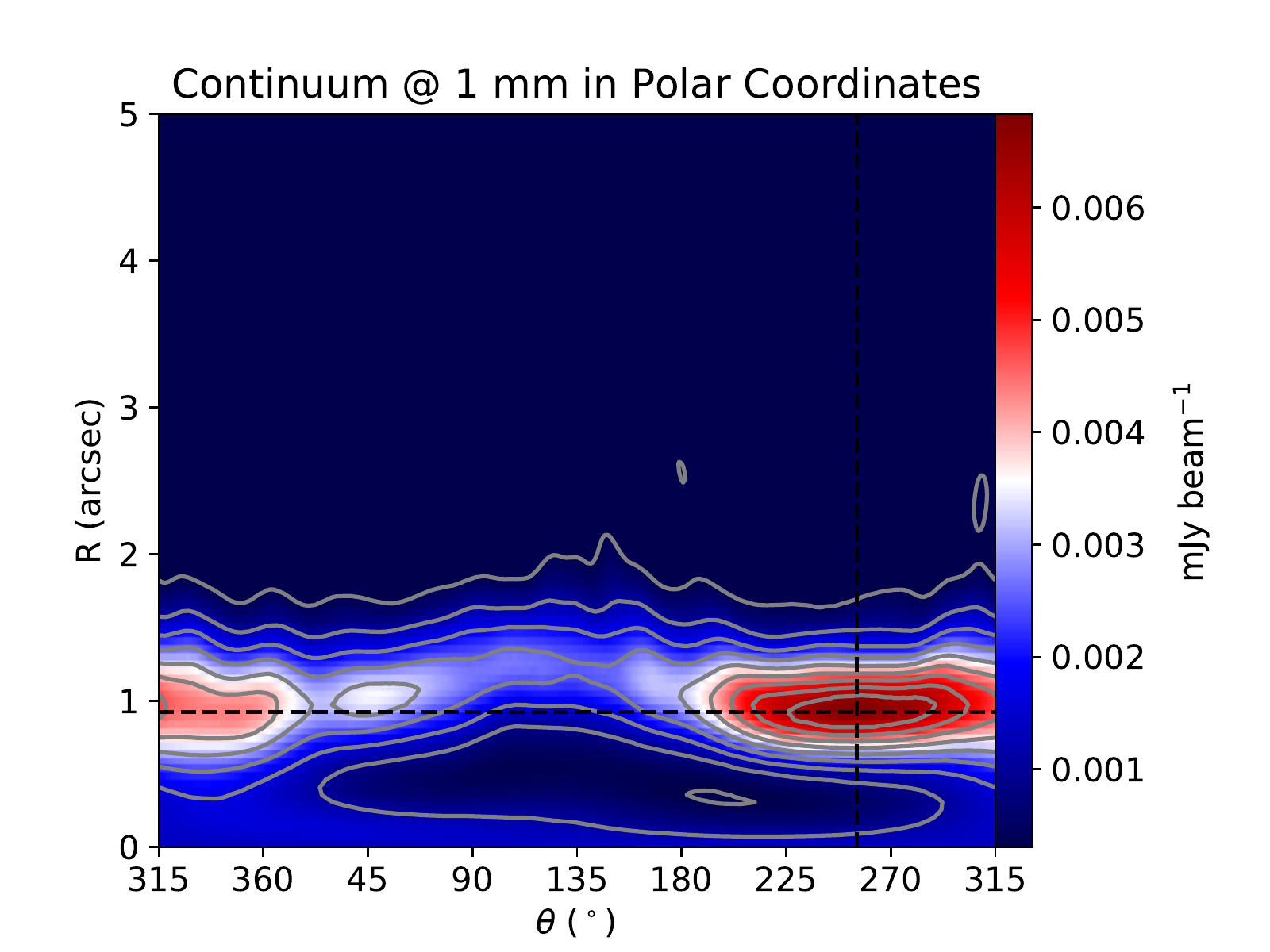}
\includegraphics[width=0.253\textwidth,trim = 0mm 0mm 0mm 0mm,clip]{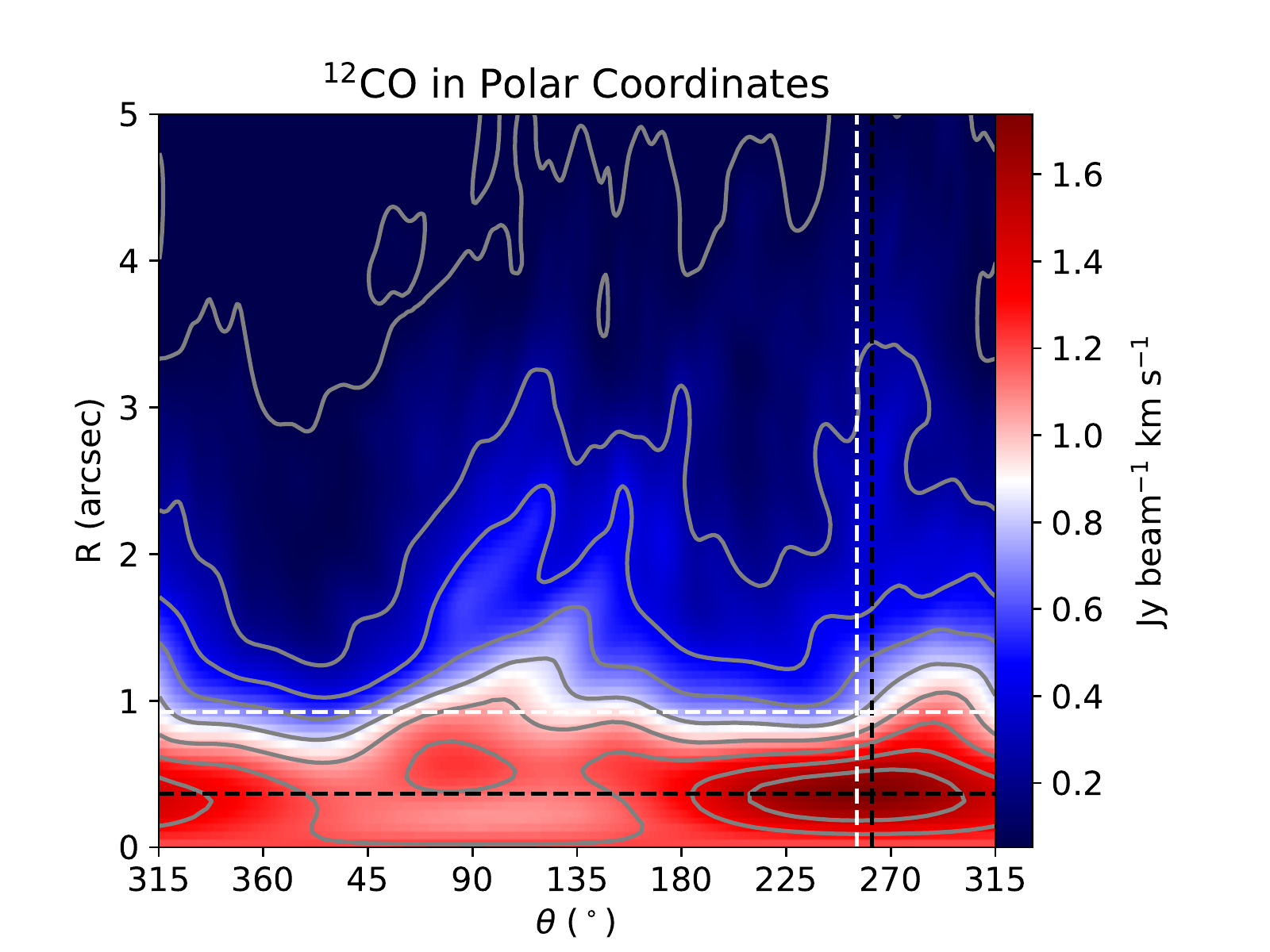}\includegraphics[width=0.253\textwidth,trim = 0mm 0mm 0mm 0mm,clip]{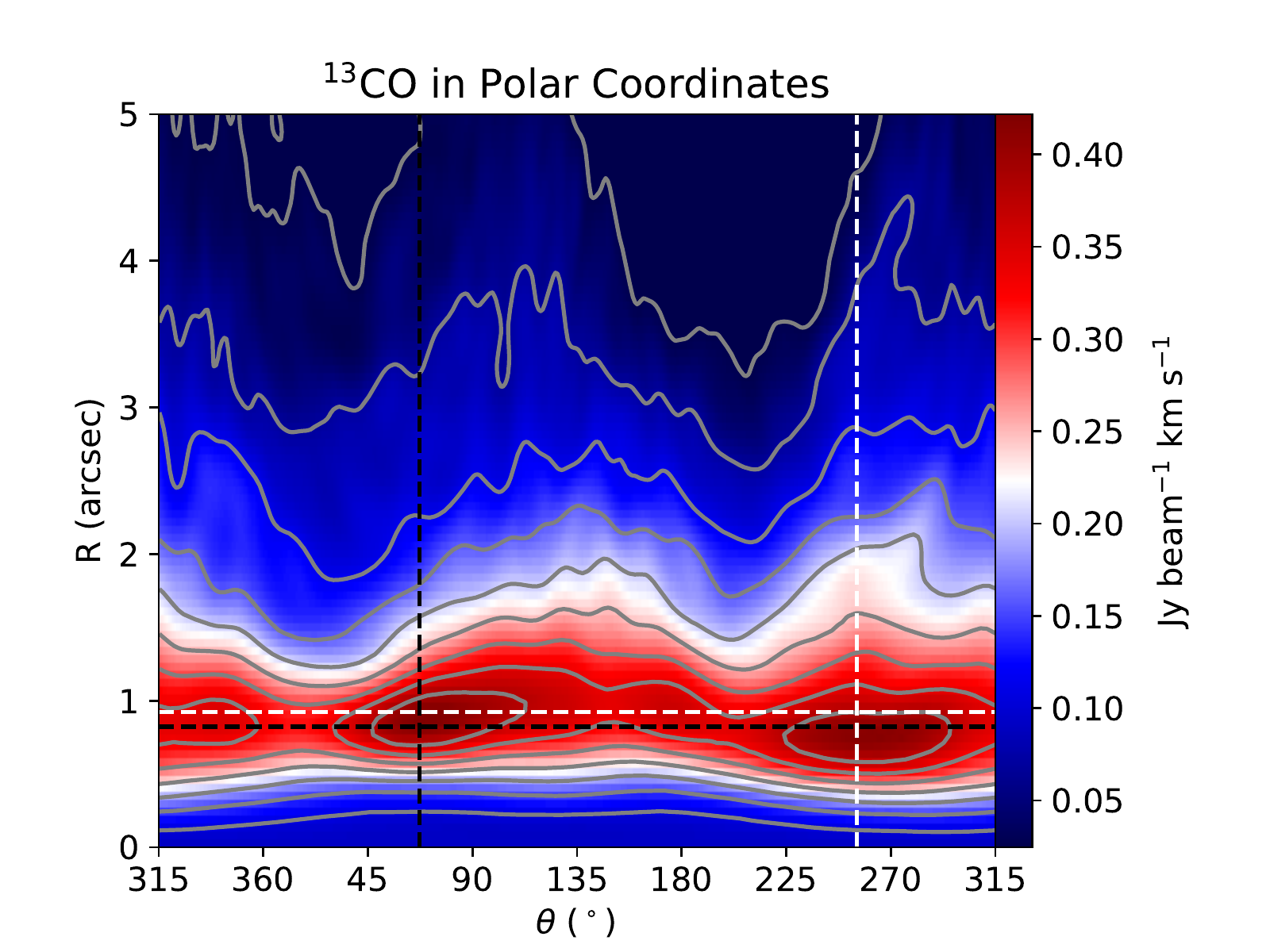}

 \includegraphics[width=0.253\textwidth,trim = 0mm 0mm 0mm 0mm,clip]{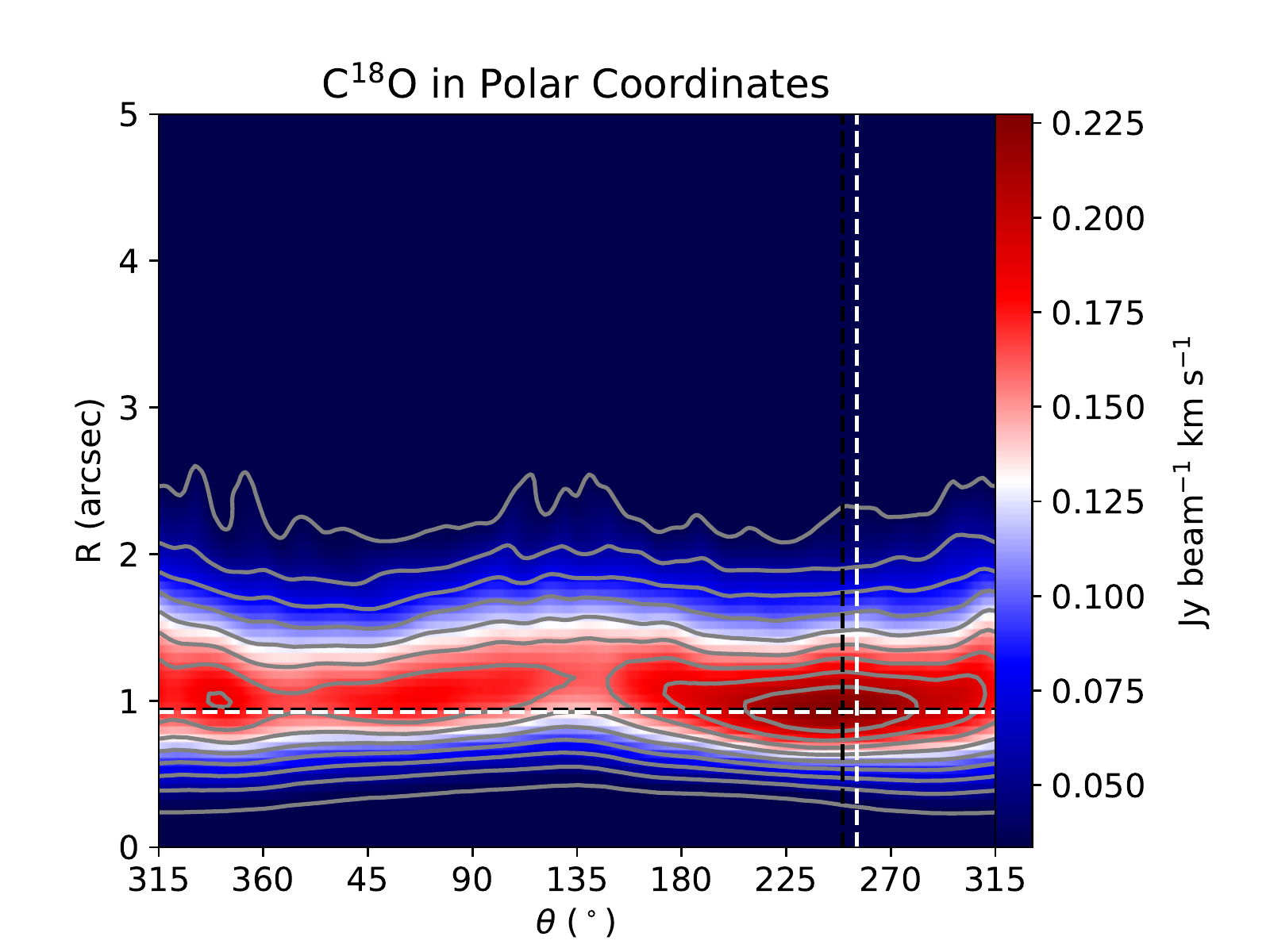}\includegraphics[width=0.253\textwidth,trim = 0mm 0mm 0mm 0mm,clip]{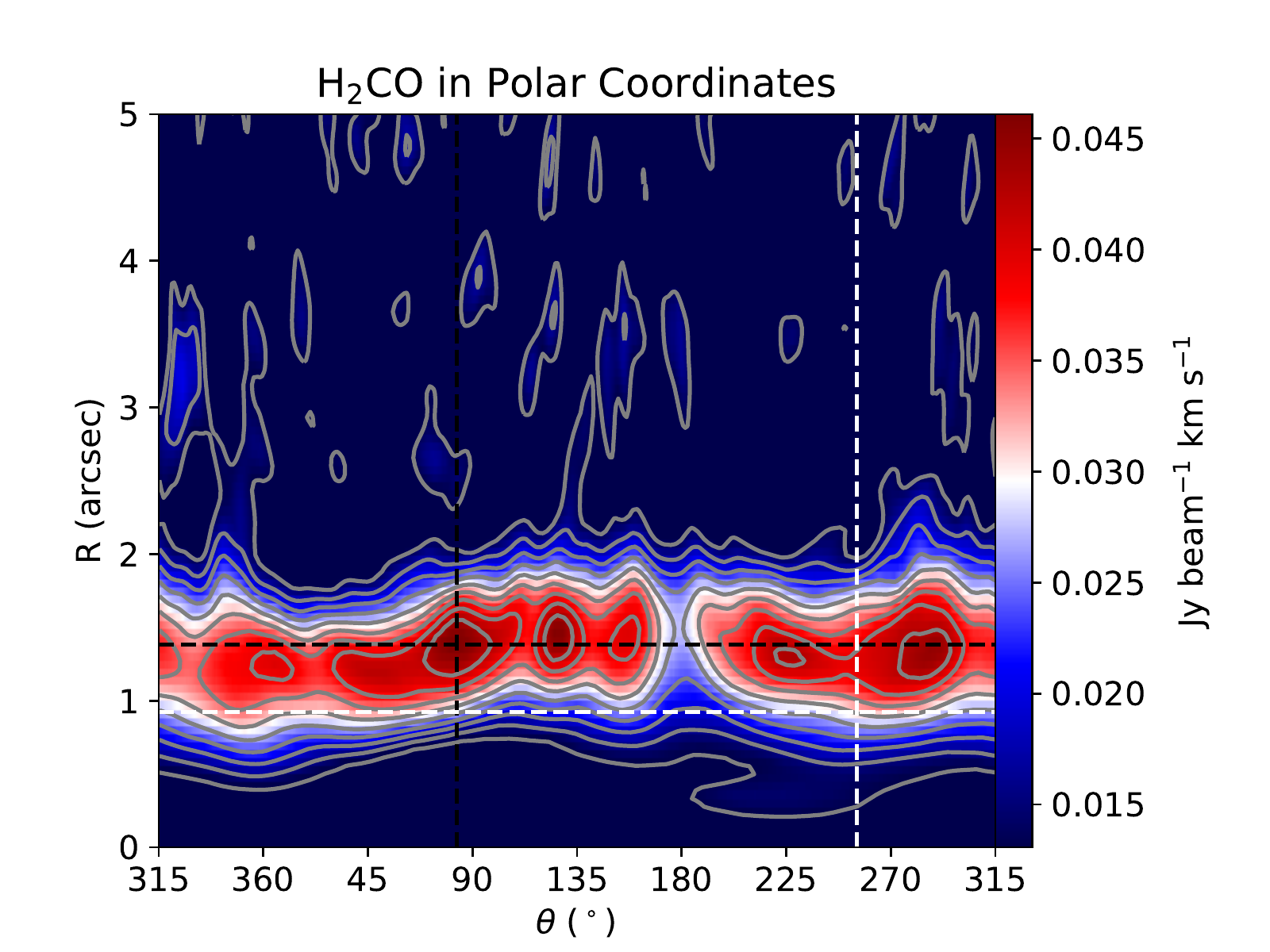}\includegraphics[width=0.253\textwidth,trim = 0mm 0mm 0mm 0mm,clip]{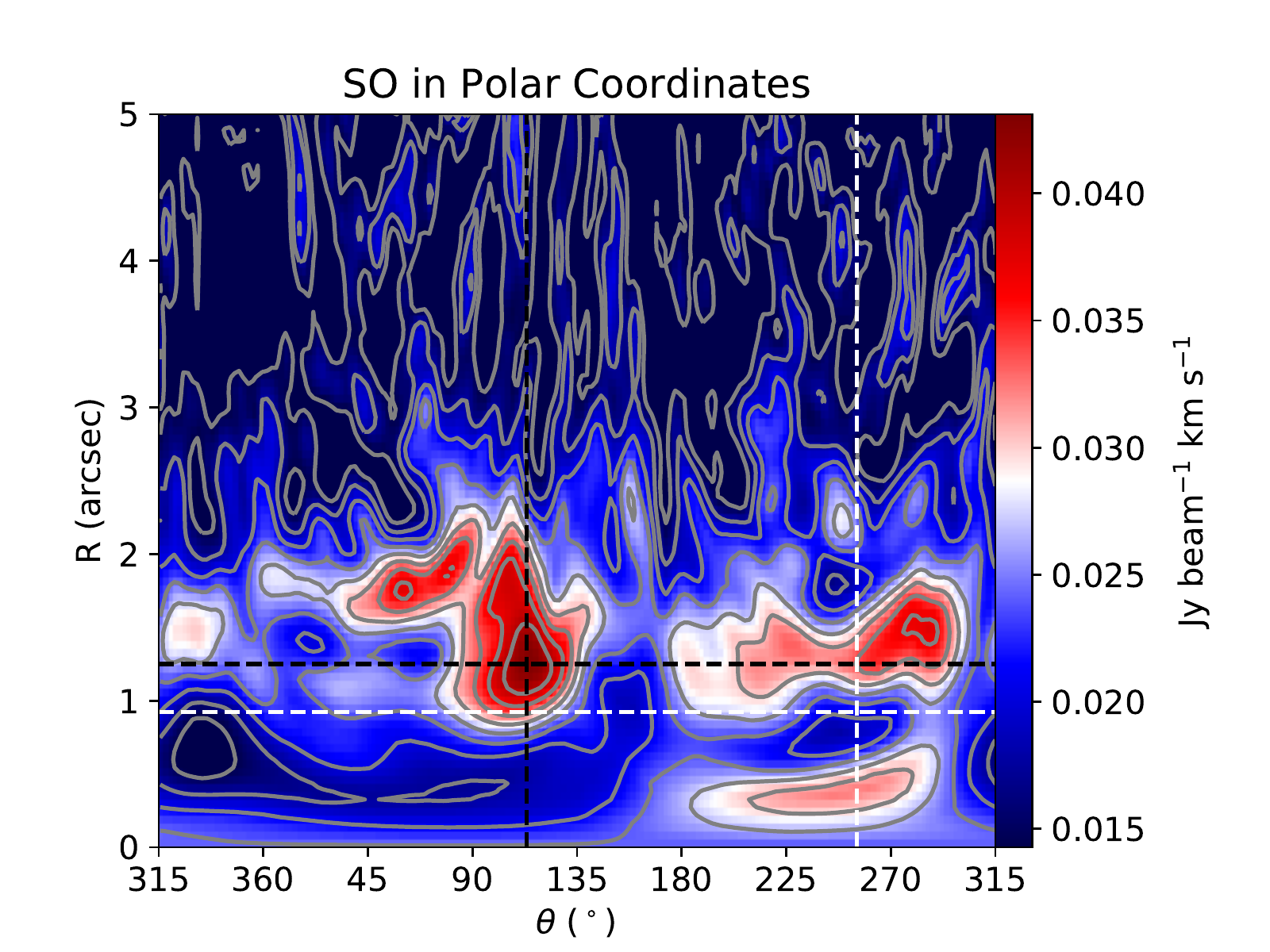} 
  \caption{De-projected integrated intensity maps in polar coordinates. Top left: Continuum emission. The black dashed lines depicts the position of the continuum peak. In the rest of the panels, the white dashed lines depict the position of the continuum peak, while the black dashed lines depict the peak of each species.}
 \label{Fig:polar_maps}
\end{center}
\end{figure*}

Spectra of the source-integrated flux of all the detected lines are shown in  Fig. \ref{Fig:stacked_spectra}. The spectra were computed inside regions with S/N$>$5$\sigma$. Spectral cubes of the  $\rm ^{12}$CO 2-1, $\rm ^{13}$CO 2-1, C$\rm ^{18}$O 2-1, H$_2$CO 3$\rm _{03}$-2$\rm _{02}$, and the SO  5$_6$-4$_5$ lines were constructed to explore the morphology and kinematics of the gaseous disk. The zero-, first-, and second-moment maps of  the   $^{12}$CO 2$\rightarrow$1, $^{13}$CO  2$\rightarrow$1, C$^{18}$O 2$\rightarrow$1, H$_2$CO 3$\rm _{03}$$\rightarrow$2$\rm _{02}$,  and  SO  5$_5$$\rightarrow$4$_4$  lines are shown in Fig. \ref{Fig:moment_maps}. For an easier comparison, we also show in Fig. \ref{Fig:polar_maps} the integrated intensity maps of the different species, as well as the continuum intensity map, in polar coordinates.

As Fig. \ref{Fig:moment_maps} shows, the shape of the emission dramatically changes from the  $^{12}$CO to the C$^{18}$O map where the emission shows a ring-like structure. The $\rm ^{12}$CO and $\rm ^{13}$CO  lines are optically thick, and their emission is more extended than that of C$\rm ^{18}$O. The emission of $\rm ^{12}$CO is elongated in the southwest to northeast direction. Along the ring, the $\rm ^{12}$CO integrated intensity map shows a peak at $\theta \sim$ 270$\rm ^{\circ}$, close to the position of the dust trap, but at a closer distance, r$\rm \sim 0.254 \arcsec$ ($\sim$55 au). The map also has a local maximum almost at the opposite side of the disk at $\theta \sim$ 90 $\rm ^{\circ}$. Diffuse 5$\sigma$ emission is detected as far as 5$\arcsec$ (814 au) from the center. The central regions of the map present a depression in flux, $\sim$1 Jy beam$\rm ^{-1}$ km s$\rm ^{-1}$, compared to $\sim$1.7 Jy beam$\rm ^{-1}$ km s$\rm ^{-1}$ in the peak. $^{13}$CO shows an elongated shape around a central ring-like structure, similar to the ring observed in C$^{18}$O emission. Azimuthal asymmetries are observed for the three isotopologs. In the case of C$^{18}$O, the shape of the emission resembles that observed in the continuum \citep{Pietu2005,Tang2012,Fuente2017}. Emission at the 5$\sigma$ level is observed from $\sim$0.$\arcsec$4 (65 au) to $\sim$2$\arcsec$ (326 au), with a bright emission ring at 0.$\arcsec$65 (106 au). In cooler disks around T Tauri stars the expectation is that CO will be more severely depleted in regions with large dust densities, such as a dust trap. The fact that C$\rm ^{18}$O emission in AB Aur follows the dust emission 
suggests that CO is not heavily depleted in this disk and supports pressure trap theory.

\begin{figure}[h!]
\begin{center}
 \includegraphics[width=0.45\textwidth,trim = 0mm 0mm 0mm 0mm,clip]{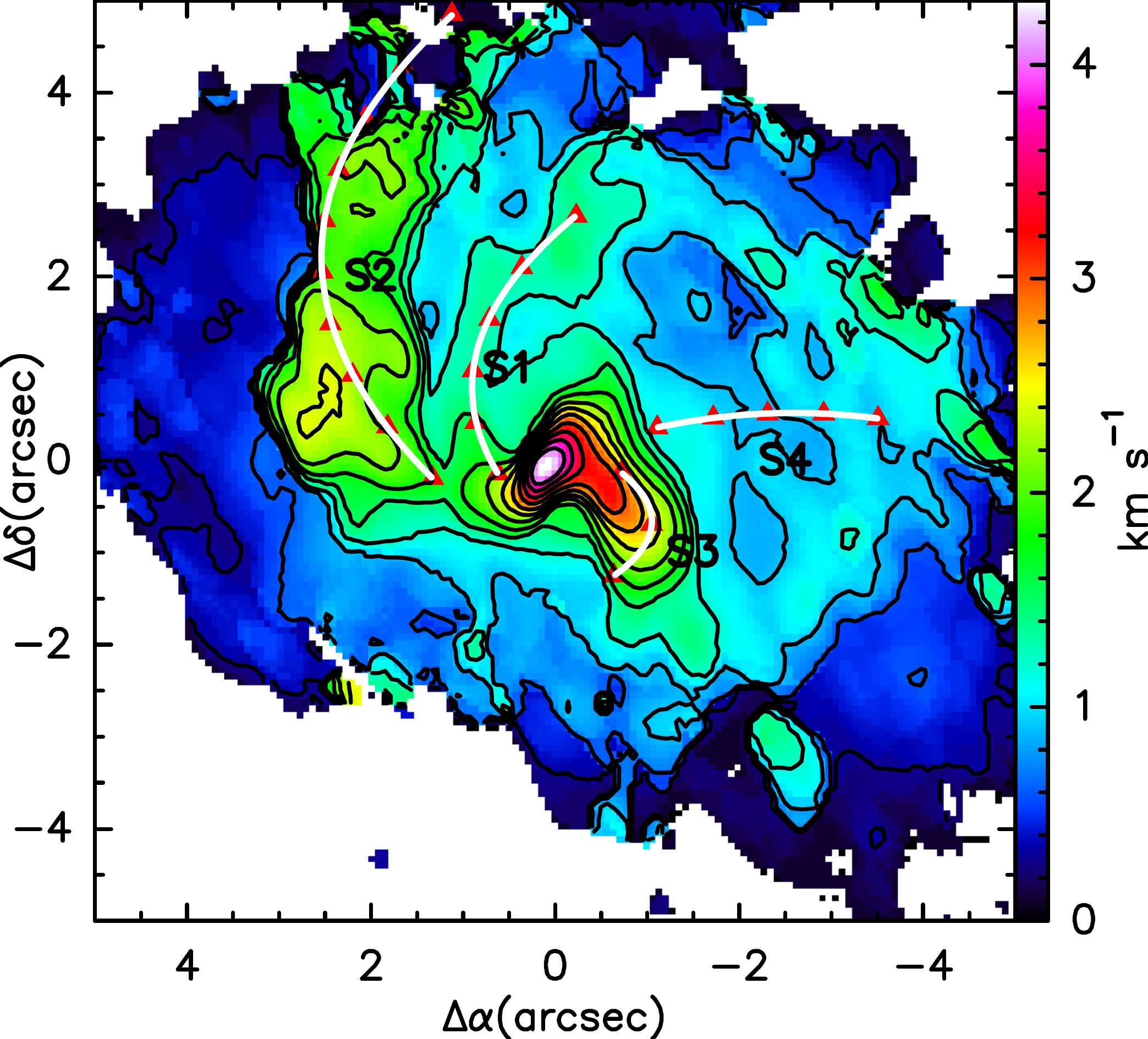}
  \caption{$^{12}$CO second-moment map showing the position of the spiral arms (solid white lines). The individual positions where spectra were extracted following \cite{Tang2012} are shown as red triangles. The name of the spirals from  \cite{Tang2012} are also included.}
 \label{Fig:CO_spiral_positions}
\end{center}
\end{figure}

\begin{figure*}[h!]
\begin{center}
 \includegraphics[width=1\textwidth, trim = 30mm 10mm 30mm 0mm, clip]{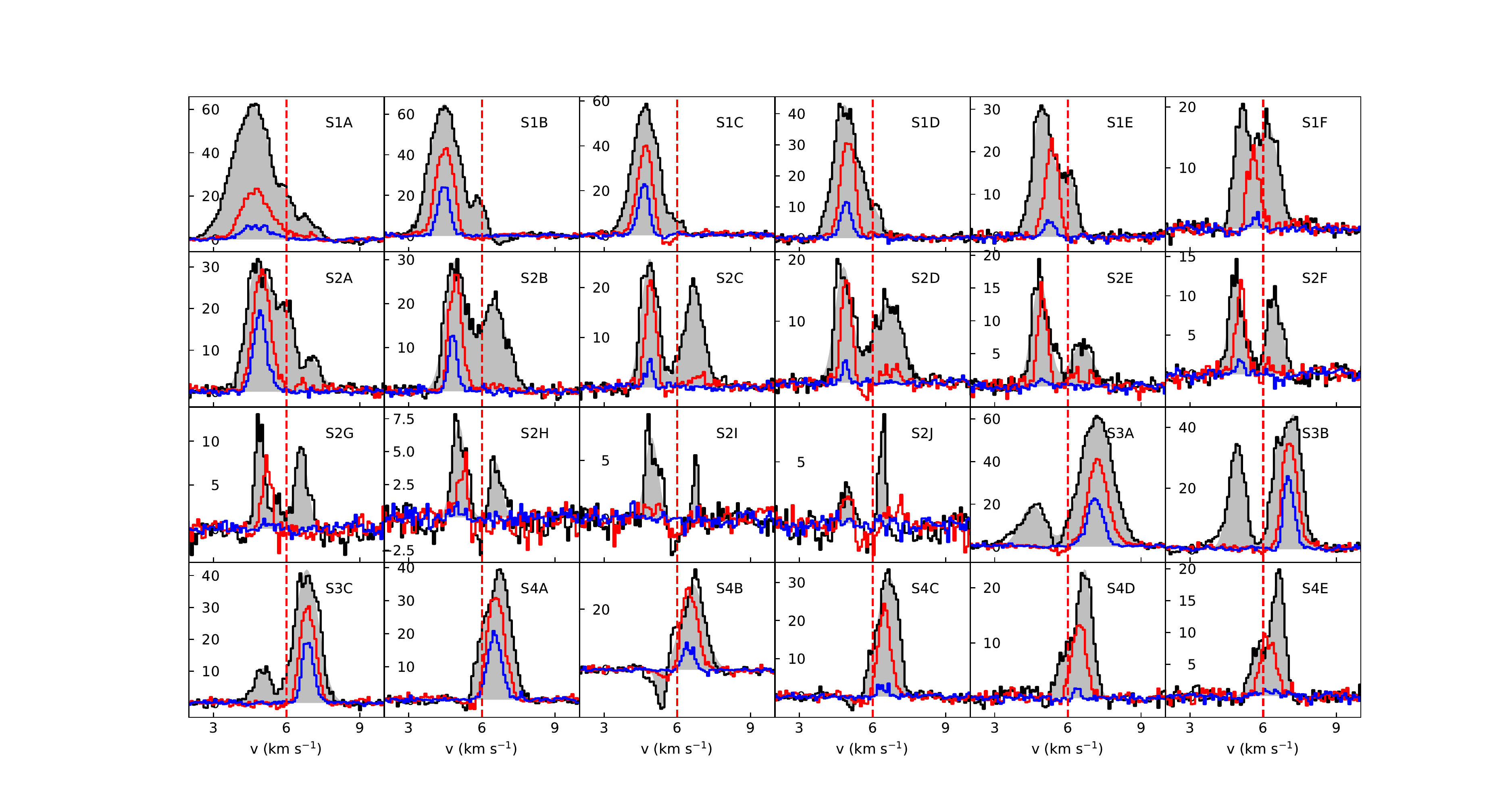}
  \caption{Spectra along the different spiral arms detected by \cite{Tang2012}; see red triangles in Fig. \ref{Fig:CO_spiral_positions}. We show $^{12}$CO in black, $^{13}$CO in red, and C$^{18}$O in blue. The grey shaded areas are Gaussian fits to the observed $^{12}$CO spectra}
 \label{Fig:CO_spiral arms_spectra}
\end{center}
\end{figure*}

The outermost parts of the disk show  twisted structures suggestive of spiral arms. \cite{Tang2012} identified four spiral arms and labeled them S1 to S4. We can only identify S1, S2, and S3. Such structures are especially apparent in the $\rm ^{12}$CO second-moment map (see Fig. \ref{Fig:CO_spiral_positions}) because the spiral arms add velocity components resulting in larger line widths. These structures can also be identified as excess emission in the integrated intensity maps of $\rm ^{12}$CO and $\rm ^{13}$CO, and as deviations from rotation in the first-moment maps. We do not observe any special structures associated with the spiral arm S4. \cite{Tang2012}  marked a series of positions along the different spiral arms to study in detail the associated kinematics. We repeated the same strategy using the same positions as in  \cite{Tang2012}; see triangles in Fig. \ref{Fig:CO_spiral_positions}. The resulting spectra are shown in Fig. \ref{Fig:CO_spiral arms_spectra}. The velocity components due to the spiral arms are detected in $\rm ^{12}CO$ and in $\rm ^{13}CO$ when they are prominent enough, but are never detected in C$\rm ^{18}O$, which points to an origin out of the disk midplane.  The presence of multiple components results in an artificial broadening of line widths, as observed in the second-moment map of $\rm ^{12}$CO (see Fig. \ref{Fig:moment_maps}, top right).

The zeroth-, first-, and second-moment maps of the H$_2$CO J=3$\rm _{03}$-2$\rm _{02}$ line are shown in Fig. \ref{Fig:moment_maps}, since this is the only H$_2$CO line with sufficiently high S/N to produce moments maps. The emission shows a ring-like shape extending from $\rm \sim$0.$\arcsec$7 to $\rm \sim$1.$\arcsec$6, plus a diffuse emission extending further out ($\rm \sim$4 $\arcsec$). Interestingly, the emission peaks further from the center than C$\rm ^{18}$O. We co-added the images of the $3\rm _{22}-2\rm _{21}$ and 3$\rm _{21}-2\rm _{20}$ lines in order to improve the S/N of our data but the quality of the final image was not good enough for detailed mapping. Instead, we decided to fit the source-integrated spectra (see Fig. \ref{Fig:stacked_spectra}).

Of all the SO lines observed, only the J$_N$=5$\rm _6$-4$\rm _5$ line was intense enough to build the zeroth, first-, and second-moment maps. The emission shows strong azimuthal variations. While $\rm ^{13}$CO, C$\rm ^{18}$O and p-H$\rm _2$CO show circular cavities in their inner regions, SO shows emission also in the inner regions, and not only in the outer disk. More interesting is the fact that SO emission shows its peak in a position that is displaced almost 180$\rm ^{\circ}$ in azimuth with respect to the dust trap (see Fig. \ref{Fig:moment_maps}, bottom panels). Similar behavior was observed in CS towards HD 142527 \citep{vanDerPlas2014}, with a maximum in the CS 7-6 integrated intensity map on the opposite side of the disk with respect to the dust trap. Our SO map further shows an interesting feature at PA $\rm \sim$ 270$\rm ^{\circ}$ that extends from the ring till the innermost regions, similar to the bridge reported by \cite{Riviere2019} for HCO$\rm ^+$ emission.

\begin{table}
\caption{Peak position in radial profiles, and angle peak in azimuthal profiles of the continuum and the different species surveyed before de-projection. The last column gives the binding energy of the different species surveyed.}
\label{Tab:profile_peaks}
\begin{tabular}{lllll}
\hline \hline
Species & Peak  & Peak & Peak.  & E$\rm _D$   \\
              & position & position & angle  &  \\
 --           &  ($\arcsec$)    & (au) &  ($^{\circ}$) & (K)   \\
\hline
Cont \@ 1mm & 0.96 & 157 & 269  &  -- \\
$\rm ^{12}$CO &  0.252 & 52 & 279  &  1575$^1$  \\
$\rm ^{13}$CO &  0.80  &130 & 279 & 1575$^1$ \\
C$\rm ^{18}$O & 0.89 & 146 & 275  & 1575$^1$ \\
p$-$H$\rm _2$CO  & 1.20 & 195 & 279  & 3260$^2$  \\
SO  & 1.41 &  229 & 105  &  2600 $^3$\\
HCO$\rm ^+$ &  0.00 & 0.0 & 92  & -- \\
HCN & 1.02  & 166 & 99 & 2050$^3$  \\
\hline
\end{tabular} \\
\noindent
Refs: $^1$ \cite{Fayolle2016}; $^2$ \cite{Noble2012}; $^3$ \cite{Garrod2006}
\end{table}

\section{Gas kinetic temperature}\label{Sec:kin_temp}

The gas kinetic temperature determines the thermal pressure and is therefore a key parameter to describe the disk shape and its stability. The gas kinetic temperature is also important from the chemical point of view since many reaction rates are highly dependent on gas temperature. Models computing  the gas temperature in protoplanetary disks typically find that the thermal coupling of  gas and dust is a good approximation over most of the disk and 
that this assumption breaks down at the surface layers of the disk, where the value of the visual extinction A$_V$ in the vertical outward direction becomes lower than 1 \citep{KampDullemond2004, Woitke2009, Walsh2010}. However, recent studies \citep{Akimkin2013, Facchini2017} have shown that the gas and dust temperatures can diverge at much higher densities ($\rm n_H \sim 10^{6}-10^{7}~cm^{-3}$). We assume that the the gas kinetic temperature can be used as a tracer of the dust temperature in the disk layers from which the bulk of the molecular emission originates, where densities are expected to be larger than $10^{7}~cm^{-3}$.

\begin{figure}[h!]
\begin{center}
 \includegraphics[width=0.5\textwidth,trim = 9mm 0mm 0mm 0mm,clip]{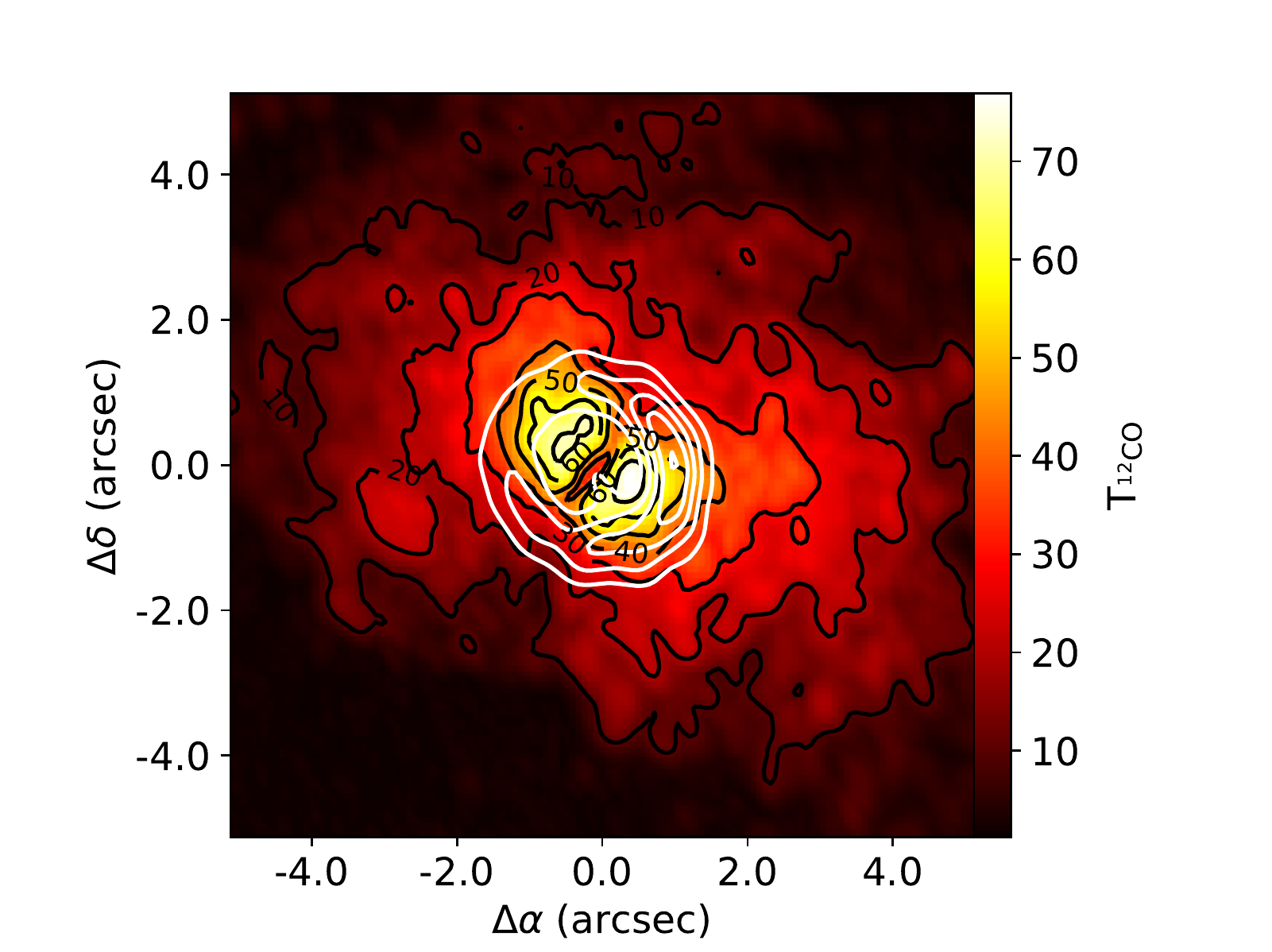}
 \caption{Peak temperature map of $\rm ^{12}$CO. The white solid contours depict the continuum emission, starting at 5$\sigma$ and ending at the map maximum of 0.68 Jy/beam.}
 \label{Fig:12CO_peak_temp}
\end{center}
\end{figure}

To observationally determine the gas kinetic temperature in the different layers of the disk is not an easy task and different molecular probes are needed. Based on our data we use two different approaches. First we estimate the gas kinetic temperature, $T_k$, from the intensity of the optically thick CO 2-1 line. In the case of optically thick emission, $T_b \sim h \nu /k (\exp(-h \nu/kT_{ex}) -1)^{-1}$. The critical density of the J=2$\rightarrow$1 line of $^{12}$CO is as low as n(H$_2$)$\sim$a few 10$^3$ cm$^{-3}$, and one would expect that the emission is thermalized, $T_{ex}$ = $T_k$, even at the disk surface. Therefore, the $^{12}$CO line is a good tracer of the gas kinetic temperature in the surface molecular layer with densities of $>$ 10$^3$ cm$^{-3}$. In Fig. \ref{Fig:12CO_peak_temp}, we show the peak intensity map of the $^{12}$CO 2$\rightarrow$1 line. The values of  $T_b$ range from $\sim$ 70 K in the vicinity of the star to $\sim$ 10 K at a distance of 4$\arcsec$ from the star, which is further away than the dusty ring, in the protostellar envelope. The kinetic temperature is $\sim$15 K in the envelope; it increases to 55 K in the dust ring, and reaches a value of $\sim$70 K in positions close to the star. These temperatures correspond to the $\tau \sim 1$ layer in the disk or envelope surface. Indeed, hotter gas as warm as $\sim$1000 K is expected in the diffuse disk atmosphere and close to the star, but these regions are not expected to contribute to the molecular emission we are observing. We note that \cite{Semenov2005} reported $\rm T > 15~K$ for the outer disk and envelope using single-dish observations. The discrepancy could be explained if the emission is not highly optically thick at these distances, resulting in $\rm T_{ex} < T_{kin}$, thus making the $\rm ^{12}CO$ peak temperature a poor proxy of the overall gas temperature in the outer regions of the system.

An independent estimate of the gas kinetic temperature can be obtained by fitting  the p-H$\rm _2$CO lines. Taking into account the higher critical density of these transitions and the fact that p-H$\rm_2$CO is expected to be abundant in a layer closer to the disk midplane than CO, the temperature derived for the  p-H$\rm _2$CO is expected to be lower than that derived from $^{12}$CO. The $T_b (3 _{22}-2 _{21} +  3_{21}-2 _{20})/T_b  (3 _{03}-2 _{02})$ line ratio is known to be a good gas thermometer \citep[see e.g., ][]{Tang2018}.  To derive the kinetic temperatures from this ratio, we computed $\rm T_b(3\rm _{22}-2\rm _{21})$, $\rm T_b(3\rm _{21}-2\rm _{20})$, and $\rm T_b(3\rm _{03}-2\rm _{02})$ for a grid of kinetic temperatures, assuming n(H$\rm _2) = 10^{8}~cm^{-3}$ and N(H$_2$CO) = 3$\times$10$\rm ^{12}~cm^{-2}$. Using the  channels with emission over 3$\sigma$ of the resulting spectra we derive temperatures in the range 29 to 71 K,  with a mean value of 39 K over the disk. This temperature range is compatible with the dust temperature estimated by \cite{Pacheco2016} using the RADMC code \citep{Dullemond2004} for the disk midplane at R$\sim$200 au, where the p-H$\rm _2$CO radial emission peaks, suggesting gas-dust thermalization in the  p-H$\rm _2$CO emitting region. Since we also detected two SO lines, we can use them to derive another temperature estimate. However, we note that the dependence of the SO ratio on the temperature is steeper than that on H$_2$CO. The mean temperature derived from SO over the disk is 37 K, again in good agreement with previous estimates, and with the value derived from the H$_2$CO ratio.
 
One interesting result is that the disk around this Herbig Ae/Be star is significantly warmer than the disks around T Tauri stars. For instance,  \cite{Henning2010} derived gas kinetic temperatures of $\sim$10 K based on multi-transition analysis of C$_2$H for the T Tauri disks DM Tau and LkCa 15 using interferometric data.  This is not unexpected because of the higher stellar irradiation of Herbig Ae-Be stars (see also \citealp{Agundez2018}). The AB Aur disk is also warmer than the disk around the Herbig Ae/Be star HD 163296 for which  \cite{Guzman2018} reported an excitation temperature of $\sim$24 K from the H$_2$CO lines.  It then appears that AB Aur is  well suited to studying the gas chemistry in a warm disk, similar to MWC 480 \citep{Loomis2020}. An interesting consequence is that contrary to what happens to T Tauri disks and HD 163296, CO is not expected to be significantly depleted in AB Aur and can be used as a reliable tracer of gas mass.

\section{Gas-to-dust ratio}\label{Sec:gtd_ratio}
High-spatial-resolution images of gas and dust in protoplanetary disks have shown that the spatial distribution of both phases can be dramatically different, implying that the gas-to-dust ratio  varies along the radial and azimuthal directions. Studying the spatial variations of this ratio is therefore of paramount importance for understanding protoplanetary disk evolution and planet formation theories.

In this section, we use our observations of $^{13}$CO, C$^{18}$O, and continuum at 1 mm to estimate the gas-to-dust ratio along the dust ring. To that aim, we compute the $^{13}$CO column densities using the standard local thermodynamic equilibrium (LTE) formula, 

\begin{equation}
N({\rm ^{13}CO}) =\frac{8 \pi \nu_{ij}^{3} \Delta V Q_{rot} \tau}{c^{3} A_{ij} g_{up}}e^{\frac{E_{up}}{k T_{rot}}} \left( e^{\frac{h \nu_{ij}}{k T_{rot}}} - 1 \right)^{-1}
,\end{equation}
where $\nu_{ij}$ is the frequency of the $\rm ^{13}CO$ 2-1 transition, $\Delta$V is
the line width, $Q_{rot}$ is the partition function, $\tau$ is the line optical depth, c is the speed of light, $A_{ij}$  is the Einstein coefficient for spontaneous transitions, $g_{up}$ is the degeneracy coefficient, $E_{up}$ is the energy of the upper level, k is
the Boltzmann constant, $T_{rot}$ is the rotational temperature, and h is the Planck constant. For $T_{rot}$ we assumed
$T_{rot}$= 39~K, the value derived from the H$_2$CO ratio, $\Delta$V can be obtained from the second moment maps, and we
only needed to derive the line opacity assuming abundance ratios $\rm ^{12}$CO/$\rm ^{13}$CO = 60  \citep{Savage2002}, and 
 $\rm ^{12}$CO/C$\rm ^{18}$O = 550 \citep{Wilson1994}. The total gas column density can be derived assuming
a CO (or C$^{18}$O) abundance with respect to H$_2$. The observed abundances of C$^{18}$O in the interstellar medium (ISM)
are  in the range of 1-3 $\times$10$^{-7}$ \citep{Frerking1982, Frerking1987, Trevino2019}, with
the lowest values being found in cold and dense regions.
For our calculations, we adopt X(C$\rm ^{18}$O)=(1.7$\pm$0.7)$\times$10$^{-7}$ which is the averaged  value in molecular clouds.
 
The dust column density was derived from the continuum maps using:

\begin{equation}
N_{dust} = \frac{S_{\nu}}{B_{\nu}(T_{dust})\kappa}
,\end{equation}
with $\kappa_\nu$ = 1 g cm$\rm ^{-2}$ \citep{Ossenkopf1994}, where we assumed $T_{dust}$ = $T_{gas}$=39~K. The dust opacity $\kappa_\nu$ depends on the amount of ice in the mantle, as well as on the gas density. We have adopted the value estimated for ice-coated grains in dense regions which is the standard choice for protoplanetary disks.

\begin{figure}[h!]
\begin{center}
 \includegraphics[width=0.5\textwidth,trim = 0mm 0mm 0mm 0mm,clip]{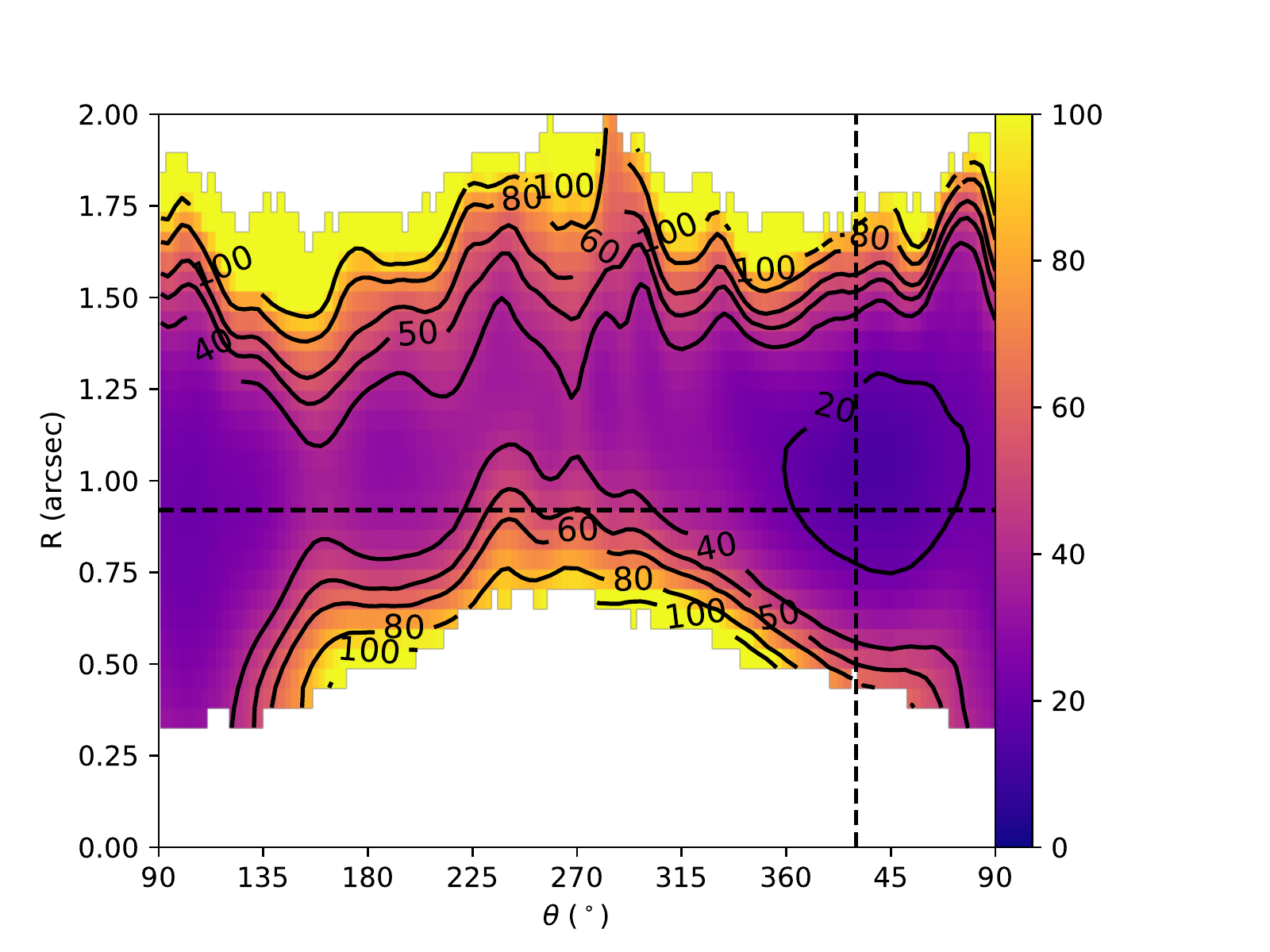}
  \caption{Deprojected gas-to-dust ratio map in polar coordinates. The horizontal and vertical black dashed lines mark the position of the continuum peak.}
 \label{Fig:gas-to-dust_ratio_polar}
\end{center}
\end{figure}

Finally, column densities were converted into gas and dust masses and the ratio of the two was computed. We
show the resulting map in polar coordinates in Fig. \ref{Fig:gas-to-dust_ratio_polar}. We masked pixels with a S/N$<$5 in order to exclude unreliable results.
Taking into account the uncertainty in the assumed value of the
CO abundance, we estimate that the average gas-to-dust ratio in the disk is in the range $\sim$30$-$70, with a minimum 
of $\sim$8$-$16 towards the dust trap. These values are significantly smaller than the canonical value of 100 
representative of the ISM. Recent studies of the gas-to-dust ratio in individual sources have
arrived at similarly low values \citep{Osorio2014, Boehler2017, Wu2018, Soon2019, Miley2019}. Indeed, the
statistical studies by \cite{Williams2014, Ansdell2016, Miotello2017}, and \cite{Long2017} were
dominated by low gas-to-dust ratios. There are different explanations for the low gas-to-dust ratios observed.
First, protoplanetary disks can have intrinsically low gas masses. Alternatively, a larger-than-expected CO depletion can lead
to artificially low gas masses. The second mechanism is expected to be more important for the coldest disks towards T Tauri disks
than towards AB Aur. In Sect. \ref{Sec:astrochem_model}, we carry out the complete chemical modeling of the AB Aur disk in order to provide a more precise determination of the gas-to-dust ratio. In summary, we note that several assumptions are made when computing gas-to-dust ratios. Such assumptions (opacity law, relative CO abundance) can lead to artificially low gas-to-dust ratios.  Nevertheless, the low gas-to-dust ratio derived in this section agrees with the results from the more sophisticated models computed in Section \ref{Sec:astrochem_model}.

Interestingly, the minimum in the gas-to-dust ratio is reached at the position of the continuum peak, indicating that the dust 
trap is particularly dust rich. The same result was obtained by \cite{Boehler2017} in HD~142527. Dust trapping in local pressure maxima has been proposed to overcome the radial drift of solid particles, and hence favor planet formation in transitions disks. In particular, \cite{Fuente2017} proposed the existence of a dust trap in AB Aur to explain the 1.1mm and 2.2mm continuum maps. Our observational results confirm this scenario and provide valuable constraints for the two-fluid hydrodynamical simulations of this prototypical disk.

\section{Chemical segregation}
\begin{figure*}[h!]
\begin{center}
  \includegraphics[width=1.0\textwidth,trim = 0mm 0mm 0mm 0mm,clip]{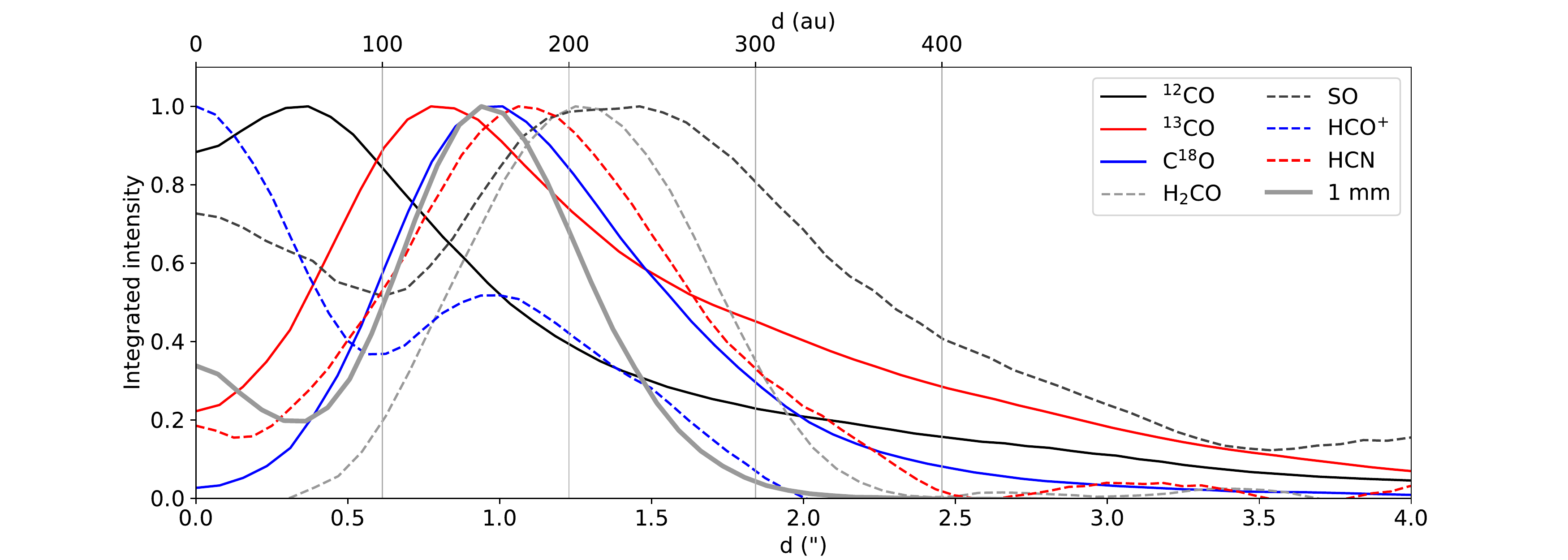}
  \caption{Radial profiles of de-projected integrated intensity maps (moment zero maps) normalized to the peak of the distribution, assuming $i=24.9^{\circ}$ and $PA=37^{\circ}$. We also include for comparison the radial profile of the continuum intensity.}
 \label{Fig:rad_prof}
\end{center}
\end{figure*}

AB Aur is one of the few transitional disks that have been imaged at high spatial resolution in various molecular species, providing a valuable opportunity to study the chemistry in transition disks. In Fig. \ref{Fig:rad_prof} we show the integrated intensity radial profiles of the continuum and the different molecular transitions observed at high spatial resolution after de-projection with i = 26$^{\circ}$ and PA = -37$^{\circ}$ (see Sect. \ref{Sec:cont_emission}). In addition to the data presented, we have included the profiles of the HCO$\rm ^+$ 3$\rightarrow$2 and HCN  3$\rightarrow$2 maps reported by \cite{Riviere2019}. 

Overall, we observe strong radial segregation between the different species, 
with differences as large as $\rm \sim$100 au in their radial emission peaks. The only species that peaks
toward the center is HCO$\rm ^+$, while $\rm ^{12}$CO peaks at $\sim$0.$\arcsec$37 (60 au). HCO$\rm ^+$ is an important ion in the disk intermediate layers, and its abundance seems to increase toward the more heavily UV-irradiated regions of the disk \citep{Sternberg1995, Fuente2003, Riviere2019}. The other species peak at positions ranging from 0.$\arcsec$8 (130 au, $\rm ^{13}$CO) to  1.$\arcsec$5 (244 au, SO). Continuum emission at 1 mm peaks at 0.$\arcsec$94 (153 au), close to the C$\rm ^{18}$O peak, and overlaps with C$\rm ^{18}$O almost perfectly in the inner regions, while C$\rm ^{18}$O is more extended in the outer parts. The HCN spatial distribution from \cite{Riviere2019} is almost coincident with C$\rm ^{18}$O emission. The most extended species are $\rm ^{12}$CO and $\rm ^{13}$CO, which still show emission at distances $\rm > 4\arcsec$. Again, this is compatible with emission from the remnant envelope in the outer parts of the system. The high densities and cold temperatures prevailing  in the midplane regions produce a rapid and efficient adsorption of gas-phase molecules onto dust grains.  One might postulate that this layered structure is related to the snow lines of the different species, which are determined by their binding energies. For comparison, we list the binding energies of the different species in Table \ref{Tab:profile_peaks}. If thermal desorption were the main desorption mechanism in the midplane,  the radius of the midplane snow line would be expected to anti-correlate with the binding energy, that is, the smaller the binding energy, the further away the snow line. The distribution of radial emission peaks of the molecules observed in AB Aur does not follow this trend (see Fig. \ref{Fig:rad_prof} and Table \ref{Tab:profile_peaks}). Indeed, the farthest emission peak radius corresponds to SO which is the molecule with the second-largest binding energy. This suggests that thermal desorption is not driving the chemical composition of this disk. Full gas-grain chemical models show that the proportionality between the binding energy and the condensation temperature is in the range 30$-$50 K \citep{Hollenbach2009,  MartinDomenech2014, Agundez2018}. Following this rule, the condensation temperatures for CO, HCN, SO, and H$_2$CO are $\approx$30 K,  $\approx$40 K, $\approx$50 K, and $\approx$ 65 K, respectively. Assuming that the temperature derived from H$_2$CO is close to the midplane temperature, this would imply that the temperature in the dusty ring is lower than the condensation temperature for all species except for CO and HCN. We note that most of the emission from the observed species arises from the warm molecular layer where freeze-out is not efficient. Furthermore, freeze-out timescales in the outer disk could be of the order of the disk lifetime. Finally, nonthermal processes such as shocks could counter the effect of freeze-out,  further lowering its role on the chemistry of this warm disk.

\begin{figure*}[h!]  
\begin{center}
  \includegraphics[width=\textwidth,trim = 0mm 10mm 0mm 0mm,clip]{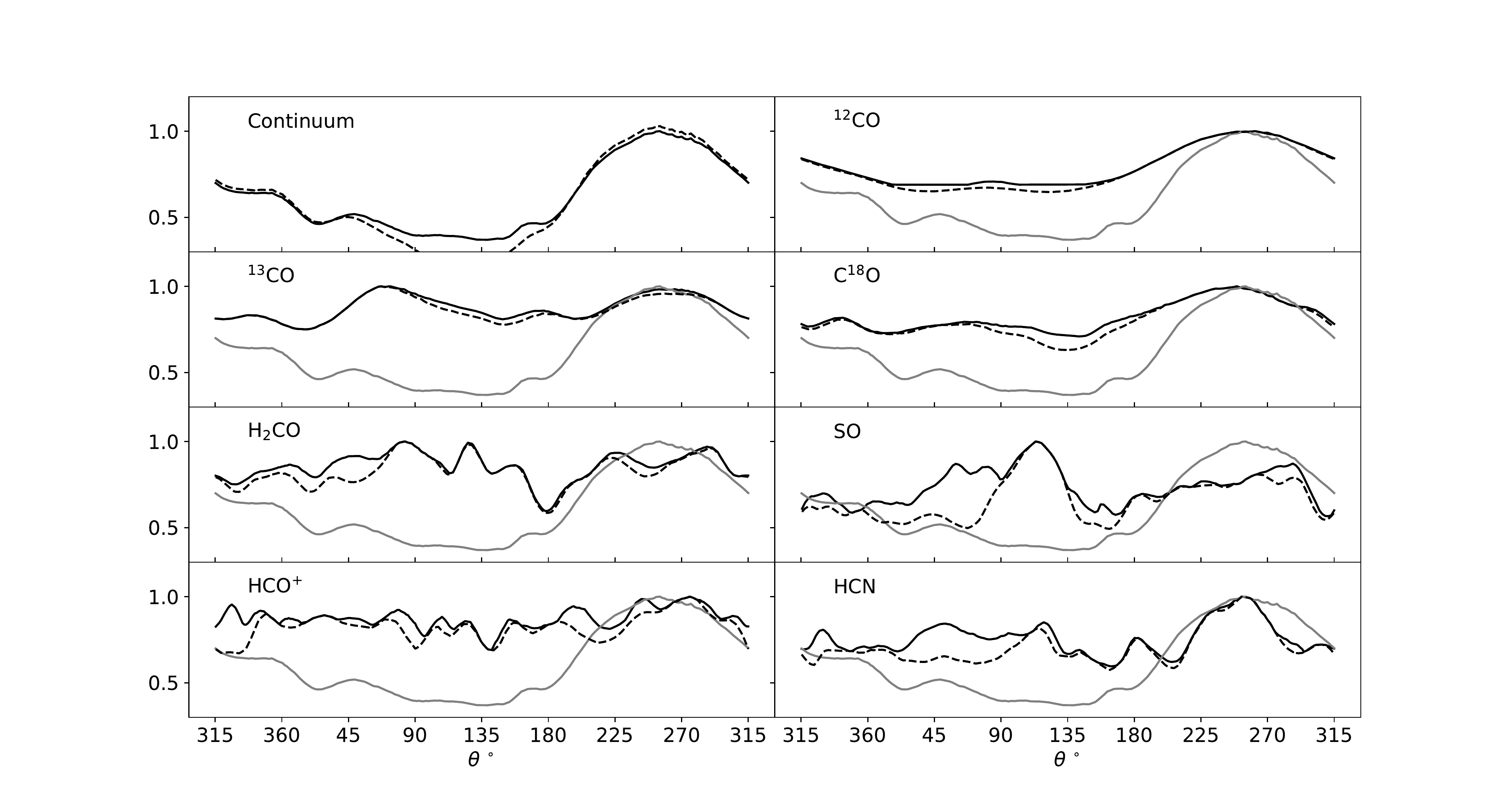}
  \caption{Cuts along the azimuth of the intensity normalized to its maximum value for the continuum emission and the different species surveyed after de-projection. The black solid line depicts the cut at the radial distance of the emission peak. The black dashed line depicts the cut along the crescent (see text). The grey line depicts the continuum azimuthal cut along the maximum.}
 \label{Fig:az_prof}
\end{center}
\end{figure*}

Significant differences are also found in the spatial distribution of the different molecules in azimuth. This is not surprising taking into account the asymmetric distribution of the solid particles.  We show in Fig. \ref{Fig:az_prof} the azimuthal profiles of the different species detected and the positions of the peaks in the azimuth are given in Table \ref{Tab:profile_peaks}. Only the emission of C$\rm ^{18}$O follows that of the continuum emission with its emission peak close to the position of the dust trap ($\sim$315$^{\circ}$). In the  case of $^{13}$CO, we observe two peaks with similar intensity at $\sim$315$^{\circ}$ and $\sim$135$^{\circ}$ which roughly corresponds to the maximum (dust trap) and minimum (counter-dust trap) in the continuum emission. Most striking is the case of SO which peaks at  $\sim$180$^{\circ}$ (in a position nearly opposite to the dust trap). The emission of HCN, H$_2$CO, and HCO$^+$ is relatively flat along the ring.

It is tempting to think that differences in the grain size distribution and gas-to-dust ratio would produce changes in the chemical composition of the gas. \cite{Pacheco2016} carried out some chemical calculations to investigate the influence of gas-to-dust ratio, gas density, and grain size on the abundances of H$_2$CO and SO. These latter authors concluded that the gas-to-dust ratio and grain size have a moderate impact on the abundances of these species, which are mainly determined by the gas density and time evolution as long as the dust temperature is below the condensation temperature. Furthermore, they find that the abundance of SO is the most sensitive to the gas density, and its abundance decreases to values $<$10$^{-12}$ in less than 0.1 Myr, the typical age of the dust trap, for  high densities (n(H$_2$)$>$10$^7$ cm$^{-3}$). Below  the SO evaporation temperature, the depletion of SO is proportional to the gas density and azimuthal variation in the SO abundance might reflect changes in the average gas density along the disk.  This would explain why SO does not peak towards the dust trap. This model assumed that the molecules are well shielded from the stellar UV radiation and the same temperature for small and large grains. The equilibrium temperature of small grains depends on their size and grain composition \citep{Ysard2019}. Small particles are expected to be warmer than large particles, avoiding the freeze out of the most volatile molecular species. This effect would contribute to increase the dust temperature in the counter-dust trap, hence boosting the SO abundance. It would also explain why the $\rm ^{13}$CO emission  has a secondary peak towards a position opposite the dust trap.

\begin{figure*}[h!]
\begin{center}
 \includegraphics[width=0.4\textwidth,trim = 0mm 0mm 0mm 0mm,clip]{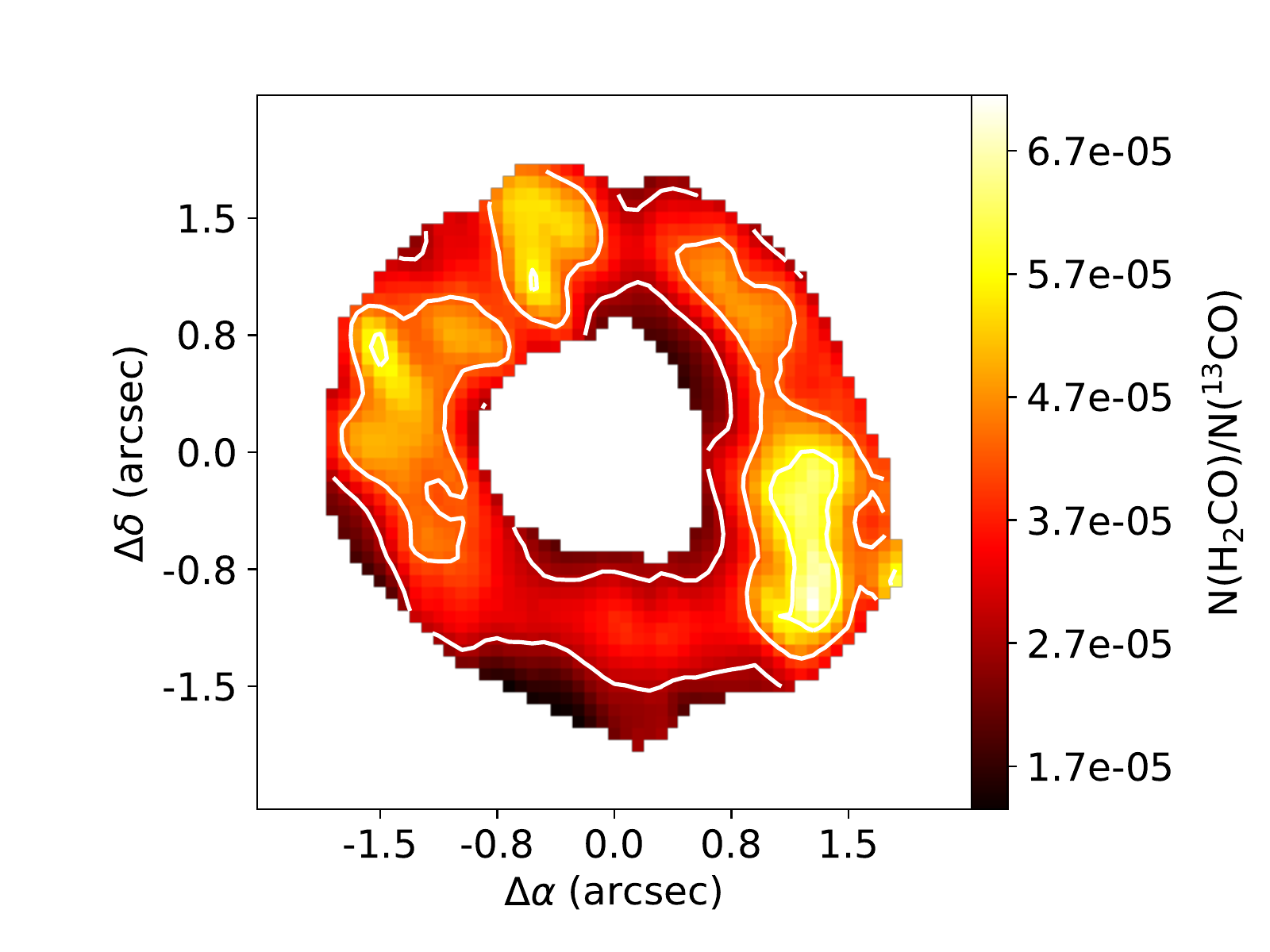}\includegraphics[width=0.4\textwidth,trim = 0mm 0mm 0mm 0mm,clip]{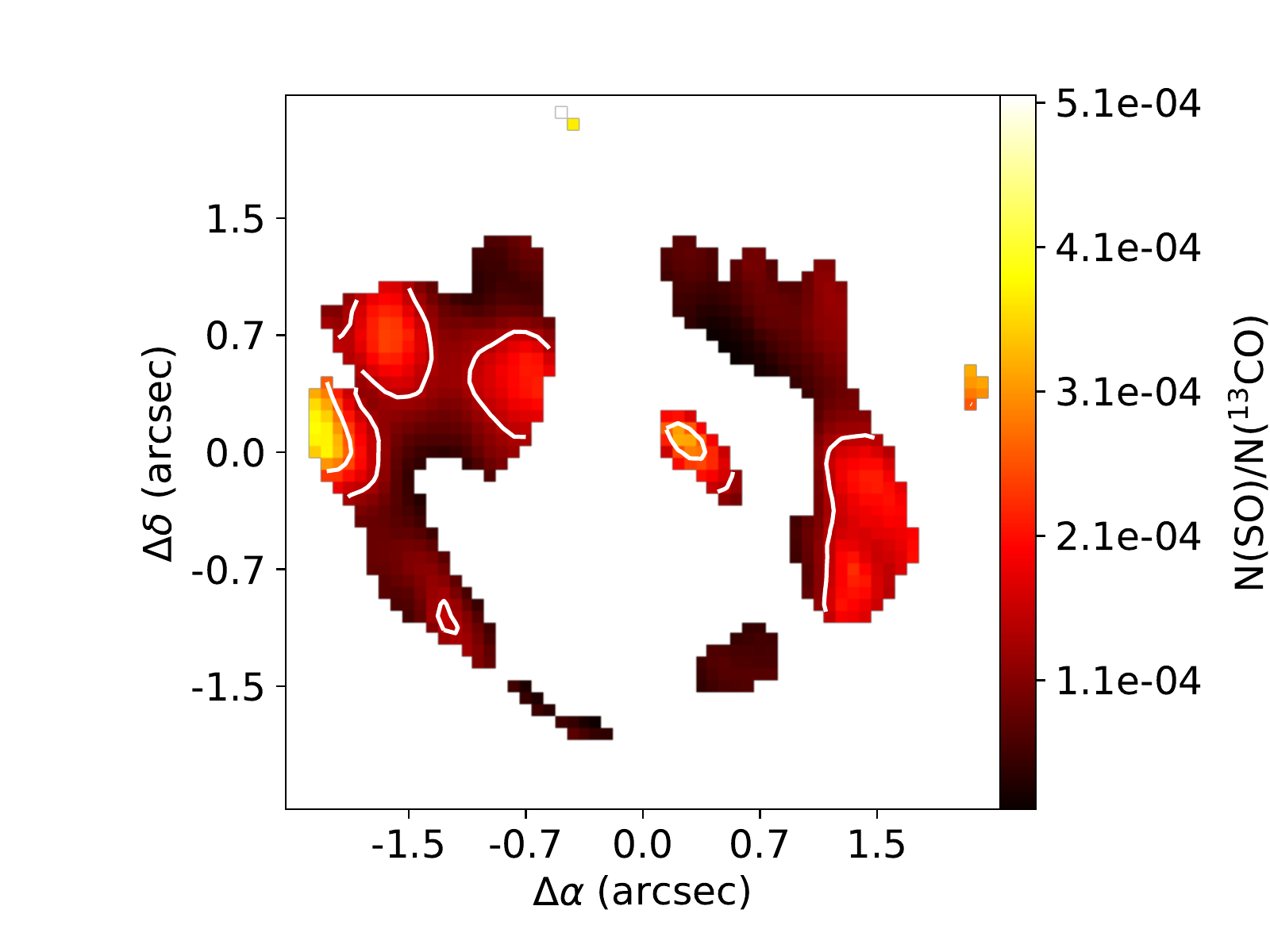}

\includegraphics[width=0.4\textwidth,trim = 0mm 0mm 0mm 0mm,clip]{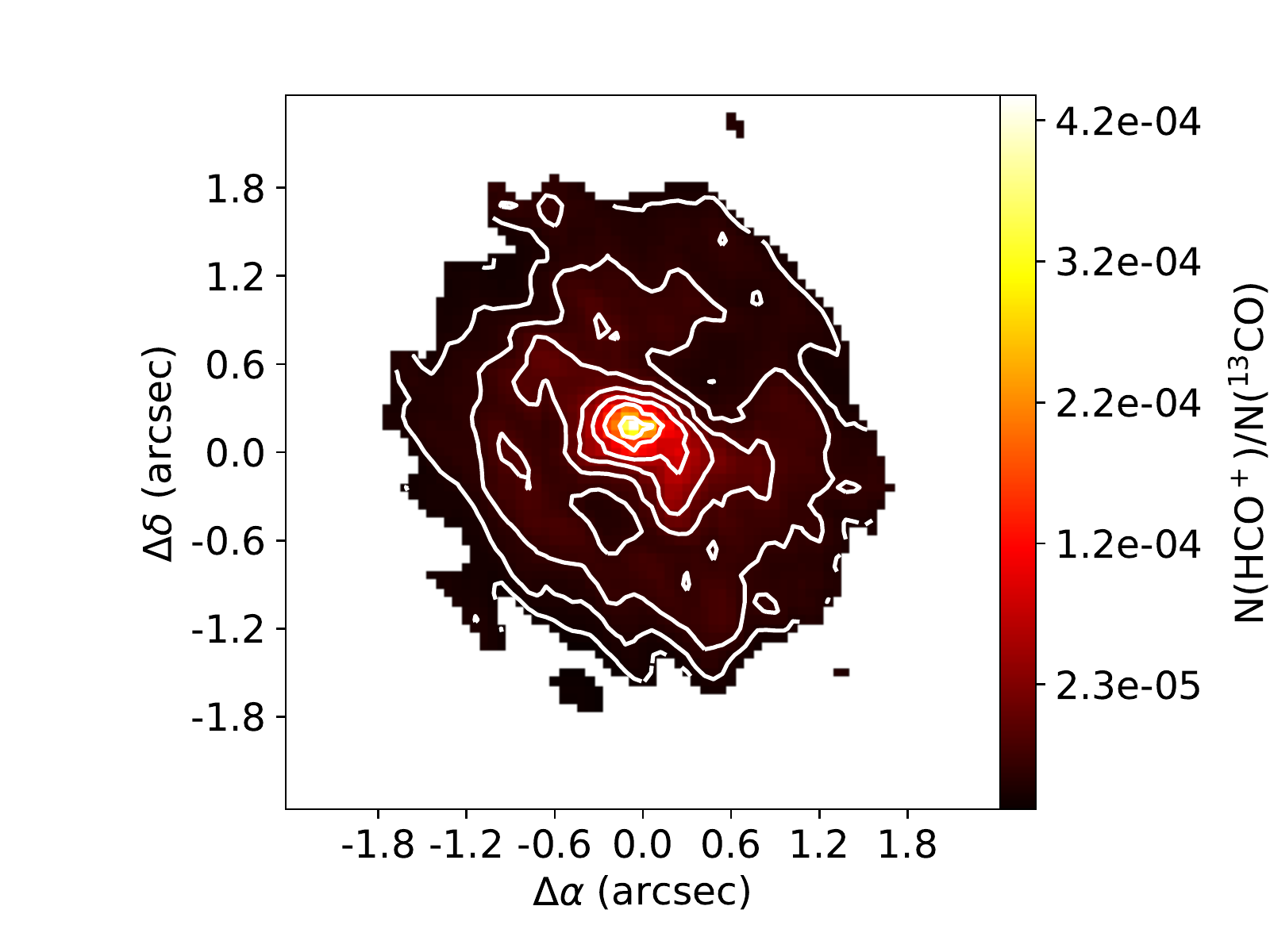}\includegraphics[width=0.4\textwidth,trim = 0mm 0mm 0mm 0mm,clip]{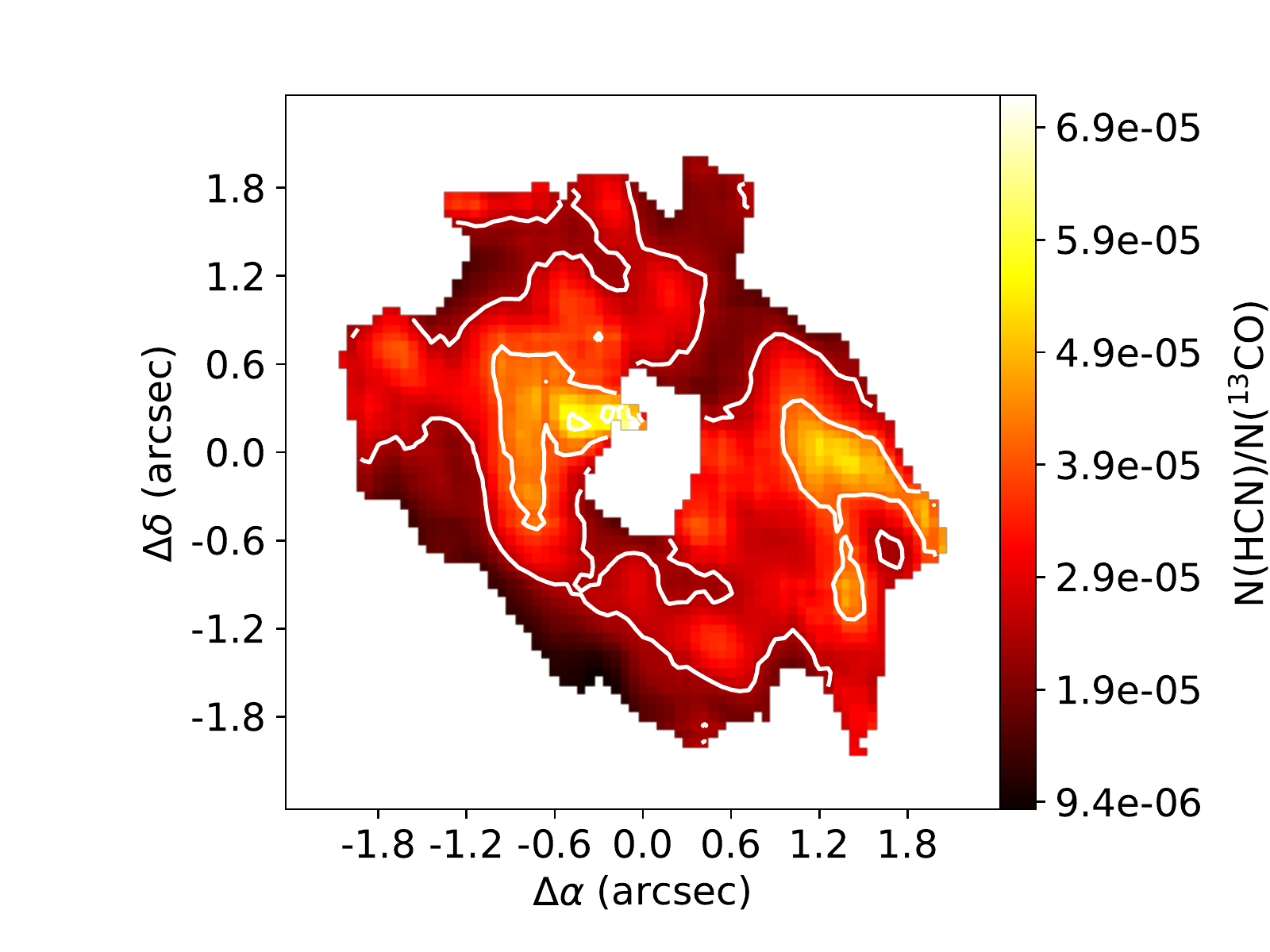}
 \caption{Abundance maps of H$_2$CO (top left), SO (top right), HCO$^+$ (bottom left), and HCN (bottom right). The white contours depict five abundances equally spaced between the map minimum and the map maximum.}
 \label{Fig:abundance_maps}
\end{center}
\end{figure*}

\begin{figure}[h!]
\begin{center}
  \includegraphics[width=0.5\textwidth,trim = 0mm 0mm 0mm 0mm,clip]{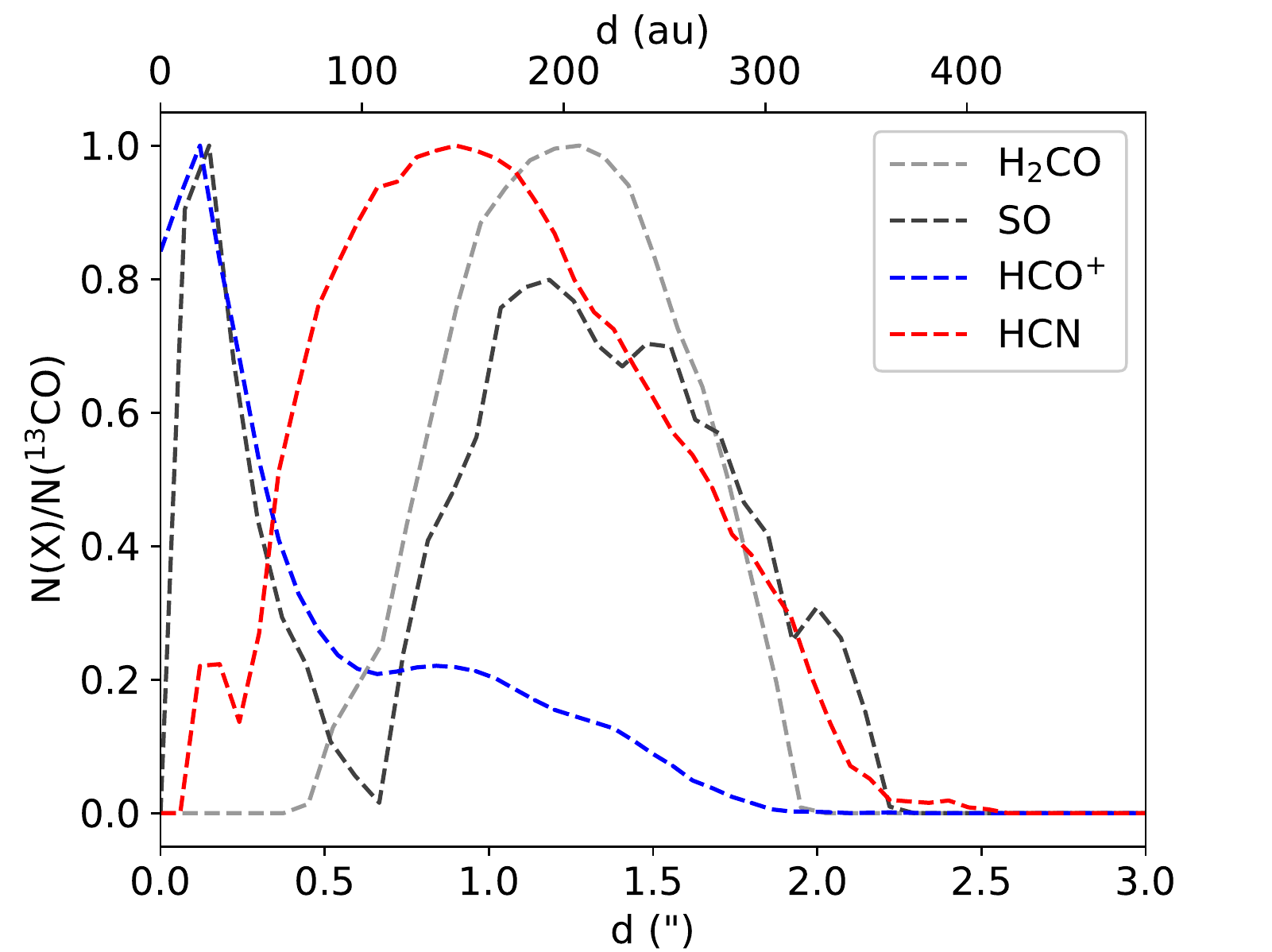}
  \caption{Abundance radial profiles as a function of distance.}
 \label{Fig:ab_rad_prof}
\end{center}
\end{figure}

\section{Molecular column densities and abundances}

We estimated the column densities of CO, H$\rm _2$CO, and SO assuming LTE, and the
mean kinetic temperature derived from the H$\rm _2$CO line ratios. We also computed column density maps for HCO$\rm ^+$ and HCN using the data published by \cite{Riviere2019}, and following the same procedure. Table \ref{Tab:col_dens} summarizes the results for the different species. As a first step to estimate molecular abundances,  we use the $^{13}$CO column densities derived in Section \ref{Sec:gtd_ratio} to compute N(X)/N($\rm ^{13}$CO) ratios, where X refers to the different molecules. The resulting maps are shown in Fig. \ref{Fig:abundance_maps}. Pixels with relative errors larger than 20\% have been masked. The abundance radial profiles derived from these maps are shown in Fig. \ref{Fig:ab_rad_prof}. The molecular ion HCO$^+$ is the only one which presents emission towards the star position. As discussed by \cite{Riviere2019}, this is better understood as the consequence of photo-chemistry.  In contrast, the N-bearing species HCN is probing the high-density gas within the dusty ring. With $\rm T_{eff} \sim 10^4$ K, AB Aur stellar emission peaks at the soft UV (around 3000 $\rm \AA$), which readily dissociates HCN, and therefore HCN appears where it is shielded by the dust. In contrast to HCN and HCO$\rm ^+$, the peaks in the abundance of H$_2$CO and SO are located beyond the dusty ring, in the outer disk. In the following, we discuss the chemistry of the different
surveyed species in more detail.

\subsection{H$_2$CO}
Formaldehyde is one of the most abundant  molecules in circumstellar disks \citep{Oberg2010, Fuente2010,  Oberg2011, Qi2013, Pacheco2015, Pacheco2016, Guilloteau2016, Carney2017, Oberg2017, Guzman2018, Pegues2020}. As an organic species that can lead to the formation of more complex organic compounds, it has received a lot of attention with dedicated single-dish and interferometric surveys. In particular, H$_2$CO was detected in 10 out of the 20 disks of the spectral survey carried out by \cite{Guilloteau2016} using the IRAM 30m telescope. More recently, an ALMA survey of H$_2$CO was performed by \cite{Pegues2020} with 13 detections out of a sample of 15 disks.  Restricting the sample to Herbig Ae/Be stars, the millimeter lines of formaldehyde have been detected towards HD142527  (\cite{Oberg2011}), AB Aur \citep{Fuente2010, Pacheco2015, Pacheco2016}, HD 163296 \citep{Qi2013, Carney2017, Guzman2018},
and MWC 480 \citep{Pegues2020}.

The  map of the H$_2$CO abundance towards AB Aur  is shown in Fig. \ref{Fig:abundance_maps}. The p-H$\rm _2$CO abundance varies within a factor  of approximately seven along the disk with the radial abundance peak located at  R$\sim$200 au, beyond the dusty ring traced by the 1.3mm emission. This kind of radial profile, with the peak of H$_2$CO abundance beyond the dusty ring, was previously observed in HD 163296 \citep{Qi2013, Carney2017, Guzman2018} and TW Hya \citep{Qi2013, Oberg2017}, and has been interpreted in terms of surface chemistry. Beyond the CO snow line, formaldehyde is formed  by hydrogenation of CO on the grain surfaces and is then released to the gas phase mainly by photo-desorption. \cite{Guzman2018} determined an ortho-to-para H$_2$CO ratio of 1.8$-$2.8 corresponding to a spin temperature of 11 $-$ 22 K, and provides further support to a low-temperature origin for the H$_2$CO molecules in HD~163296, consistent with formation on grain surfaces. This interpretation cannot be directly extrapolated to the case of AB Aur. The disk around AB Aur is significantly warmer than those around T Tauri stars. Indeed, \cite{Qi2013} derived a H$_2$CO rotation temperature  below 20~K for a sample of T Tauri disks.  The disk around AB Aur is also quite different
from the well-known full gas-rich disk HD 163296. For comparison, the mass of the HD~163296 disk
is one order of magnitude larger than that of AB Aur \citep{Woitke2019}. Furthermore,  the AB Aur disk is warmer.
\cite{Woitke2019} derived a  mean gas temperature, T$_g$=29~K, and mean dust temperature, T$_d$=27~K, for HD~163296. Using the same methodology, these authors derived mean temperatures  of 54~K and 36~K for the gas and dust in AB Aur, respectively. In the case of the warm disk around AB Aur, having an H$\rm _2$CO that extends further out than CO most likely implies that it is formed through gas-phase reactions, and not through CO surface hydrogenation. A reaction of the following kind could be the main formation mechanism:

\begin{equation}
CH_3 + O \rightarrow H_2CO + H
.\end{equation}

 Assuming X($^{13}$CO)=1.7$\times$10$^{-6}$, we estimate that the peak  p-H$_2$CO abundance is $\sim$10$^{-10}$. 
 We adopt the equilibrium value of the ortho-to-para ratio at 39~K (ortho-to-para ratio = 3), and derive a peak  H$_2$CO abundance  of $\sim$4$\times$10$^{-10}$.  Based on higher angular resolution ALMA data,  \cite{Carney2017} derived abundances of 4$-$8$\times$10$^{-12}$ for HD 163296.  Even assuming an uncertainty of a factor of five in our estimate of the H$_2$CO abundance because of the assumed value of X($^{13}$CO), our data prove that the abundance of H$_2$CO in AB Aur is more than 50 times higher than in HD~163296.

\subsection{SO}
Sulfur monoxide is a useful tool to search for the warm (T$_k$$\sim$60$-$100 K)  disks in Class 0 stars \citep{Maret2020}. However, contrary to H$_2$CO, sulfur monoxide has been detected in very few protoplanetary disks. In fact, the detection of SO in AB Aur was the first one in a protoplanetary disk \citep{Fuente2010}. Later, \cite{Guilloteau2016} detected SO in only 4 out of the 20 protoplanetary disks of their sample using the IRAM 30m telescope.  High-angular-resolution observations of SO in Class I objects reveal that SO is coming from a narrow ring located at the interface between the disk and the molecular cloud \citep{Sakai2014, Podio2015, Sakai2016}. These works suggest that the SO abundance is enhanced in the centrifugal barrier and outer parts of the disk due to 
possible accretion shocks which heat the gas to temperatures higher than 60~K. The SO abundance seems to decrease after the main accretion phase when the circumstellar disk becomes colder. The detection and high abundance of SO in AB Aur together with the high H$_2$CO abundance derived from our high-spatial-resolution data proves that this warm disk hosts a differentiated chemistry which merits a dedicated study. In Sect. \ref{Sec:astrochem_model}, we model the chemistry of this interesting disk.

\begin{table}
\caption{Column density statistics for the different species surveyed.}
\label{Tab:col_dens}
\centering
\begin{tabular}{llll}
\hline \hline
Species & N(X)$\rm _{min}$  &  N(X)$\rm _{max}$ &  N(X)$\rm _{mean}$  \\
 --           &  (cm$\rm ^{-2}$)    & (cm$\rm ^{-2}$) &  (cm$\rm ^{-2}$)   \\
\hline
$\rm ^{13}$CO &  8.8$\rm \times 10^{15}$ & 2.5$\rm \times 10^{17}$ & 9.8$\rm \times 10^{16}$ \\
p-H$\rm _2$CO &  3.1$\rm \times 10^{12}$ & 8.3$\rm \times 10^{12}$ & 5.3$\rm \times 10^{12}$ \\
SO &  2.1$\rm \times 10^{13}$ & 3.4$\rm \times 10^{13}$ & 2.5$\rm \times 10^{13}$ \\
HCO$\rm ^+$ &  5.3$\rm \times 10^{12}$ & 3.9$\rm \times 10^{13}$ & 1.1$\rm \times 10^{13}$ \\
HCN &  2.3$\rm \times 10^{12}$ & 6.4$\rm \times 10^{12}$ & 3.6$\rm \times 10^{12}$ \\
\hline
\end{tabular}
\end{table}

\section{Astrochemical modeling}\label{Sec:astrochem_model}
Protoplanetary disks constitute the link between the ISM and planetary systems, and the study of sulfur species in these objects is of paramount importance to understand the chemical composition of the Solar System and comets. Searches for  S-bearing molecules in protoplanetary disks have provided very few detections. Thus far, only one S-species, CS, has been widely detected in protoplanetary disks. The chemically related compound H$_2$CS was detected by \cite{LeGal2019} in MWC 480 and, tentatively, LkCa 15.  \cite{Phuong2018} reported the detection of H$_2$S in  GG Tau. 

Recent spectral line surveys have increased the number of detected interstellar sulfur molecules to a dozen species in prestellar cores \citep{Vastel2018}, protostellar envelopes \citep{Drozdovskaya2018}, and photon-dominated regions \cite[PDRs][]{Riviere2019}. Together with these new detections, new models have been developed to advance our understanding of the ISM gas-phase S-chemistry. Recently, new {\it ab initio} quantum calculations of key reaction rates have improved the reliability of the SO abundance predictions \citep{Fuente2016, Fuente2019}. A complete revision of the sulfur surface chemistry has been carried out by \cite{Laas2019}. \cite{Navarro2020} developed an up-to-date sulfur chemical network based on the Kinetic Database for Astrochemistry (KIDA) including the most recent updates \citep{Fuente2016, Fuente2019, LeGal2019, Laas2019}.  In the following we apply this updated network to the case of the AB Aur disk.

To further characterize the gas in the AB Aur protoplanetary disk we computed a set of 1+1D Nautilus simulations \citep{Wakelam2016}, following the prescriptions used by \cite{LeGal2019}, in order to model the disks around \object{LkCa 15} and \object{MWC 480} and adapting the physical parameters to match our observations of AB Aur, as listed in Table \ref{Tab:model_parameters}. In the following, we provide a summary of the model equations and parameters. A more thorough description can be found in \cite{LeGal2019}.

\subsection{Physical structure}
In order to simulate the chemical processes in the AB Aur disk, we need to make some assumptions on its physical structure. The vertical temperature profile at a given radius will be assumed to follow the modified prescription of \cite{Dartois2003} used in \cite{Rosenfeld2013}, which reads:

\begin{equation}
\small
T(z) = \left\{
    \begin{array}{ll}
        T_{\rm{mid}}+(T_{\rm{atm}}-T_{\rm{mid}})\left[ \sin \left(\frac{\pi z}{2z_q}\right) \right]^{2\delta}&\mbox{if} \, z<z_q\\
        T_{\rm{atm}}&\mbox{if} \, z\ge z_q,
    \end{array}
\right.
\label{eq:T_z}
\end{equation}
with the temperature in the disk midplane and in the atmosphere varying with radii as
\begin{eqnarray}
T_{\rm{mid}}=T_{\mathrm{mid},R_c}\,\left(\frac{r}{R_{\rm{c}}}\right)^{-q},\\
T_{\rm{atm}}=T_{\mathrm{atm},R_c}\,\left(\frac{r}{R_{\rm{c}}}\right)^{-q}
,\end{eqnarray}
where $R_c$ is a characteristic radius, and $z_q=4H$, where the pressure scale height H is described by
\begin{equation}
H=\sqrt{\frac{k_{\rm B} \, T_{\rm mid} \,r^3}{\mu \,m_{\rm H}\, G \,M_{\star}}},
\end{equation}
where $k_{\rm{B}}$ is the Boltzmann constant, $\mu=2.4$ is the mean molecular weight of the gas, $m_{\rm{H}}$ is the proton mass, and $M_\star$ is the mass of the central star. The midplane temperature $T_{\rm mid}$ follows the equation
\begin{equation}
    T_{\rm{mid}}(r)\approx \left(\frac{\varphi L_\star}{8\pi r^2 \sigma_{\rm{SB}}}\right)^{1/4},
    \label{eq:Tmid_Rc}
\end{equation}
where $L_\star$ is the stellar luminosity, $\varphi$ is the flaring index, and $\sigma_{\rm{SB}}$ is the Stefan-Boltzman constant.

The dust is assumed to have the same temperature as the gas, which is true for most of the disk (only in the uppermost layers are different temperatures expected). The vertical density profile is derived solving 

\begin{equation}
\rho (z) =\rho_0 e^{-\frac{z^2}{2H^{2}}}
,\end{equation}
where $\rho_0$ is the midplane density of the gas. 

A key parameter for astrochemical models of protoplanetary systems is the UV flux impinging the disk. The UV flux coming from the central star is assumed to be a multiple of the ISM radiation field, an assumption which holds true for HAe stars \citep{Chapillon2008}. The UV flux at a given radius is the combination of the photons coming directly from the star plus those scattered downwards by small dust grains in the disk atmosphere, and is described by equation
\begin{equation}
f_{\rm{UV}}=\frac{0.5 f_{\rm{UV},R_c}}{\left(\frac{r}{R_c}\right)^2+\left(\frac{4\rm{H}}{R_c}\right)^2}.    
\end{equation}

\begin{table*}[]
\caption{Model parameters}
\label{Tab:model_parameters}
\centering
\begin{tabular}{lll}
\hline \hline
Parameter nane & Parameter value & Ref.  \\
\hline
Stellar mass, $M_\star$ & 2.4 M$_\odot$ & 1 \\
Characteristic radius, $R_{\rm c}$ & 98 au & 2 \\
Temperature power-law index, $q$ & 0.1 & 3 \\
Mid-plane temperature at $R_{\rm c}$, $T_{\rm mid}$ & 42 K & 2 \\
Atmosphere temperature at $R_{\rm c}$, $T_{\rm mid}$ & 70 K & 2\\
Surface density power-law index, $\gamma$ & 2.15 & 3 \\
Surface density at  $R_{\rm c}$ & 0.5 g cm$\rm ^{-2}$ & 2 \\
Outer radius, $R_{\rm out}$ & 700 au & 2 \\
UV reference flux, $f_{\rm UV, R_c}$ in Draine units & 10$\rm ^5$ & 2\\
\hline
\end{tabular}
\noindent
\tablefoot{
(1) \cite{Riviere2019}; 
(2) this work; 
(3) \cite{Pietu2005};
}
\end{table*}

\subsection{Chemical network}
 Nautilus includes grain surface reactions, which are crucial to explain chemical evolution in high-density environments such as protoplanetary disks. It uses a three-phase scheme that includes  gas phase, grain surface, and grain mantle reactions, and the interaction between them \citep{Ruaud2016}.
 
To take into account chemical inheritance from previous stages, we first simulate the chemical evolution of a starless dense molecular cloud up to a characteristic age of 1~Myr. For this 0D model, typical constant physical conditions were used: grain and gas temperatures of 10~K, a gas density n$_H$ = n(H) + 2n(H$_2$) = 2$\times$10$^4$ cm$^{-3}$  , and a cosmic-ray molecular hydrogen ionization rate of 1.3$\times$10$^{-17}$ s$^{-1}$ ; this parent molecular cloud is also considered to be shielded from external UV photons by a visual extinction of 30 mag. For this first simulation stage, we consider that initially all the elements are in atomic form (see Table \ref{Table:elemental_ab}) except for hydrogen which is assumed to be initially already fully molecular. The elemental gas phase sulfur abundance is still a controversial issue (see e.g.,\citealp{Fuente2016, Vidal2017, Vastel2018, Fuente2019, Navarro2020}). In our models, we considered the high-S abundance case, with the sulfur elemental abundance equal to the solar one \citep[i.e. 1.5$\rm \times 10^{-5}$,][]{Asplund2005}, and the low-S case  with S/H$\sim$8$\times$10$^{-8}$ which is the value usually adopted to fit the abundances of S-bearing species in dark clouds \citep{Agundez2013}. We compare  the output of our 1+1D simulation after 1 Myr with observations. 
 
\begin{table}
\centering
\caption{Initial elemental abundances}
\label{Table:elemental_ab}
\begin{tabular}{lcc}
\hline\hline
Species & $n_i/n_{\text{H}}$    & Reference\\
\hline
H$_2$   & 0.5\\
He      & 9.0$\times$10$^{-2}$              & 1\\
C$^+$   & 1.7$\times$10$^{-4}$              & 2 \\
N       & 6.2$\times$10$^{-5}$              & 2\\  
O       & 2.4$\times$10$^{-4}$              & 3\\ 
S$^+$   & 8.0$\times$10$^{-8}$   & 4 \\
Si$^+$  & 8.0$\times$10$^{-9}$              & 4 \\
Fe$^+$  & 3.0$\times$10$^{-9}$              & 4 \\
Na$^+$  & 2.0$\times$10$^{-9}$              & 4 \\
Mg$^+$  & 7.0$\times$10$^{-9}$              & 4 \\
P$^+$   & 2.0$\times$10$^{-10}$             & 4 \\
Cl$^+$  & 1.0$\times$10$^{-9}$              & 4 \\
F$^+$   & 6.7$\times$10$^{-9}$             & 5 \\
\hline
\label{tab:elemental_ab}
\end{tabular}
\noindent
\tablefoot{
(1) \cite{Wakelam2008}; (2) \cite{Jenkins2009};
(3) \cite{Hincelin2011};
(4) \cite{Graedel1982};
(5) \cite{Neufeld2015}}

\end{table}

\subsection{Model setup}
The physical parameters assumed for AB Aur, summarized in Table \ref{Tab:model_parameters}, are fed into Nautilus to compute the astrochemical model. We modeled a gaseous disk extending from 10 to 700 au surrounding a 2.4 M$_\odot$ star. The midplane temperature of 42 K at 98 au is used to match the value of $\sim$ 39 K at $r \sim$ 200 au that we derived in Section \ref{Sec:kin_temp}. The atmospheric temperature (i.e., T at z = 4H), $T_{\rm atm}$ was set to 70 K. The exponent of the temperature power-law index, q, was set to 0.1 following the value derived by \cite{Pietu2005} using $\rm ^{13}$CO. The surface density at the reference radius (98 au) was set to 0.5 g cm$\rm ^{-2}$, and the exponent of the surface-density power law was set to 2.15 following \cite{Pietu2005}. 

We computed six models (see Table \ref{Tab:model_setup} for an overview) to test the impact of some important parameters in our study: the gas-to-dust ratio, the sulfur abundance, and the C/O abundance ratio. To explore the impact of the gas-to-dust ratio, we computed models with the standard gas-to-dust ratio of the ISM (i.e., 100), as well as models with a gas-to-dust ratio of 40, in agreement with  the value we derived in Section \ref{Sec:gtd_ratio}. For each of the two gas-to-dust ratios, we computed two sets of models: the first with an ISM standard sulfur abundance of 1.5$\rm \times 10^{-5}$, and the second with a depleted sulfur abundance of 8$\rm \times 10^{-8}$.  Finally, to test the impact of the C/O ratio, we computed two additional models with a gas-to-dust ratio of 40, a depleted O abundance resulting in C/O = 1, and the high- and low-sulfur abundances. In total, we discuss six models, which are summarized in Table. \ref{Tab:model_setup}.
 
 \begin{table}[]
\caption{Model setup}
\label{Tab:model_setup}
\centering
\begin{tabular}{llll}
\hline \hline
Model ID & Gas-to-dust ratio & $[S/H]$ & C/O \\
\hline
1 & 40 & 1.5$\rm \times 10^{-5}$ & 0.7 \\
2 & 40 & 8$\rm \times 10^{-8}$ & 0.7 \\
3 & 100 & 1.5$\rm \times 10^{-5}$ & 0.7 \\
4 & 100 & 8$\rm \times 10^{-8}$ &  0.7 \\
5 & 40 & 1.5$\rm \times 10^{-5}$ & 1 \\
6 & 40 & 8$\rm \times 10^{-8}$ & 1 \\
\hline
\end{tabular}
\end{table}
 
\begin{figure*}[h!]
\begin{center}
 \includegraphics[width=1.1\textwidth,trim =30mm 0mm 0mm 0mm,clip]{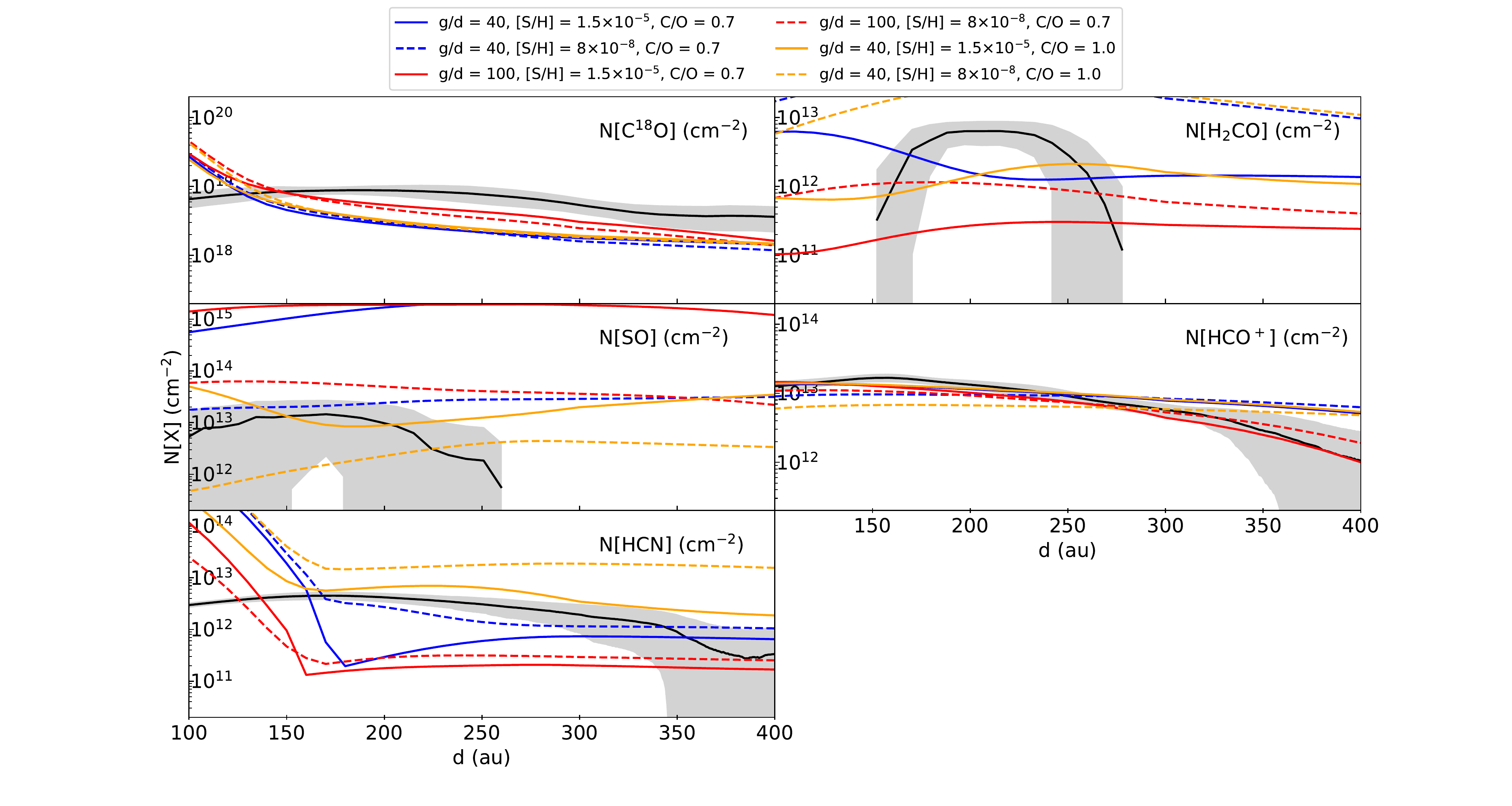}
 \caption{Comparison with models. The black solid line shows the derived radial profile of the column density for each species, with the light-grey shaded region showing the 1$\sigma$ uncertainty. The model parameters for each model are shown in the figure legend.}
 \label{Fig:model_comparison}
\end{center}
\end{figure*}

\subsection{Model results}

Figure \ref{Fig:model_comparison} shows  the results of the six models described in the previous section. Our goal is to 
explore the molecular chemistry in the warm AB Aur disk. In addition, as explained below, our chemical study can help to constrain the gas-to-dust ratio, the initial sulfur abundance, and the C/O ratio in AB Aur.

One of the main results of our study is the estimation of the gas-to-dust ratio in the protoplanetary disk in AB Aur.
For this aim, we used spectroscopic observations of the CO isotopologs and the dust continuum emission at 1.1mm
to determine the gas and dust mass, respectively. Using canonical values for the dust emissivity and the
C$^{18}$O abundance, we determined that the mean gas-to-dust ratio in the dusty ring is $\sim$40$_{-10}^{+30}$.  
We ran models with gas-to-dust ratio equal to 40 and 100 in order to explore the impact of this parameter on the gas chemistry and further corroborate our result. Our models show that the gas-to-dust ratio has little to no impact on the CO and HCO$^+$ column densities. The column densities of SO, H$\rm _2$CO, and HCN are more sensitive to the gas-to-dust ratio. The high abundance of H$_2$CO observed in AB Aur clearly favors models with a gas-to-dust ratio close to 40. 

The elemental abundance ratios within a disk are of paramount importance to set the composition of the gaseous atmospheres of giant planets. Special attention should be given to the gas-phase elemental carbon-to-oxygen  elemental ratio (hereafter
C/O) which has a key role in establishing the chemistry in planetary atmospheres \citep[e.g.,][]{Oberg2011b, Piso2015, Mordasini2016, Espinoza2017, Brewer2017, Madhusudhan2017, Cridland2019}. In this regard, there is evidence that protoplanetary disks can have C/O  close to 1.0, that is, larger than the solar value (C/O $\rm \sim$ 0.7). 
\cite{Semenov2018} proposed that C/O should be $>1$ in DM Tau in order to explain the observed lower limit to the CS/SO abundance ratio. Based on ALMA observations of CO, C$_2$H, and HCN, \cite{Cleeves2018} estimated C/O$\sim$0.8
in the proto-planetary disk IM Lup. High values of C/O have also been proposed by \cite{LeGal2019} to account for the observed abundance of nitriles in proto-planetary disks. We ran models with C/O=0.7 and C/O=1 in order to investigate the 
value of C/O in AB Aur. Out of all the species observed, the abundance of HCN is very sensitive to the value of C/O, and the observed abundance is better adjusted assuming C/O = 1. Furthermore, formaldehyde (H$\rm _2$CO) and SO also favor models with C/O = 1.

Our data would also allow us to constrain the Sulfur elemental abundance in the AB Aur disk based on the SO column density. One main problem to determine S/H is that the abundance of SO strongly depends on the C/O ratio \citep[see e.g.][]{Fuente2016, Semenov2018, Fuente2019}. In fact, the SO abundance is better reproduced by models with $\rm [$S/H$\rm ]$ = 8$\times$10$^{-8}$ and C/O=0.7, pointing to a sulfur depletion of more than two orders of magnitude in the parent cloud. In fact, this model reproduces the abundances of all the observed species except HCN. However, if we adopt a C/O  of 1 to match chemical predictions with the observed HCN column densities, we need to assume the solar sulfur abundance to reproduce the observed SO column density. This would imply that the elemental sulfur abundance is close to the solar value in molecular clouds, as previously suggested by some authors \citep{Fuente2019, Navarro2020}. However, this latter model underestimates the H$_2$CO column density. One possibility is that C/O  changes along the disk, and H$_2$CO and HCN are probing regions with different physical and chemical species. High spatial resolution observations of other sulfur-bearing species such as CS, H$_2$CS, and H$_2$S, would help to determine both values (C/O and S/H) along the disk.

\section{Summary and conclusions}
The Herbig star AB Aur is a widely studied system hosting a transitional disk. The presence of spiral arms, high levels of accretion and outflow activity, and 
 a dust trap make this target an ideal candidate to study the dynamical and chemical evolutions of the gas in proto-planetary disks. This paper is part of a long-term study of which the main aim is to characterize the dust and gas of this prototypical disk \citep{Fuente2010, Pacheco2015, Pacheco2016, Fuente2017,Riviere2019}. On the basis of high-spatial-resolution NOEMA observations of  $^{12}$CO, $^{13}$CO, C$^{18}$O, H$_2$CO, and SO millimeter lines, we investigated the physical and chemical properties of the molecular gas. 
 Our results can be summarized as follows: \\

\noindent
1.- We detect intense emission of the 
 J=2$\rightarrow$1 lines of $\rm ^{12}$CO, $\rm ^{13}$CO, C$\rm ^{18}$O, and the H$_2$CO 3$\rm _{03}$-2$\rm _{02}$ and SO  5$_6$-4$_5$ lines. High-angular-resolution images (beam $<$0.9$"$) were produced for these lines. The H$_2$CO 3$\rm _{22}$-2$\rm _{21}$,  H$_2$CO 3$\rm _{21}$-2$\rm _{20}$ and SO  5$_5$-4$_4$ were detected at the 3$\times$$\sigma$ level, precluding synthesis imaging. We also use previous images of the HCO$^+$ 3-2 and HCN 3-2 published by \cite{Riviere2019} for our chemical study. \\

\noindent
2.- Our data reveal that different molecular species are probing different disk and envelope regions.  While  $^{12}$CO is tracing the disk surface and the low-density remnant envelope, the emission of the rarer isotopolog C$^{18}$O 2-1 is coming from the dusty ring, since its radial emission peak is spatially coincident with the peak of the 1.3mm continuum emission. The same is true for the emission of the HCN 3-2 line. Within the dust cavity, the gas chemistry is likely driven by UV radiation and the only detected species  is $^{12}$CO and HCO$^+$. The volatile species H$_2$CO and SO are more intense in the outer disk, with their radial emission peaking further away than the dusty ring. \\

\noindent
3.- Significant differences were found in the azimuthal distribution of the different molecules. The C$^{18}$O emission peaks close to the position of the dust trap ($\sim$269$^{\circ}$). In the  case of $^{13}$CO, we observe two peaks with similar intensity at $\sim$269$^{\circ}$ and $\sim$100$^{\circ}$ which roughly
correspond to the maximum (dust trap) and minimum ($\sim$180$^{\circ}$ away from dust trap) in the continuum emission. Most striking is the case of SO which only 
peaks at  $\sim$100$^{\circ}$ ($\sim$180$^{\circ}$ away from dust trap). The emissions of HCN, H$_2$CO, and HCO$^+$ are quite flat along the ring. \\

\noindent
4.- The detection of several lines of H$_2$CO and SO  allows us to carry out a multi-transition study. We estimate a mean rotation temperature of $\sim$ 39 K from the H$_2$CO lines. This temperature is higher than those measured with this species in T Tauri disks and HD 163296, which confirms the higher temperature of the AB Aur disk. Furthermore, this is consistent with the temperature of $\sim$37 K that we obtained using SO lines.  \\

\noindent
5.- The derived gas-to-dust ratios along the central parts of the dust ring range from $\rm \sim 10$ to $\rm \sim 40$, and reach 200 only in the inner and outer edges. As expected, the minimum gas-to-dust ratio is measured towards the dust trap. This value of the gas-to-dust ratio is in agreement with values typically found in T Tauri and Herbig Ae/Be disks. One exception would be HD 163297 which is known
to be a gas-rich disk with a gas-to-dust ratio of $\sim$100. \\

\noindent
6.- Assuming the equilibrium value of the ortho-to-para ratio at 39 K (ortho-to-para ratio = 3), we derive a mean  H$_2$CO abundance with respect to $\rm ^{13}CO$ of 3.7$\times$10$^{-10}$ which is higher by more than a factor of ten than that derived in HD 163296.  We interpret the high H$_2$CO abundance in AB Aur as a consequence of the warm chemistry in this more evolved disk.
 The derived mean SO abundance is $\sim$ 4$\times$10$^{-10}$, similar to that of H$_2$CO. \\

\noindent
7.- To further investigate the chemistry in AB Aur, we developed a dedicated 1+1D disk astrochemical model based on the model developed by \cite{LeGal2019}. Our best-fit model is obtained for a gas-to-dust ratio=40, and a depleted total amount of sulfur ($\rm [S/H]$=8$\rm \times 10^{-8}$). Interestingly, we find that an elevated C/O ratio better reproduces the abundances of  HCN, H$\rm _2$CO, and SO, in agreement with previous studies claiming that evolved disks are oxygen depleted \citep{Bergin2016, Cleeves2018, LeGal2019a, Miotello2019}. \\

Our data show that AB Aur hosts a peculiar transition disk characterized by a high gas and dust temperature and a low gas-to-dust ratio, with a large sulfur depletion. Therefore, studying the dynamics and chemistry of the gas in this disk might provide important clues for understanding the poorly known disks associated with high-mass stars. 

\begin{acknowledgements}
We thank the anonymous referee for a thorough report that helped us to improve the quality of the paper.
PRM and AF thank the Spanish MINECO for funding support from AYA2016-75066-C2-1/2-P, AYA2017-85111-P. SPTM acknowledges to the European Union's Horizon 2020 research and innovation program for funding support given under agreement No 639450 (PROMISE). MO acknowledges financial support from the State Agency for Research of the Spanish MCIU through the AYA2017-84390-C2-1-R grant (co-funded by FEDER) and through the Center of Excellence Severo Ochoa award for the Instituto de Astrof\'\i sica de Andaluc\'\i a (SEV/2017/0709).
\end{acknowledgements}

 \bibliographystyle{aa} 
\bibliography{biblio}

\begin{thebibliography}{100}
\expandafter\ifx\csname natexlab\endcsname\relax\def\natexlab#1{#1}\fi

\bibitem[{{Ag{\'u}ndez} {et~al.}(2018){Ag{\'u}ndez}, {Roueff}, {Le Petit}, \&
  {Le Bourlot}}]{Agundez2018}
{Ag{\'u}ndez}, M., {Roueff}, E., {Le Petit}, F., \& {Le Bourlot}, J. 2018,
  \aap, 616, A19

\bibitem[{{Ag{\'u}ndez} \& {Wakelam}(2013)}]{Agundez2013}
{Ag{\'u}ndez}, M. \& {Wakelam}, V. 2013, Chemical Reviews, 113, 8710

\bibitem[{{Akimkin} {et~al.}(2013){Akimkin}, {Zhukovska}, {Wiebe}, {Semenov},
  {Pavlyuchenkov}, {Vasyunin}, {Birnstiel}, \& {Henning}}]{Akimkin2013}
{Akimkin}, V., {Zhukovska}, S., {Wiebe}, D., {et~al.} 2013, \apj, 766, 8

\bibitem[{{Ansdell} {et~al.}(2016){Ansdell}, {Williams}, {van der Marel},
  {Carpenter}, {Guidi}, {Hogerheijde}, {Mathews}, {Manara}, {Miotello},
  {Natta}, {Oliveira}, {Tazzari}, {Testi}, {van Dishoeck}, \& {van
  Terwisga}}]{Ansdell2016}
{Ansdell}, M., {Williams}, J.~P., {van der Marel}, N., {et~al.} 2016, \apj,
  828, 46

\bibitem[{{Asplund} {et~al.}(2005){Asplund}, {Grevesse}, \&
  {Sauval}}]{Asplund2005}
{Asplund}, M., {Grevesse}, N., \& {Sauval}, A.~J. 2005, in Astronomical Society
  of the Pacific Conference Series, Vol. 336, Cosmic Abundances as Records of
  Stellar Evolution and Nucleosynthesis, ed. T.~G. {Barnes}, III \& F.~N.
  {Bash}, 25

\bibitem[{{Baruteau} {et~al.}(2019){Baruteau}, {Barraza}, {P{\'e}rez},
  {Casassus}, {Dong}, {Lyra}, {Marino}, {Christiaens}, {Zhu}, {Carmona},
  {Debras}, \& {Alarcon}}]{Baruteau2019}
{Baruteau}, C., {Barraza}, M., {P{\'e}rez}, S., {et~al.} 2019, \mnras, 486, 304

\bibitem[{{Bergin} {et~al.}(2016){Bergin}, {Du}, {Cleeves}, {Blake}, {Schwarz},
  {Visser}, \& {Zhang}}]{Bergin2016}
{Bergin}, E.~A., {Du}, F., {Cleeves}, L.~I., {et~al.} 2016, \apj, 831, 101

\bibitem[{{Boccaletti} {et~al.}(2020){Boccaletti}, {Di Folco}, {Pantin},
  {Dutrey}, {Guilloteau}, {Tang}, {Pi{\'e}tu}, {Habart}, {Milli}, {Beck}, \&
  {Maire}}]{Boccaletti2020}
{Boccaletti}, A., {Di Folco}, E., {Pantin}, E., {et~al.} 2020, \aap, 637, L5

\bibitem[{{Boehler} {et~al.}(2017){Boehler}, {Weaver}, {Isella}, {Ricci},
  {Grady}, {Carpenter}, \& {Perez}}]{Boehler2017}
{Boehler}, Y., {Weaver}, E., {Isella}, A., {et~al.} 2017, \apj, 840, 60

\bibitem[{{Brewer} {et~al.}(2017){Brewer}, {Fischer}, \&
  {Madhusudhan}}]{Brewer2017}
{Brewer}, J.~M., {Fischer}, D.~A., \& {Madhusudhan}, N. 2017, \aj, 153, 83

\bibitem[{{Carney} {et~al.}(2017){Carney}, {Hogerheijde}, {Loomis}, {Salinas},
  {{\"O}berg}, {Qi}, \& {Wilner}}]{Carney2017}
{Carney}, M.~T., {Hogerheijde}, M.~R., {Loomis}, R.~A., {et~al.} 2017, \aap,
  605, A21

\bibitem[{{Chapillon} {et~al.}(2008){Chapillon}, {Guilloteau}, {Dutrey}, \&
  {Pi{\'e}tu}}]{Chapillon2008}
{Chapillon}, E., {Guilloteau}, S., {Dutrey}, A., \& {Pi{\'e}tu}, V. 2008, \aap,
  488, 565

\bibitem[{{Cleeves} {et~al.}(2018){Cleeves}, {{\"O}berg}, {Wilner}, {Huang},
  {Loomis}, {Andrews}, \& {Guzman}}]{Cleeves2018}
{Cleeves}, L.~I., {{\"O}berg}, K.~I., {Wilner}, D.~J., {et~al.} 2018, \apj,
  865, 155

\bibitem[{{Cridland} {et~al.}(2019){Cridland}, {van Dishoeck}, {Alessi}, \&
  {Pudritz}}]{Cridland2019}
{Cridland}, A.~J., {van Dishoeck}, E.~F., {Alessi}, M., \& {Pudritz}, R.~E.
  2019, \aap, 632, A63

\bibitem[{{Dartois} {et~al.}(2003){Dartois}, {Dutrey}, \&
  {Guilloteau}}]{Dartois2003}
{Dartois}, E., {Dutrey}, A., \& {Guilloteau}, S. 2003, \aap, 399, 773

\bibitem[{{Dong} {et~al.}(2018){Dong}, {Liu}, {Eisner}, {Andrews}, {Fung},
  {Zhu}, {Chiang}, {Hashimoto}, {Liu}, {Casassus}, {Esposito}, {Hasegawa},
  {Muto}, {Pavlyuchenkov}, {Wilner}, {Akiyama}, {Tamura}, \&
  {Wisniewski}}]{Dong2018}
{Dong}, R., {Liu}, S.-y., {Eisner}, J., {et~al.} 2018, \apj, 860, 124

\bibitem[{{Drozdovskaya} {et~al.}(2018){Drozdovskaya}, {van Dishoeck},
  {J{\o}rgensen}, {Calmonte}, {van der Wiel}, {Coutens}, {Calcutt},
  {M{\"u}ller}, {Bjerkeli}, {Persson}, {Wampfler}, \&
  {Altwegg}}]{Drozdovskaya2018}
{Drozdovskaya}, M.~N., {van Dishoeck}, E.~F., {J{\o}rgensen}, J.~K., {et~al.}
  2018, \mnras, 476, 4949

\bibitem[{{Dullemond} \& {Dominik}(2004)}]{Dullemond2004}
{Dullemond}, C.~P. \& {Dominik}, C. 2004, \aap, 417, 159

\bibitem[{{Espinoza} {et~al.}(2017){Espinoza}, {Fortney}, {Miguel},
  {Thorngren}, \& {Murray-Clay}}]{Espinoza2017}
{Espinoza}, N., {Fortney}, J.~J., {Miguel}, Y., {Thorngren}, D., \&
  {Murray-Clay}, R. 2017, \apjl, 838, L9

\bibitem[{{Facchini} {et~al.}(2017){Facchini}, {Birnstiel}, {Bruderer}, \& {van
  Dishoeck}}]{Facchini2017}
{Facchini}, S., {Birnstiel}, T., {Bruderer}, S., \& {van Dishoeck}, E.~F. 2017,
  \aap, 605, A16

\bibitem[{{Fayolle} {et~al.}(2016){Fayolle}, {Balfe}, {Loomis}, {Bergner},
  {Graninger}, {Rajappan}, \& {{\"O}berg}}]{Fayolle2016}
{Fayolle}, E.~C., {Balfe}, J., {Loomis}, R., {et~al.} 2016, \apjl, 816, L28

\bibitem[{{Frerking} {et~al.}(1982){Frerking}, {Langer}, \&
  {Wilson}}]{Frerking1982}
{Frerking}, M.~A., {Langer}, W.~D., \& {Wilson}, R.~W. 1982, \apj, 262, 590

\bibitem[{{Frerking} {et~al.}(1987){Frerking}, {Langer}, \&
  {Wilson}}]{Frerking1987}
{Frerking}, M.~A., {Langer}, W.~D., \& {Wilson}, R.~W. 1987, \apj, 313, 320

\bibitem[{{Fuente} {et~al.}(2017){Fuente}, {Baruteau}, {Neri}, {Carmona},
  {Ag{\'u}ndez}, {Goicoechea}, {Bachiller}, {Cernicharo}, \&
  {Bern{\'e}}}]{Fuente2017}
{Fuente}, A., {Baruteau}, C., {Neri}, R., {et~al.} 2017, \apjl, 846, L3

\bibitem[{{Fuente} {et~al.}(2010){Fuente}, {Cernicharo}, {Ag{\'u}ndez},
  {Bern{\'e}}, {Goicoechea}, {Alonso-Albi}, \& {Marcelino}}]{Fuente2010}
{Fuente}, A., {Cernicharo}, J., {Ag{\'u}ndez}, M., {et~al.} 2010, \aap, 524,
  A19

\bibitem[{{Fuente} {et~al.}(2016){Fuente}, {Cernicharo}, {Roueff}, {Gerin},
  {Pety}, {Marcelino}, {Bachiller}, {Lefloch}, {Roncero}, \&
  {Aguado}}]{Fuente2016}
{Fuente}, A., {Cernicharo}, J., {Roueff}, E., {et~al.} 2016, \aap, 593, A94

\bibitem[{{Fuente} {et~al.}(2019){Fuente}, {Navarro}, {Caselli}, {Gerin},
  {Kramer}, {Roueff}, {Alonso-Albi}, {Bachiller}, {Cazaux}, {Commercon},
  {Friesen}, {Garc{\'\i}a-Burillo}, {Giuliano}, {Goicoechea}, {Gratier},
  {Hacar}, {Jim{\'e}nez-Serra}, {Kirk}, {Lattanzi}, {Loison}, {Malinen},
  {Marcelino}, {Mart{\'\i}n-Dom{\'e}nech}, {Mu{\~n}oz-Caro}, {Pineda},
  {Tafalla}, {Tercero}, {Ward-Thompson}, {Trevi{\~n}o-Morales},
  {Rivi{\'e}re-Marichalar}, {Roncero}, {Vidal}, \& {Ballester}}]{Fuente2019}
{Fuente}, A., {Navarro}, D.~G., {Caselli}, P., {et~al.} 2019, \aap, 624, A105

\bibitem[{{Fuente} {et~al.}(2003){Fuente}, {Rodr{\i}guez-Franco},
  {Garc{\i}a-Burillo}, {Mart{\i}n-Pintado}, \& {Black}}]{Fuente2003}
{Fuente}, A., {Rodr{\i}guez-Franco}, A., {Garc{\i}a-Burillo}, S.,
  {Mart{\i}n-Pintado}, J., \& {Black}, J.~H. 2003, \aap, 406, 899

\bibitem[{{Fukagawa} {et~al.}(2004){Fukagawa}, {Hayashi}, {Tamura}, {Itoh},
  {Hayashi}, {Oasa}, {Takeuchi}, {Morino}, {Murakawa}, {Oya}, {Yamashita},
  {Suto}, {Mayama}, {Naoi}, {Ishii}, {Pyo}, {Nishikawa}, {Takato}, {Usuda},
  {Ando}, {Iye}, {Miyama}, \& {Kaifu}}]{Fukagawa2004}
{Fukagawa}, M., {Hayashi}, M., {Tamura}, M., {et~al.} 2004, \apjl, 605, L53

\bibitem[{{Gaia Collaboration}(2018)}]{GAIA2018}
{Gaia Collaboration}. 2018, VizieR Online Data Catalog, I/345

\bibitem[{{Garcia Lopez} {et~al.}(2006){Garcia Lopez}, {Natta}, {Testi}, \&
  {Habart}}]{GarciaLopez2006}
{Garcia Lopez}, R., {Natta}, A., {Testi}, L., \& {Habart}, E. 2006, \aap, 459,
  837

\bibitem[{{Garrod} \& {Herbst}(2006)}]{Garrod2006}
{Garrod}, R.~T. \& {Herbst}, E. 2006, \aap, 457, 927

\bibitem[{{Grady} {et~al.}(1999){Grady}, {Woodgate}, {Bruhweiler}, {Boggess},
  {Plait}, {Lindler}, {Clampin}, \& {Kalas}}]{Grady1999}
{Grady}, C.~A., {Woodgate}, B., {Bruhweiler}, F.~C., {et~al.} 1999, \apjl, 523,
  L151

\bibitem[{{Graedel} {et~al.}(1982){Graedel}, {Langer}, \&
  {Frerking}}]{Graedel1982}
{Graedel}, T.~E., {Langer}, W.~D., \& {Frerking}, M.~A. 1982, \apjs, 48, 321

\bibitem[{{Guilloteau} {et~al.}(2016){Guilloteau}, {Reboussin}, {Dutrey},
  {Chapillon}, {Wakelam}, {Pi{\'e}tu}, {Di Folco}, {Semenov}, \&
  {Henning}}]{Guilloteau2016}
{Guilloteau}, S., {Reboussin}, L., {Dutrey}, A., {et~al.} 2016, \aap, 592, A124

\bibitem[{{Guzm{\'a}n} {et~al.}(2018){Guzm{\'a}n}, {{\"O}berg}, {Carpenter},
  {Le Gal}, {Qi}, \& {Pagues}}]{Guzman2018}
{Guzm{\'a}n}, V.~V., {{\"O}berg}, K.~I., {Carpenter}, J., {et~al.} 2018, \apj,
  864, 170

\bibitem[{{Hashimoto} {et~al.}(2011){Hashimoto}, {Tamura}, {Muto}, {Kudo},
  {Fukagawa}, {Fukue}, {Goto}, {Grady}, {Henning}, {Hodapp}, {Honda},
  {Inutsuka}, {Kokubo}, {Knapp}, {McElwain}, {Momose}, {Ohashi}, {Okamoto},
  {Takami}, {Turner}, {Wisniewski}, {Janson}, {Abe}, {Brandner}, {Carson},
  {Egner}, {Feldt}, {Golota}, {Guyon}, {Hayano}, {Hayashi}, {Hayashi}, {Ishii},
  {Kandori}, {Kusakabe}, {Matsuo}, {Mayama}, {Miyama}, {Morino}, {Moro-Martin},
  {Nishimura}, {Pyo}, {Suto}, {Suzuki}, {Takato}, {Terada}, {Thalmann},
  {Tomono}, {Watanabe}, {Yamada}, {Takami}, \& {Usuda}}]{Hashimoto2011}
{Hashimoto}, J., {Tamura}, M., {Muto}, T., {et~al.} 2011, \apjl, 729, L17

\bibitem[{{Henning} {et~al.}(2010){Henning}, {Semenov}, {Guilloteau}, {Dutrey},
  {Hersant}, {Wakelam}, {Chapillon}, {Launhardt}, {Pi{\'e}tu}, \&
  {Schreyer}}]{Henning2010}
{Henning}, T., {Semenov}, D., {Guilloteau}, S., {et~al.} 2010, \apj, 714, 1511

\bibitem[{{Hincelin} {et~al.}(2011){Hincelin}, {Wakelam}, {Hersant},
  {Guilloteau}, {Loison}, {Honvault}, \& {Troe}}]{Hincelin2011}
{Hincelin}, U., {Wakelam}, V., {Hersant}, F., {et~al.} 2011, \aap, 530, A61

\bibitem[{{Hollenbach} {et~al.}(2009){Hollenbach}, {Kaufman}, {Bergin}, \&
  {Melnick}}]{Hollenbach2009}
{Hollenbach}, D., {Kaufman}, M.~J., {Bergin}, E.~A., \& {Melnick}, G.~J. 2009,
  \apj, 690, 1497

\bibitem[{{Jenkins}(2009)}]{Jenkins2009}
{Jenkins}, E.~B. 2009, \apj, 700, 1299

\bibitem[{{Kamp} \& {Dullemond}(2004)}]{KampDullemond2004}
{Kamp}, I. \& {Dullemond}, C.~P. 2004, \apj, 615, 991

\bibitem[{{Laas} \& {Caselli}(2019)}]{Laas2019}
{Laas}, J.~C. \& {Caselli}, P. 2019, \aap, 624, A108

\bibitem[{{Le Gal} {et~al.}(2019{\natexlab{a}}){Le Gal}, {Brady}, {{\"O}berg},
  {Roueff}, \& {Le Petit}}]{LeGal2019a}
{Le Gal}, R., {Brady}, M.~T., {{\"O}berg}, K.~I., {Roueff}, E., \& {Le Petit},
  F. 2019{\natexlab{a}}, \apj, 886, 86

\bibitem[{{Le Gal} {et~al.}(2019{\natexlab{b}}){Le Gal}, {{\"O}berg}, {Loomis},
  {Pegues}, \& {Bergner}}]{LeGal2019}
{Le Gal}, R., {{\"O}berg}, K.~I., {Loomis}, R.~A., {Pegues}, J., \& {Bergner},
  J.~B. 2019{\natexlab{b}}, \apj, 876, 72

\bibitem[{{Long} {et~al.}(2017){Long}, {Herczeg}, {Pascucci}, {Drabek-Maunder},
  {Mohanty}, {Testi}, {Apai}, {Hendler}, {Henning}, {Manara}, \&
  {Mulders}}]{Long2017}
{Long}, F., {Herczeg}, G.~J., {Pascucci}, I., {et~al.} 2017, \apj, 844, 99

\bibitem[{{Loomis} {et~al.}(2020){Loomis}, {{\"O}berg}, {Andrews}, {Bergin},
  {Bergner}, {Blake}, {Cleeves}, {Czekala}, {Huang}, {Le Gal}, {M{\'e}nard},
  {Pegues}, {Qi}, {Walsh}, {Williams}, \& {Wilner}}]{Loomis2020}
{Loomis}, R.~A., {{\"O}berg}, K.~I., {Andrews}, S.~M., {et~al.} 2020, \apj,
  893, 101

\bibitem[{{Madhusudhan} {et~al.}(2017){Madhusudhan}, {Bitsch}, {Johansen}, \&
  {Eriksson}}]{Madhusudhan2017}
{Madhusudhan}, N., {Bitsch}, B., {Johansen}, A., \& {Eriksson}, L. 2017,
  \mnras, 469, 4102

\bibitem[{{Maret} {et~al.}(2020){Maret}, {Maury}, {Belloche}, {Gaudel},
  {Andr{\'e}}, {Cabrit}, {Codella}, {Lef{\`e}vre}, {Podio}, {Anderl}, {Gueth},
  \& {Hennebelle}}]{Maret2020}
{Maret}, S., {Maury}, A.~J., {Belloche}, A., {et~al.} 2020, arXiv e-prints,
  arXiv:2001.06355

\bibitem[{{Mart{\'\i}n-Dom{\'e}nech} {et~al.}(2014){Mart{\'\i}n-Dom{\'e}nech},
  {Mu{\~n}oz Caro}, {Bueno}, \& {Goesmann}}]{MartinDomenech2014}
{Mart{\'\i}n-Dom{\'e}nech}, R., {Mu{\~n}oz Caro}, G.~M., {Bueno}, J., \&
  {Goesmann}, F. 2014, \aap, 564, A8

\bibitem[{{Miley} {et~al.}(2019){Miley}, {Pani{\'c}}, {Haworth}, {Pascucci},
  {Wyatt}, {Clarke}, {Richards}, \& {Ratzka}}]{Miley2019}
{Miley}, J.~M., {Pani{\'c}}, O., {Haworth}, T.~J., {et~al.} 2019, \mnras, 485,
  739

\bibitem[{{Miotello} {et~al.}(2019){Miotello}, {Facchini}, {van Dishoeck},
  {Cazzoletti}, {Testi}, {Williams}, {Ansdell}, {van Terwisga}, \& {van der
  Marel}}]{Miotello2019}
{Miotello}, A., {Facchini}, S., {van Dishoeck}, E.~F., {et~al.} 2019, \aap,
  631, A69

\bibitem[{{Miotello} {et~al.}(2017){Miotello}, {van Dishoeck}, {Williams},
  {Ansdell}, {Guidi}, {Hogerheijde}, {Manara}, {Tazzari}, {Testi}, {van der
  Marel}, \& {van Terwisga}}]{Miotello2017}
{Miotello}, A., {van Dishoeck}, E.~F., {Williams}, J.~P., {et~al.} 2017, \aap,
  599, A113

\bibitem[{{Mordasini} {et~al.}(2016){Mordasini}, {van Boekel}, {Molli{\`e}re},
  {Henning}, \& {Benneke}}]{Mordasini2016}
{Mordasini}, C., {van Boekel}, R., {Molli{\`e}re}, P., {Henning}, T., \&
  {Benneke}, B. 2016, \apj, 832, 41

\bibitem[{{Navarro-Almaida} {et~al.}(2020){Navarro-Almaida}, {Le Gal, R.},
  {Fuente, A.}, {Rivi\`ere-Marichalar, P.}, {Wakelam, V.}, {Cazaux, S.},
  {Caselli, P.}, {Laas, J. C.}, {Alonso-Albi, T.}, {Loison, J. C.}, {Gerin,
  M.}, {Kramer, C.}, {Roueff, E.}, {Bachiller, R.}, {Commer\c{c}on, B.},
  {Friesen, R.}, {Garc\'{\i}a-Burillo, S.}, {Goicoechea, J. R.}, {Giuliano, B.
  M.}, {Jim\'enez-Serra, I.}, {Kirk, J. M.}, {Lattanzi, V.}, {Malinen, J.},
  {Marcelino, N.}, {Mart\'{\i}n-Dom\`enech, R.}, {Mu\~noz Caro, G. M.},
  {Pineda, J.}, {Tercero, B.}, {Trevi\~no-Morales, S. P.}, {Roncero, O.},
  {Hacar, A.}, {Tafalla, M.}, \& {Ward-Thompson, D.}}]{Navarro2020}
{Navarro-Almaida}, D., {Le Gal, R.}, {Fuente, A.}, {et~al.} 2020, A\&A, 637,
  A39

\bibitem[{{Neufeld} {et~al.}(2015){Neufeld}, {Godard}, {Gerin}, {Pineau des
  For{\^e}ts}, {Bernier}, {Falgarone}, {Graf}, {G{\"u}sten}, {Herbst},
  {Lesaffre}, {Schilke}, {Sonnentrucker}, \& {Wiesemeyer}}]{Neufeld2015}
{Neufeld}, D.~A., {Godard}, B., {Gerin}, M., {et~al.} 2015, \aap, 577, A49

\bibitem[{{Noble} {et~al.}(2012){Noble}, {Theule}, {Mispelaer}, {Duvernay},
  {Danger}, {Congiu}, {Dulieu}, \& {Chiavassa}}]{Noble2012}
{Noble}, J.~A., {Theule}, P., {Mispelaer}, F., {et~al.} 2012, \aap, 543, A5

\bibitem[{{{\"O}berg} {et~al.}(2017){{\"O}berg}, {Guzm{\'a}n}, {Merchantz},
  {Qi}, {Andrews}, {Cleeves}, {Huang}, {Loomis}, {Wilner}, {Brinch}, \&
  {Hogerheijde}}]{Oberg2017}
{{\"O}berg}, K.~I., {Guzm{\'a}n}, V.~V., {Merchantz}, C.~J., {et~al.} 2017,
  \apj, 839, 43

\bibitem[{{{\"O}berg} {et~al.}(2011{\natexlab{a}}){{\"O}berg}, {Murray-Clay},
  \& {Bergin}}]{Oberg2011b}
{{\"O}berg}, K.~I., {Murray-Clay}, R., \& {Bergin}, E.~A. 2011{\natexlab{a}},
  \apjl, 743, L16

\bibitem[{{{\"O}berg} {et~al.}(2010){{\"O}berg}, {Qi}, {Fogel}, {Bergin},
  {Andrews}, {Espaillat}, {van Kempen}, {Wilner}, \& {Pascucci}}]{Oberg2010}
{{\"O}berg}, K.~I., {Qi}, C., {Fogel}, J. K.~J., {et~al.} 2010, \apj, 720, 480

\bibitem[{{{\"O}berg} {et~al.}(2011{\natexlab{b}}){{\"O}berg}, {Qi}, {Fogel},
  {Bergin}, {Andrews}, {Espaillat}, {Wilner}, {Pascucci}, \&
  {Kastner}}]{Oberg2011}
{{\"O}berg}, K.~I., {Qi}, C., {Fogel}, J. K.~J., {et~al.} 2011{\natexlab{b}},
  \apj, 734, 98

\bibitem[{{Osorio} {et~al.}(2014){Osorio}, {Anglada}, {Carrasco-Gonz{\'a}lez},
  {Torrelles}, {Mac{\'\i}as}, {Rodr{\'\i}guez}, {G{\'o}mez}, {D'Alessio},
  {Calvet}, {Nagel}, {Dent}, {Quanz}, {Reggiani}, \&
  {Mayen-Gijon}}]{Osorio2014}
{Osorio}, M., {Anglada}, G., {Carrasco-Gonz{\'a}lez}, C., {et~al.} 2014, \apjl,
  791, L36

\bibitem[{{Ossenkopf} \& {Henning}(1994)}]{Ossenkopf1994}
{Ossenkopf}, V. \& {Henning}, T. 1994, \aap, 291, 943

\bibitem[{{Pacheco-V{\'a}zquez} {et~al.}(2015){Pacheco-V{\'a}zquez}, {Fuente},
  {Ag{\'u}ndez}, {Pinte}, {Alonso-Albi}, {Neri}, {Cernicharo}, {Goicoechea},
  {Bern{\'e}}, {Wiesenfeld}, {Bachiller}, \& {Lefloch}}]{Pacheco2015}
{Pacheco-V{\'a}zquez}, S., {Fuente}, A., {Ag{\'u}ndez}, M., {et~al.} 2015,
  \aap, 578, A81

\bibitem[{{Pacheco-V{\'a}zquez} {et~al.}(2016){Pacheco-V{\'a}zquez}, {Fuente},
  {Baruteau}, {Bern{\'e}}, {Ag{\'u}ndez}, {Neri}, {Goicoechea}, {Cernicharo},
  \& {Bachiller}}]{Pacheco2016}
{Pacheco-V{\'a}zquez}, S., {Fuente}, A., {Baruteau}, C., {et~al.} 2016, \aap,
  589, A60

\bibitem[{{Pegues} {et~al.}(2020){Pegues}, {{\"O}berg}, {Bergner}, {Loomis},
  {Qi}, {Le Gal}, {Cleeves}, {Guzm{\'a}n}, {Huang}, {J{\o}rgensen}, {Andrews},
  {Blake}, {Carpenter}, {Schwarz}, {Williams}, \& {Wilner}}]{Pegues2020}
{Pegues}, J., {{\"O}berg}, K.~I., {Bergner}, J.~B., {et~al.} 2020, \apj, 890,
  142

\bibitem[{{Phuong} {et~al.}(2018){Phuong}, {Chapillon}, {Majumdar}, {Dutrey},
  {Guilloteau}, {Pi{\'e}tu}, {Wakelam}, {Diep}, {Tang}, {Beck}, \&
  {Bary}}]{Phuong2018}
{Phuong}, N.~T., {Chapillon}, E., {Majumdar}, L., {et~al.} 2018, \aap, 616, L5

\bibitem[{{Pi{\'e}tu} {et~al.}(2005){Pi{\'e}tu}, {Guilloteau}, \&
  {Dutrey}}]{Pietu2005}
{Pi{\'e}tu}, V., {Guilloteau}, S., \& {Dutrey}, A. 2005, \aap, 443, 945

\bibitem[{{Piso} {et~al.}(2015){Piso}, {{\"O}berg}, {Birnstiel}, \&
  {Murray-Clay}}]{Piso2015}
{Piso}, A.-M.~A., {{\"O}berg}, K.~I., {Birnstiel}, T., \& {Murray-Clay}, R.~A.
  2015, \apj, 815, 109

\bibitem[{{Podio} {et~al.}(2015){Podio}, {Codella}, {Gueth}, {Cabrit},
  {Bachiller}, {Gusdorf}, {Lee}, {Lefloch}, {Leurini}, {Nisini}, \&
  {Tafalla}}]{Podio2015}
{Podio}, L., {Codella}, C., {Gueth}, F., {et~al.} 2015, \aap, 581, A85

\bibitem[{{Qi} {et~al.}(2013){Qi}, {{\"O}berg}, \& {Wilner}}]{Qi2013}
{Qi}, C., {{\"O}berg}, K.~I., \& {Wilner}, D.~J. 2013, \apj, 765, 34

\bibitem[{{Rivi{\`e}re-Marichalar} {et~al.}(2019){Rivi{\`e}re-Marichalar},
  {Fuente}, {Baruteau}, {Neri}, {Trevi{\~n}o-Morales}, {Carmona},
  {Ag{\'u}ndez}, \& {Bachiller}}]{Riviere2019}
{Rivi{\`e}re-Marichalar}, P., {Fuente}, A., {Baruteau}, C., {et~al.} 2019,
  \apjl, 879, L14

\bibitem[{{Rodr{\'\i}guez} {et~al.}(2014){Rodr{\'\i}guez}, {Zapata}, {Dzib},
  {Ortiz-Le{\'o}n}, {Loinard}, {Mac{\'\i}as}, \& {Anglada}}]{Rodriguez2014}
{Rodr{\'\i}guez}, L.~F., {Zapata}, L.~A., {Dzib}, S.~A., {et~al.} 2014, \apjl,
  793, L21

\bibitem[{{Rosenfeld} {et~al.}(2013){Rosenfeld}, {Andrews}, {Wilner},
  {Kastner}, \& {McClure}}]{Rosenfeld2013}
{Rosenfeld}, K.~A., {Andrews}, S.~M., {Wilner}, D.~J., {Kastner}, J.~H., \&
  {McClure}, M.~K. 2013, \apj, 775, 136

\bibitem[{{Ruaud} {et~al.}(2016){Ruaud}, {Wakelam}, \& {Hersant}}]{Ruaud2016}
{Ruaud}, M., {Wakelam}, V., \& {Hersant}, F. 2016, \mnras, 459, 3756

\bibitem[{{Sakai} {et~al.}(2016){Sakai}, {Oya}, {L{\'o}pez-Sepulcre},
  {Watanabe}, {Sakai}, {Hirota}, {Aikawa}, {Ceccarelli}, {Lefloch}, {Caux},
  {Vastel}, {Kahane}, \& {Yamamoto}}]{Sakai2016}
{Sakai}, N., {Oya}, Y., {L{\'o}pez-Sepulcre}, A., {et~al.} 2016, \apjl, 820,
  L34

\bibitem[{{Sakai} {et~al.}(2014){Sakai}, {Oya}, {Sakai}, {Watanabe}, {Hirota},
  {Ceccarelli}, {Kahane}, {Lopez-Sepulcre}, {Lefloch}, {Vastel}, {Bottinelli},
  {Caux}, {Coutens}, {Aikawa}, {Takakuwa}, {Ohashi}, {Yen}, \&
  {Yamamoto}}]{Sakai2014}
{Sakai}, N., {Oya}, Y., {Sakai}, T., {et~al.} 2014, \apjl, 791, L38

\bibitem[{{Salyk} {et~al.}(2013){Salyk}, {Herczeg}, {Brown}, {Blake},
  {Pontoppidan}, \& {van Dishoeck}}]{Salyk2013}
{Salyk}, C., {Herczeg}, G.~J., {Brown}, J.~M., {et~al.} 2013, \apj, 769, 21

\bibitem[{{Savage} {et~al.}(2002){Savage}, {Apponi}, {Ziurys}, \&
  {Wyckoff}}]{Savage2002}
{Savage}, C., {Apponi}, A.~J., {Ziurys}, L.~M., \& {Wyckoff}, S. 2002, \apj,
  578, 211

\bibitem[{{Schreyer} {et~al.}(2008){Schreyer}, {Guilloteau}, {Semenov},
  {Bacmann}, {Chapillon}, {Dutrey}, {Gueth}, {Henning}, {Hersant}, {Launhardt},
  {Pety}, \& {Pi{\'e}tu}}]{Schreyer2008}
{Schreyer}, K., {Guilloteau}, S., {Semenov}, D., {et~al.} 2008, \aap, 491, 821

\bibitem[{{Semenov} {et~al.}(2018){Semenov}, {Favre}, {Fedele}, {Guilloteau},
  {Teague}, {Henning}, {Dutrey}, {Chapillon}, {Hersant}, \&
  {Pi{\'e}tu}}]{Semenov2018}
{Semenov}, D., {Favre}, C., {Fedele}, D., {et~al.} 2018, \aap, 617, A28

\bibitem[{{Semenov} {et~al.}(2005){Semenov}, {Pavlyuchenkov}, {Schreyer},
  {Henning}, {Dullemond}, \& {Bacmann}}]{Semenov2005}
{Semenov}, D., {Pavlyuchenkov}, Y., {Schreyer}, K., {et~al.} 2005, \apj, 621,
  853

\bibitem[{{Soon} {et~al.}(2019){Soon}, {Momose}, {Muto}, {Tsukagoshi},
  {Kataoka}, {Hanawa}, {Fukagawa}, {Saigo}, \& {Shibai}}]{Soon2019}
{Soon}, K.-L., {Momose}, M., {Muto}, T., {et~al.} 2019, \pasj, 111

\bibitem[{{Sternberg} \& {Dalgarno}(1995)}]{Sternberg1995}
{Sternberg}, A. \& {Dalgarno}, A. 1995, \apjs, 99, 565

\bibitem[{{Tang} {et~al.}(2018){Tang}, {Henkel}, {Menten}, {Wyrowski},
  {Brinkmann}, {Zheng}, {Gong}, {Lin}, {Esimbek}, {Zhou}, {Yuan}, {Li}, \&
  {He}}]{Tang2018}
{Tang}, X.~D., {Henkel}, C., {Menten}, K.~M., {et~al.} 2018, \aap, 609, A16

\bibitem[{{Tang} {et~al.}(2017){Tang}, {Guilloteau}, {Dutrey}, {Muto}, {Shen},
  {Gu}, {Inutsuka}, {Momose}, {Pietu}, {Fukagawa}, {Chapillon}, {Ho}, {di
  Folco}, {Corder}, {Ohashi}, \& {Hashimoto}}]{Tang2017}
{Tang}, Y.-W., {Guilloteau}, S., {Dutrey}, A., {et~al.} 2017, \apj, 840, 32

\bibitem[{{Tang} {et~al.}(2012){Tang}, {Guilloteau}, {Pi{\'e}tu}, {Dutrey},
  {Ohashi}, \& {Ho}}]{Tang2012}
{Tang}, Y.~W., {Guilloteau}, S., {Pi{\'e}tu}, V., {et~al.} 2012, \aap, 547, A84

\bibitem[{{Trevi{\~n}o-Morales} {et~al.}(2019){Trevi{\~n}o-Morales}, {Fuente},
  {S{\'a}nchez-Monge}, {Kainulainen}, {Didelon}, {Suri}, {Schneider},
  {Ballesteros-Paredes}, {Lee}, {Hennebelle}, {Pilleri},
  {Gonz{\'a}lez-Garc{\'\i}a}, {Kramer}, {Garc{\'\i}a-Burillo}, {Luna},
  {Goicoechea}, {Tremblin}, \& {Geen}}]{Trevino2019}
{Trevi{\~n}o-Morales}, S.~P., {Fuente}, A., {S{\'a}nchez-Monge}, {\'A}.,
  {et~al.} 2019, \aap, 629, A81

\bibitem[{{van der Plas} {et~al.}(2014){van der Plas}, {Casassus},
  {M{\'e}nard}, {Perez}, {Thi}, {Pinte}, \& {Christiaens}}]{vanDerPlas2014}
{van der Plas}, G., {Casassus}, S., {M{\'e}nard}, F., {et~al.} 2014, \apjl,
  792, L25

\bibitem[{{Vastel} {et~al.}(2018){Vastel}, {Qu{\'e}nard}, {Le Gal}, {Wakelam},
  {Andrianasolo}, {Caselli}, {Vidal}, {Ceccarelli}, {Lefloch}, \&
  {Bachiller}}]{Vastel2018}
{Vastel}, C., {Qu{\'e}nard}, D., {Le Gal}, R., {et~al.} 2018, \mnras, 478, 5514

\bibitem[{{Vidal} {et~al.}(2017){Vidal}, {Loison}, {Jaziri}, {Ruaud},
  {Gratier}, \& {Wakelam}}]{Vidal2017}
{Vidal}, T. H.~G., {Loison}, J.-C., {Jaziri}, A.~Y., {et~al.} 2017, \mnras,
  469, 435

\bibitem[{{Wakelam} \& {Herbst}(2008)}]{Wakelam2008}
{Wakelam}, V. \& {Herbst}, E. 2008, \apj, 680, 371

\bibitem[{{Wakelam} {et~al.}(2016){Wakelam}, {Ruaud}, {Hersant}, {Dutrey},
  {Semenov}, {Majumdar}, \& {Guilloteau}}]{Wakelam2016}
{Wakelam}, V., {Ruaud}, M., {Hersant}, F., {et~al.} 2016, \aap, 594, A35

\bibitem[{{Walsh} {et~al.}(2010){Walsh}, {Millar}, \& {Nomura}}]{Walsh2010}
{Walsh}, C., {Millar}, T.~J., \& {Nomura}, H. 2010, \apj, 722, 1607

\bibitem[{{Williams} \& {Best}(2014)}]{Williams2014}
{Williams}, J.~P. \& {Best}, W. M.~J. 2014, \apj, 788, 59

\bibitem[{{Wilson} \& {Rood}(1994)}]{Wilson1994}
{Wilson}, T.~L. \& {Rood}, R. 1994, \araa, 32, 191

\bibitem[{{Woitke} {et~al.}(2019){Woitke}, {Kamp}, {Antonellini}, {Anthonioz},
  {Baldovin-Saveedra}, {Carmona}, {Dionatos}, {Dominik}, {Greaves},
  {G{\"u}del}, {Ilee}, {Liebhardt}, {Menard}, {Min}, {Pinte}, {Rab}, {Rigon},
  {Thi}, {Thureau}, \& {Waters}}]{Woitke2019}
{Woitke}, P., {Kamp}, I., {Antonellini}, S., {et~al.} 2019, \pasp, 131, 064301

\bibitem[{{Woitke} {et~al.}(2009){Woitke}, {Kamp}, \& {Thi}}]{Woitke2009}
{Woitke}, P., {Kamp}, I., \& {Thi}, W.~F. 2009, \aap, 501, 383

\bibitem[{{Wu} {et~al.}(2018){Wu}, {Hirano}, {Takakuwa}, {Yen}, \&
  {Aso}}]{Wu2018}
{Wu}, C.-J., {Hirano}, N., {Takakuwa}, S., {Yen}, H.-W., \& {Aso}, Y. 2018,
  \apj, 869, 59

\bibitem[{{Ysard} {et~al.}(2019){Ysard}, {Koehler}, {Jimenez-Serra}, {Jones},
  \& {Verstraete}}]{Ysard2019}
{Ysard}, N., {Koehler}, M., {Jimenez-Serra}, I., {Jones}, A.~P., \&
  {Verstraete}, L. 2019, \aap, 631, A88

\end{thebibliography}

\end{document}